\documentclass[
reprint,
amsmath,
amssymb,
aip,
superscriptaddress,
twocolumn
]{revtex4-2}

\usepackage[colorlinks, breaklinks=true, linkcolor=blue, citecolor=blue, linktocpage=true]{hyperref}
\usepackage{graphicx}
\usepackage{dcolumn}
\usepackage{bm}
\usepackage[utf8]{inputenc}
\usepackage[T1]{fontenc}
\usepackage{mathptmx}

\begin{document}

\title{Axion electrodynamics in topological materials}

\author{Akihiko Sekine}
\email{akihiko.sekine@riken.jp}
\affiliation{RIKEN Center for Emergent Matter Science, Wako, Saitama 351-0198, Japan}

\author{Kentaro Nomura}
\affiliation{Institute for Materials Research, Tohoku University, Sendai 980-8577, Japan}

\date{\today}

\begin{abstract}
One of the intriguing properties characteristic to three-dimensional topological materials is the topological magnetoelectric phenomena arising from a topological term called the $\theta$ term.
Such magnetoelectric phenomena are often termed the axion electrodynamics, since the $\theta$ term has exactly the same form as the action describing the coupling between a hypothetical elementary particle, axion, and a photon.
The axion was proposed about forty years ago to solve the so-called strong CP problem in quantum chromodynamics, and is now considered as a candidate for dark matter.
In this tutorial, we overview theoretical and experimental studies on the axion electrodynamics in three-dimensional topological materials.
Starting from the topological magnetoelectric effect in three-dimensional time-reversal invariant topological insulators, we describe the basic properties of static and dynamical axion insulators whose realizations require magnetic orderings. We also discuss the electromagnetic responses of Weyl semimetals with a focus on the chiral anomaly.
We extend the concept of the axion electrodynamics in condensed matter to topological superconductors, whose responses to external fields can be described by a gravitational topological term analogous to the $\theta$ term.
\\
\end{abstract}

\maketitle

\tableofcontents


\section{Introduction}
Conventionally, metals and insulators have been distinguished by the existence of band gaps.
In 2005, a novel phase of matter that does not belong to either conventional metals or insulators, called the topological insulator, was discovered \cite{Kane2005,Kane2005a,Hasan2009,Qi2011,Ando2013}.
It is notable that topological insulators have bulk band gaps but also have gapless boundary (edge or surface) states.
Furthermore, a topological insulator phase and a trivial insulator phase cannot be connected adiabatically to each other.
In other words, bulk band-gap closing is required for the transitions between topologically nontrivial and trivial phases.
In addition, before the establishment of the concept of topological insulators, different phases of matter had usually been distinguished from each other by the order parameters which indicate spontaneous symmetry breaking.
For example, magnetism can be understood as a consequence of spontaneous spin rotational symmetry breaking.
However, from the viewpoint of symmetry analysis, time-reversal invariant topological insulators and time-reversal invariant band insulators cannot be distinguished.
The ways to distinguish such topologically nontrivial and trivial insulator phases can be divided into two types (which of course give rise to equivalent results).
One way is introducing a ``topological invariant'' such as $\mathbb{Z}_2$ invariant \cite{Kane2005,Moore2007,Fu2007,Fu2007a}, which are calculated from the Bloch-state wave function of the system.
The other way is the ``topological field theory'' \cite{Qi2008}, which describes the responses of topological phases to external fields  and is the focus of this tutorial.

\begin{figure*}[!t]
\centering
\includegraphics[width=2\columnwidth]{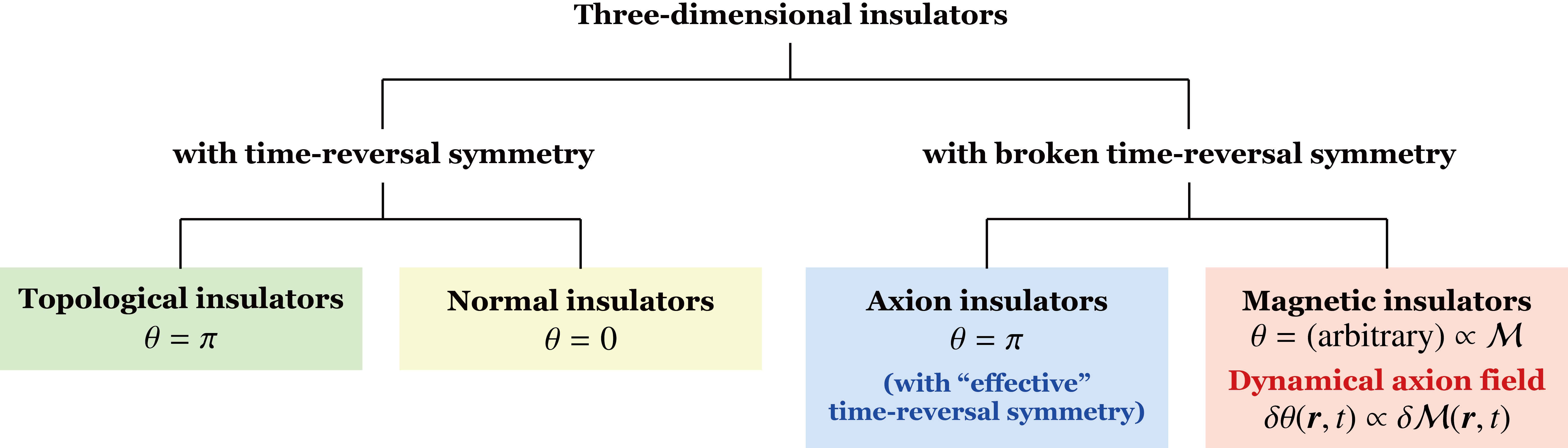}
\caption{Schematic of a classification of 3D insulators in terms of time-reversal symmetry and the orbital magnetoelectric coupling coefficient $\theta$.
In the first classification process, 3D insulators are divided into two types: insulators with or without time-reversal symmetry.
In the second classification process, 3D insulators with time-reversal symmetry are divided into types: topological insulators and normal (trivial) insulators.
Topological insulators are characterized by the topological magnetoelectric effect with the quantized coefficient $\theta=\pi$ (mod $2\pi$).
In the second classification process, 3D insulators with broken time-reversal symmetry are divided into two types: axion insulators and magnetic insulators.
In axion insulators, time-reversal symmetry is broken but an ``effective'' time-reversal symmetry represented by a combination of time-reversal and a lattice translation is present, leading to the topological magnetoelectric effect with the quantized coefficient $\theta=\pi$ (mod $2\pi$).
In magnetic insulators, the value of $\theta$ is arbitrary, including $\theta=0$.
In a class of magnetic insulators termed topological magnetic insulators, $\theta$ is proportional to their magnetic order parameters $M$ such as the N\'{e}el vector (i.e., antiferromagnetic order parameter), and the fluctuation of the order parameter realizes a dynamical axion field $\delta\theta(\bm{r},t)\propto \delta M(\bm{r},t)$ in condensed matter.
Here, note that spatial inversion symmetry must be broken in order for the value of $\theta$ to be arbitrary, i.e., in the magnetic insulators we have mentioned above, whereas its breaking is not required in the other three phases.
See also Table~\ref{Table-Symmetry-analysis} for the role of inversion symmetry.
}
\label{Fig-Introduction}
\end{figure*}
In the topological field theory, the responses of a topological phase to external fields are described by a topological term.
In two spatial dimensions, it is well known that the quantum Hall effect of a time-reversal symmetry broken phase can be described by a Chern-Simons action with the quantized coefficient given by the first Chern number \cite{Thouless1982,Zhang1992}.
In three spatial dimensions, time-reversal symmetry plays an important role.
The topological magnetoelectric effect described by the so-called $\theta$ term \cite{Qi2008} is a hallmark response of three-dimensional (3D) time-reversal invariant topological insulators to external electric and magnetic fields.
In the presence of time-reversal symmetry, the coefficient of the magnetoelectric effect $\theta$ takes a quantized value $\theta=\pi$ (mod $2\pi$) for topological insulators, while $\theta=0$ in trivial insulators.
However, in systems with broken time-reversal symmetry e.g., in magnetically ordered phases, the value of $\theta$ can be arbitrary, i.e., can deviate from the quantized value $\pi$ or $0$, which means that the value of $\theta$ can even depend on space and time as $\theta(\bm{r},t)$.
It should be noted that spatial-inversion symmetry breaking can also lead to the deviation of $\theta$ from the quantized value $\pi$ or $0$.

In field theory literature, the phenomena described by the $\theta$ term is termed the {\it axion electrodynamics} \cite{Wilczek1987}, because the $\theta$ term has exactly the same form as the action describing the coupling between a hypothetical elementary particle, axion, and a photon.
The axion was proposed about forty years ago to solve the so-called strong CP problem in quantum chromodynamics \cite{Peccei1977,Weinberg1978,Wilczek1978}.
By subsequent studies in particle physics and astrophysics, the axion is now considered as a candidate for dark matter \cite{Preskill1983,Abbott1983,Dine1983,book-axions}.
However, regardless of intensive experimental searches, the axion has not yet been found.
Since the coefficient of the $\theta$ term, $\theta(\bm{r},t)$, is a field describing the axion, observing the magnetoelectric responses in materials whose effective action is described by a $\theta$ term is equivalent to realizing the (dynamical) axion field in condensed matter \cite{Li2010}.
So far, it has been shown theoretically that in a class of magnetic insulators such as magnetically doped topological insulators the value of $\theta(\bm{r},t)$ is proportional to the antiferromagnetic order parameter (i.e., the N\'{e}el field), i.e., the antiferromagnetic spin fluctuation is identical to a dynamical axion field \cite{Li2010}.
In Fig.~\ref{Fig-Introduction}, a classification of 3D insulators in terms of the value of $\theta$ is schematically shown.

The effective action of the form of the $\theta$ term appears not only in insulator phases but also in semimetal phases.
The key in the case of topological semimetals is the breaking of time-reversal or spatial inversion symmetry, which can lead to nonzero and nonquantized expressions for $\theta$.
For example, in a time-reversal broken Weyl semimetal with two Weyl nodes, its response to external electric and magnetic fields is described by a $\theta$ term with $\theta(\bm{r},t)=2(\bm{b}\cdot\bm{r}-b_0t)$ \cite{Zyuzin2012,Son2012,Grushin2012,Wang2013,Goswami2013}, where $\bm{b}$ is the distance between the two Weyl nodes in momentum space and $b_0$ is the energy difference between the two nodes.
In contrast, in the case of topological superconductors, their topological nature is captured only by thermal responses  \cite{Read2000,Wang2011PRB,Ryu2012}, since charge and spin are not conserved.
It has been heuristically suggested that the effective action of 3D time-reversal invariant topological superconductors may be described by an action which is analogous to the $\theta$ term but is written in terms of gravitational fields corresponding to a temperature gradient and a mechanical rotation \cite{Nomura2012,Shiozaki2013}.

In this tutorial, we overview theoretical and experimental studies on the axion electrodynamics in topological materials.
In Sec.~\ref{Sec-Quantized-Magnetoelectric-Effect} we start by deriving the topological magnetoelectric effect described by a $\theta$ term in phenomenological and microscopic ways in 3D time-reversal invariant topological insulators.
We also review recent experimental studies toward observations of the quantized magnetoelectric effect.
In Sec.~\ref{Sec-Axion-Insulators} we review the basics and recent experimental realizations of the so-called axion insulators in which the value of $\theta$ is quantized due to a combined symmetry (effective time-reversal symmetry) regardless of the breaking of time-reversal symmetry, focusing on MnBi$_2$Te$_4$ family of materials.
In Sec.~\ref{Sec-Expressions-for-theta} we consider generic expressions for $\theta$ in insulators and extend the derivation of the $\theta$ term in a class of insulators with broken time-reversal and inversion symmetries whose realization requires antiferromagnetic orderings.
In Sec.~\ref{Sec-Dynamical-Axion-Field} we describe emergent dynamical phenomena from the realization of the dynamical axion field in topological antiferromagnetic insulators.
In Sec.~\ref{Sec-Weyl-Semimetal} and Sec.~\ref{Sec-Topological-Superconductor} we extend the study of the axion electrodynamics in condensed matter to Weyl semimetals and topological superconductors, respectively, whose effective action can be described by topological terms analogous to the $\theta$ term.
In Sec.~\ref{Sec-Summary} we summarize this tutorial and outlook future directions of this fascinating research field.

\section{Quantized magnetoelectric effect in 3D topological insulators \label{Sec-Quantized-Magnetoelectric-Effect}}
In this section, we describe the basics of the topological magnetoelectric effect, one of the intriguing properties characteristic to 3D topological insulators.
We derive phenomenologically and microscopically the $\theta$ term in 3D topological insulators, which is the low-energy effective action describing their responses to external electric and magnetic fields, i.e., the topological magnetoelectric effect.
We also review recent theoretical and experimental studies toward observations of the topological magnetoelectric effect.

\subsection{Overview}
\begin{figure}[!t]
\centering
\includegraphics[width=\columnwidth]{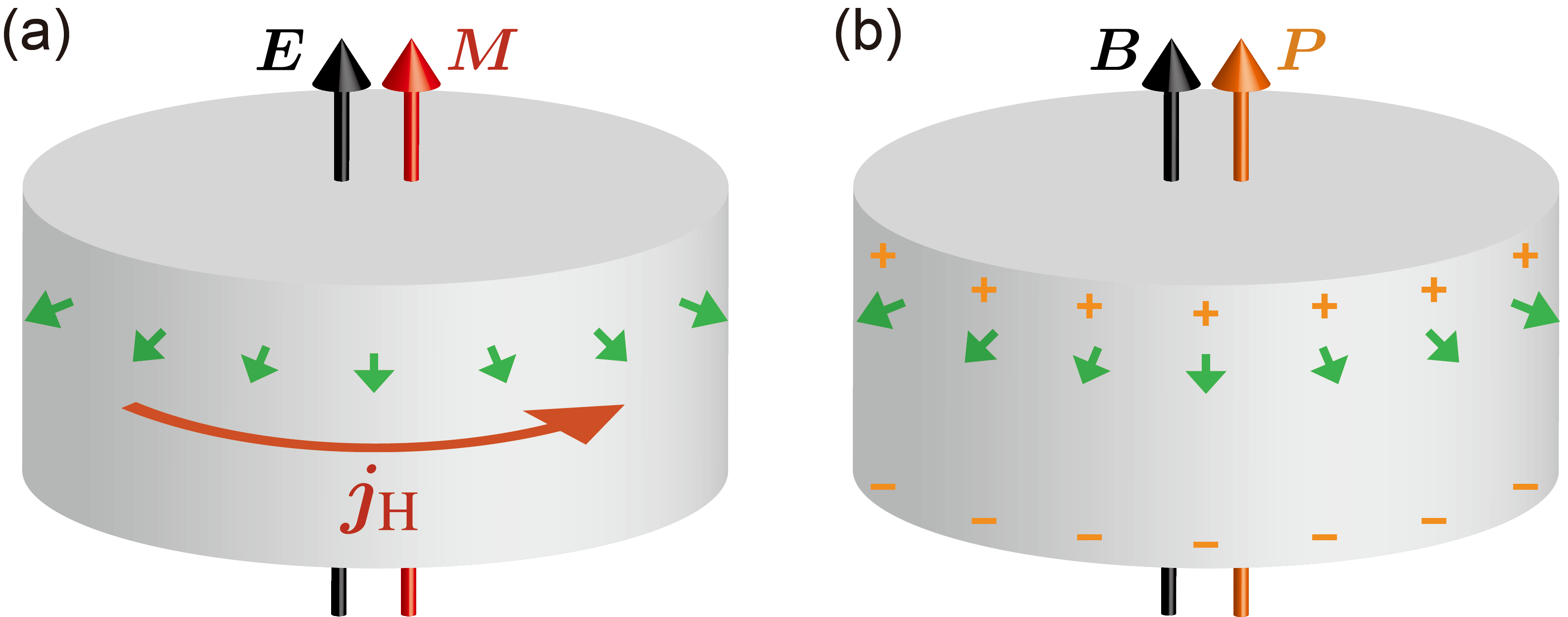}
\caption{Schematic picture of the topological magnetoelectric effect in a 3D topological insulator.
(a) Magnetization $\bm{M}$ induced by an external electric field $\bm{E}$.
$\bm{j}_{\mathrm{H}}$ is the anomalous Hall current on the side surface induced by the electric field.
(b) Electric polarization $\bm{P}$ induced by an external magnetic field $\bm{B}$.
Surface states are gapped by magnetic impurities (or a proximitized ferromagnet) whose magnetization direction is perpendicular to the surface, as indicated by green arrows.}
\label{Fig-TME}
\end{figure}

As has been briefly mentioned in the previous section, topological phases can be characterized by their response to external fields.
One of the noteworthy characters peculiar to 3D topological insulators is the topological magnetoelectric effect which is described by the so-called $\theta$ term \cite{Qi2008}.
The $\theta$ term is written as
\begin{align}
S_\theta=\int dtd^3 r\, \frac{\theta e^2}{4\pi^2\hbar c}\bm{E}\cdot\bm{B},
\label{S_theta_realtime}
\end{align}
where $h=2\pi\hbar$ is the Planck's constant, $e>0$ is the magnitude of the electron charge, $c$ is the speed of light, and $\bm{E}$ and $\bm{B}$ are external electric and magnetic fields, respectively.
From the variation of this action with respect to $\bm{E}$ and $\bm{B}$, we obtain the cross-correlated responses expressed by
\begin{align}
\bm{P}=\frac{\theta e^2}{4\pi^2\hbar c}\bm{B},\ \ \ \ \ \ \ \bm{M}=\frac{\theta e^2}{4\pi^2\hbar c}\bm{E},
\label{topological-ME-effect}
\end{align}
with $\bm{P}$ the electric polarization and $\bm{M}$ the magnetization.
We see that Eq.~(\ref{topological-ME-effect}) clearly exhibits a linear magnetoelectric effect, as schematically illustrated in Fig.~\ref{Fig-TME}.
Since $\bm{E}\cdot\bm{B}$ is odd under time reversal (i.e., $\bm{E}\cdot\bm{B}\to -\bm{E}\cdot\bm{B}$ under $t\to -t$), time-reversal symmetry requires that the action~(\ref{S_theta_realtime}) is invariant under the transformation $\theta\to -\theta$.
Then, it follows that in the presence of time-reversal symmetry $\theta$ takes a quantized value $\theta=\pi$ (mod $2\pi$) for topological insulators, while $\theta=0$ in trivial insulators.
A simple and intuitive proof of this quantization has been given \cite{Vazifeh2010}.
However, in systems with broken time-reversal symmetry e.g., in magnetically ordered phases, the value of $\theta$ can be arbitrary, i.e., can deviate from the quantized value $\pi$ or $0$ \cite{Essin2009}, which means that the value of $\theta$ can even depend on space and time as $\theta(\bm{r},t)$.
A similar argument can be applied to spatial inversion symmetry.
Namely, $\theta$ takes a quantized value $\theta=\pi$ or $\theta=0$ (mod $2\pi$) in the presence of inversion symmetry \cite{Hughes2011,Turner2012}, and inversion symmetry breaking can also lead to the deviation of $\theta$ from the quantized value, because $\bm{E}\cdot\bm{B}$ is also odd under spatial inversion.
Table~\ref{Table-Symmetry-analysis} shows the constraints on the value of $\theta$ by time-reversal and spatial-inversion symmetries.
\begin{table}[!t]
\renewcommand{\arraystretch}{1.2}
\caption{Constraints on the value of $\theta$ by time-reversal and spatial-inversion symmetries.
The mark $\checkmark$ ($\times$) indicates the presence (absence) of the symmetry.
Here, the notation of time-reversal symmetry in this table includes an ``effective'' time-reversal symmetry represented by a combination of time reversal and a lattice translation, as well as ``true'' time-reversal symmetry.
}
\begin{ruledtabular}
\begin{tabular}{ccc}
Time reversal & Inversion & Value of $\theta$ (mod $2\pi$)\\
\hline
\checkmark & \checkmark & $0$ or $\pi$ \\
\checkmark & $\times$ & $0$ or $\pi$ \\
$\times$ & \checkmark & $0$ or $\pi$ \\
$\times$ & $\times$ & arbitrary
\end{tabular}
\end{ruledtabular}
\label{Table-Symmetry-analysis}
\end{table}
%

\subsection{Symmetry analysis of the magnetoelectric coupling}
The magnetoelectric effect is the generation of bulk electric polarization (magnetization) by an external magnetic (electric) field.
The linear magnetoelectric coupling coefficient is generically described by
\begin{align}
\alpha_{ij}=\left.\frac{\partial M_j}{\partial E_i}\right|_{\bm{B}=0}=\left.\frac{\partial P_i}{\partial B_j}\right|_{\bm{E}=0},
\end{align}
where $i,j=x,y,z$ indicates spatial direction, $\bm{E}$ and $\bm{B}$ are external electric and magnetic fields, and $\bm{P}$ and $\bm{M}$ are the electric polarization and the magnetization.
In general, both time-reversal and spatial-inversion symmetries of the system must be broken, since the occurrence of nonzero $\bm{P}$ ($\bm{M}$) breaks spatial inversion (time-reversal) symmetry.
This requirement is consistent with the constraints on the value of $\theta$ by time-reversal and spatial-inversion symmetries (see Table~\ref{Table-Symmetry-analysis}).
Among several origins of the magnetoelectric effect, we are particularly interested in the orbital (i.e., electronic band) contribution to the linear magnetoelectric coupling of the form:
\begin{align}
\alpha_{ij}=\frac{e^2\theta}{4\pi^2\hbar c}\delta_{ij},
\label{ME-coupling-diagonal}
\end{align}
where $\delta_{ij}$ is the Kronecker delta.
Here, note that $\theta$ is a dimensionless constant.
Equation~(\ref{ME-coupling-diagonal}) implies the Lagrangian density $\mathcal{L}=(e^2\theta/4\pi^2\hbar c)\bm{E}\cdot\bm{B}$, since the magnetization and polarization can be derived from the free energy of the system $F$ as $\bm{M}=-\partial F/\partial \bm{B}$ and $\bm{P}=-\partial F/\partial \bm{E}$.
Notably, the susceptibility of the topological magnetoelectric effect in Eq.~(\ref{ME-coupling-diagonal}) with $\theta=\pi$ reads (in SI units)
\begin{align}
\frac{e^2}{4\pi\hbar c}\frac{1}{\mu_0^2 c}\simeq 24.3\ \mathrm{ps/m},
\end{align}
which is rather large compared to those of prototypical magnetoelectric materials, e.g., the total linear magnetoelectric susceptibility $\alpha_{xx}=\alpha_{yy}=0.7\ \mathrm{ps/m}$ of the well-known antiferromagnetic Cr$_2$O$_3$ at low temperatures \cite{Wiegelmann1994,Coh2011}.

It should be noted here that we need to take into account the presence of boundaries (i.e., surfaces) of a 3D topological insulator, when we consider the realization of the quantized magnetoelectric effect in a 3D topological insulator.
This is because, as is mentioned just above, finite $\bm{P}$ and $\bm{M}$ require the breaking of both time-reversal and spatial inversion symmetries of the whole system, whereas the bulk of the topological insulator has to respect both time-reversal and inversion symmetries.
As we will see in the following, the occurrence of the quantized magnetoelectric effect is closely related to the (half-quantized) anomalous Hall effect on the surface, which requires a somewhat special setup that breaks both time-reversal and inversion symmetries as shown in Fig.~\ref{Fig-TME}.
In this setup, time-reversal symmetry is broken due to the surface magnetization.
Inversion symmetry is also broken because the magnetization directions on a side surface and the other side surface are opposite to each other (spatial inversion does not change the direction of spin).

\subsection{Surface half-quantized anomalous Hall effect}
Before deriving the quantized magnetoelectric effect in 3D topological insulators, we briefly consider the anomalous Hall effect on the surfaces in which the Hall conductivity takes a half-quantized value $e^2/2h$.
Let us start with the effective Hamiltonian for the surface states of 3D topological insulators such as Bi$_2$Se$_3$, which is described by 2D two-component massless Dirac fermions \cite{Zhang2009}:
\begin{align}
\mathcal{H}_{\mathrm{surface}}(\bm{k})=\hbar v_{\mathrm{F}}(k_y\sigma_x-k_x\sigma_y)=\hbar v_{\mathrm{F}}(\bm{k}\times\bm{e}_z)\cdot\bm{\sigma},
\label{Surface-Dirac-fermions}
\end{align}
where $v_{\mathrm{F}}$ is the Fermi velocity of the surface state (i.e., the slope of the Dirac cone), and $\sigma_x,\sigma_y$ are the Pauli matrices for the spin degree of freedom.
The energy eigenvalues of the Hamiltonian~(\ref{Surface-Dirac-fermions}) are readily obtained as $E_{\mathrm{surface}}(\bm{k})=\pm \hbar v_{\mathrm{F}}\sqrt{k_x^2+k_y^2}$ from a simple algebra $\mathcal{H}_{\mathrm{surface}}^2=\hbar^2 v_{\mathrm{F}}^2(k_x^2+k_y^2)\bm{1}_{2\times 2}$.
The Fermi velocity of the surface states in Bi$_2$Se$_3$ is experimentally observed as $v_{\mathrm{F}}\approx 5\times 10^5$ m/s \cite{Xia2009}.

Due to the spin-momentum locking, the surface states are robust against disorder, as long as time-reversal symmetry is preserved.
Namely, the backscattering of surface electrons from $(\bm{k},\uparrow)$ to $(-\bm{k},\uparrow)$ are absent \cite{Roushan2009}.
Theoretically, it has been shown that 2D two-component massless Dirac fermions cannot be localized in the presence of nonmagnetic disorder \cite{Bardarson2007,Nomura2007}.
However, surface states are not robust against magnetic disorder which breaks time-reversal symmetry.
This is because the surface Dirac fermions described by Eq. (\ref{Surface-Dirac-fermions}) can be massive by adding a term proportional to $\sigma_z$, i.e., $m\sigma_z$, which opens a gap of $2m$ in the energy spectrum.
More precisely, such a mass term can be generated by considering the exchange interaction between the surface electrons and magnetic impurities \cite{Liu2009,Abanin2011,Nomura2011} such that $H_{\rm exch.}=J\sum_i \bm{S}_i\cdot\bm{\sigma}\delta(\bm{r}-\bm{R}_i)$, where $\bm{S}_i$ is the impurity spin at position $\bm{R}_i$.
Then, the homogeneous part of the impurity spins gives rise to the position-independent Hamiltonian
\begin{align}
\mathcal{H}_{\rm exch.}=Jn_{\mathrm{imp}}\bar{\bm{S}}_{\mathrm{imp}}\cdot\bm{\sigma}\equiv \bm{m}\cdot\bm{\sigma},
\label{Exchange-interaction}
\end{align}
where $n_{\mathrm{imp}}$ is the density of magnetic impurities and $\bar{\bm{S}}_{\mathrm{imp}}$ is the averaged spin of magnetic impurities.
Adding Eq.~(\ref{Exchange-interaction}) to the Hamiltonian~(\ref{Surface-Dirac-fermions}) leads to a gapped spectrum
\begin{align}
E_{\mathrm{surface}}(\bm{k})=\pm\sqrt{(\hbar v_{\mathrm{F}}k_x+m_y)^2+(\hbar v_{\mathrm{F}}k_y-m_x)^2+m_z^2}.
\end{align}
We see that $m_x$ and $m_y$ do not open the gap but only shift the position of the Dirac cone in momentum space.

\begin{figure}[!t]
\centering
\includegraphics[width=0.9\columnwidth]{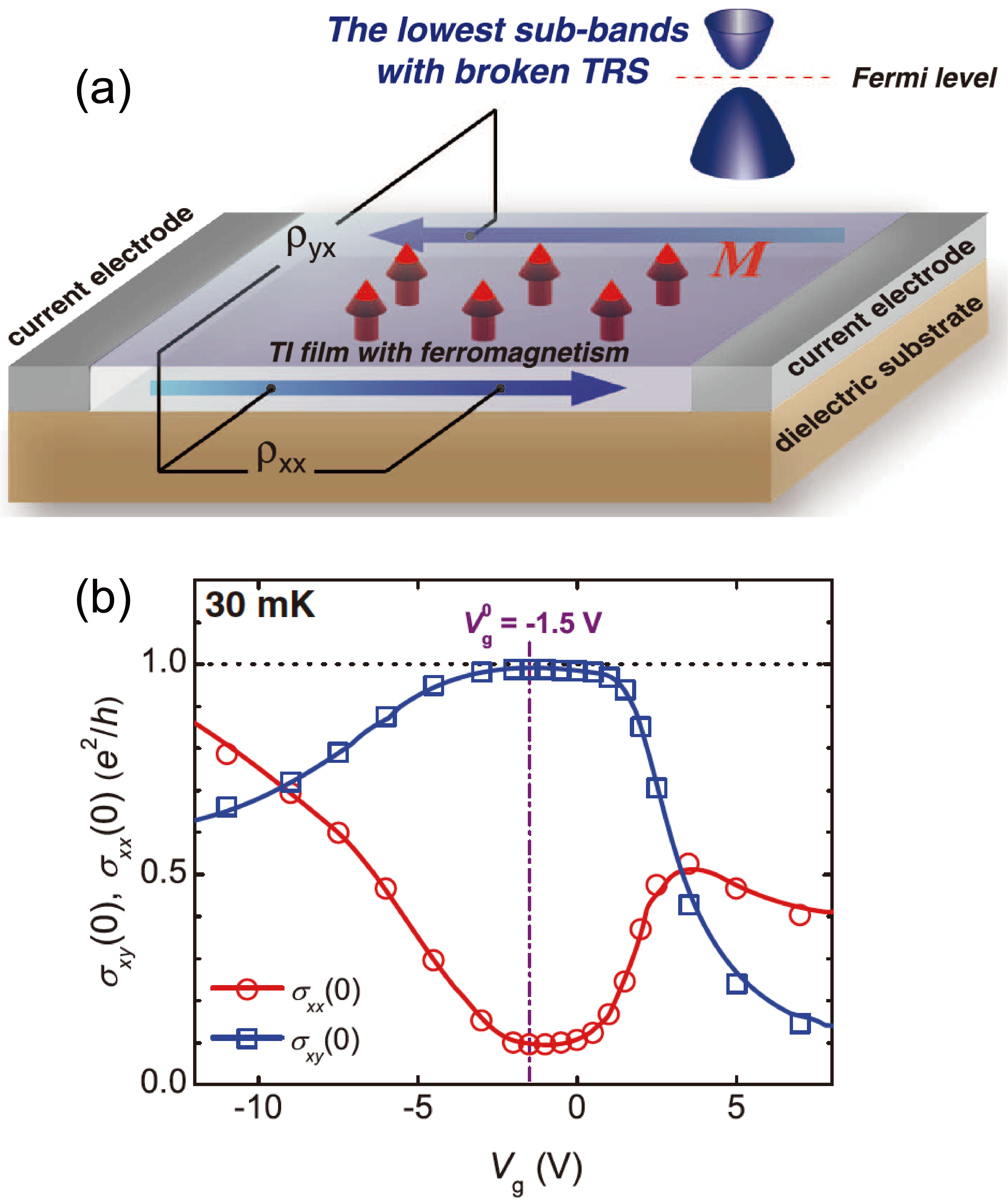}
\caption{(a) Schematic illustration of an experimental setup to detect the quantum anomalous Hall effect in a ferromagnetically ordered topological insulator thin film.
(b) Gate-voltage $V_g$ dependence of the Hall conductivity $\sigma_{xy}$ and the longitudinal conductivity $\sigma_{xx}$ in a thin film of Cr-doped (Bi,Sb)$_2$Te$_3$.
Adapted from Ref.~\onlinecite{Chang2013}.}
\label{Fig-Chang2013}
\end{figure}
Let us consider a general $2\times 2$ Hamiltonian given by $\mathcal{H}(\bm{k})=\bm{R}(\bm{k})\cdot \bm{\sigma}$.
In the case of massive Dirac fermions, $\bm{R}(\bm{k})$ is given by $\bm{R}(\bm{k})=(v_{\mathrm{F}}k_y, -v_{\mathrm{F}}k_x, m_z)$.
The Hall conductivity of the system with the Fermi level being in the gap can be calculated by \cite{Volovik-book}
\begin{align}
\sigma_{xy}&=-\frac{e^2}{h}\frac{1}{4\pi}\int dk_x dk_y\, \hat{\bm{R}}\cdot\left(\frac{\partial\hat{\bm{R}}}{\partial k_x}\times\frac{\partial\hat{\bm{R}}}{\partial k_y}\right)\nonumber\\
&=-\mathrm{sgn}(m_z)\frac{e^2}{2h},
\label{Surface-QHE}
\end{align}
where $\hat{\bm{R}}=\bm{R}(\bm{k})/|\bm{R}(\bm{k})|$ is a unit vector.
The integral is equivalent to the area where the unit vector $\hat{\bm{R}}$ moves on the unit sphere, which namely gives the winding number of $\hat{\bm{R}}$.
At $\bm{k}=0$, the unit vector $\hat{\bm{R}}$ points the north or south pole, that is, $\hat{\bm{R}}=(0,0,\mathrm{sgn}(m_z))$.
At large $\bm{k}$ with $|\bm{k}|\gg|m_z|$, $\hat{\bm{R}}$ almost points the horizontal directions.
Hence, varying $\bm{k}$, $\hat{\bm{R}}$ covers the half of the unit sphere, which gives $2\pi$.

Equation (\ref{Surface-QHE}) indicates the anomalous Hall effect occurs on the surfaces of 3D topological insulators, when magnetic impurities are doped or a magnetic film is put on the surfaces \cite{Yu2010,Nomura2011}.
The direction of the Hall current depends on the sign of $m_z$, i.e., the direction of the magnetization of magnetic impurities or proximitized magnetization.
Actually, the surface quantum anomalous Hall effect has been observed experimentally \cite{Chang2013,Checkelsky2014}.
The observed surface quantum anomalous Hall effect in a thin film of Cr-doped (Bi,Sb)$_2$Te$_3$ is shown in Fig. \ref{Fig-Chang2013}.
Note that in those systems the magnetization directions of top and bottom surfaces are the same, and thus the observed Hall conductivity is $2\times e^2/(2h)=e^2/h$.
It can be seen from Fig. \ref{Fig-Chang2013}(b) that the Hall conductivity takes the quantized value when the chemical potential lies in the surface bandgap.

\subsection{Phenomenological derivation of the $\theta$ term \label{Phenomenological-derivation}}
We have seen in the previous section that the surface states of 3D topological insulators can be gapped (i.e., the surface Dirac fermions can be massive) via the exchange interaction with magnetic impurities or proximitized magnetization which breaks time-reversal symmetry, giving rise to the surface half-quantized anomalous Hall effect.
We show phenomenologically in the following that, as a consequence of the surface half-quantized anomalous Hall effect, the topological magnetoelectric effect [Eq.~(\ref{topological-ME-effect})] emerges in 3D topological insulators.

Let us consider a case where the side surface of a cylindrical 3D topological insulator is ferromagnetically ordered due to magnetic doping or proximity effect \cite{Qi2008}, as shown in Fig. \ref{Fig-TME}.
The resulting surface Dirac fermions are massive.
When an external electric field $\bm{E}$ is applied parallel to the cylinder, the surface anomalous Hall current $\bm{j}_{\rm H}$ is induced as
\begin{align}
\bm{j}_{\rm H}=-\mathrm{sgn}(m)\frac{e^2}{2h}\hat{\bm{n}}\times\bm{E},
\end{align}
where $\hat{\bm{n}}$ is a unit vector normal to the side surface.
From the Amp\`{e}re's law, the magnetization $\bm{M}$ with $|\bm{M}|=|\bm{j}_{\rm H}|/c$ ($c$ is the speed of light) is obtained as [see Fig. \ref{Fig-TME}(a)]
\begin{align}
\bm{M}=\mathrm{sgn}(m)\frac{e^2}{2hc}\bm{E}.
\label{TME-M}
\end{align}
Similarly, when an external magnetic field $\bm{B}$ is applied parallel to the cylinder, the circulating electric field $\bm{E}^{\rm ind}$ normal to the magnetic field is induced as $\bm{\nabla}\times\bm{E}^{\rm ind}=-\partial\bm{B}/\partial t$.
Then the induced electric field $\bm{E}^{\rm ind}$ generates the surface anomalous Hall current parallel to the magnetic field as
\begin{align}
\bm{j}_{\rm H}=\mathrm{sgn}(m)\frac{e^2}{2h}\frac{\partial\bm{B}}{\partial t}.
\end{align}
On the other hand, a polarization current is equivalent to the time derivative of the electric polarization.
Finally the induced electric polarization $\bm{P}$ is given by [see Fig. \ref{Fig-TME}(b)]
\begin{align}
\bm{P}=\mathrm{sgn}(m)\frac{e^2}{2hc}\bm{B}.
\label{TME-P}
\end{align}

Equations (\ref{TME-M}) and (\ref{TME-P}) clearly show the magnetoelectric effect.
Here, recall that the magnetization and polarization can be derived from the free energy of the system $F$ as $\bm{M}=-\partial F/\partial \bm{B}$ and $\bm{P}=-\partial F/\partial \bm{E}$.
To satisfy the relations (\ref{TME-M}) and (\ref{TME-P}), the free energy must have the following form \cite{Qi2008}:
\begin{align}
F=-\int d^3r\, \frac{e^2}{2hc}\bm{E}\cdot\bm{B}=-\int d^3r\, \frac{\theta e^2}{4\pi^2\hbar c}\bm{E}\cdot\bm{B},
\label{Free-energy-TME}
\end{align}
where we have omitted $\mathrm{sgn}(m)$ for simplicity, and $\theta=\pi$.
The integrand can be regarded as the Hamiltonian density.
The equivalent action is written as
\begin{align}
S_\theta=\int d^4x\, \frac{\theta e^2}{4\pi^2\hbar c}\bm{E}\cdot\bm{B}=\int d^4x\, \frac{\theta e^2}{32\pi^2\hbar c}\epsilon^{\mu\nu\rho\lambda}F_{\mu\nu}F_{\rho\lambda},
\label{Theta-term-phenomenological}
\end{align}
where $d^4x=dtd^3r$, $F_{\mu\nu}=\partial_\mu A_\nu-\partial_\nu A_\mu$ with $A_\mu=(A_0,-\bm{A})$ being the electromagnetic four-potential, and $\epsilon^{\mu\nu\rho\lambda}$ is the Levi-Civita symbol with the convention $\epsilon^{0123}=1$.
Here, the electric field and magnetic field are given respectively by $\bm{E}=-\nabla A_0-\partial \bm{A}/\partial t$ and $\bm{B}=\nabla\times\bm{A}$.
Note that $e^2/\hbar c$ ($\simeq 1/137$) is the fine-structure constant.
Equation~(\ref{Theta-term-phenomenological}) is indeed the $\theta$ term [Eq.~(\ref{S_theta_realtime})].
Under time reversal ($t\to -t$), electric and magnetic fields are transformed as $\bm{E}\to\bm{E}$ and $\bm{B}\to -\bm{B}$, respectively.
Similarly, under spatial inversion ($\bm{r}\to -\bm{r}$), electric and magnetic fields are transformed as $\bm{E}\to -\bm{E}$ and $\bm{B}\to \bm{B}$, respectively.
Hence, the term $\bm{E}\cdot\bm{B}$ is odd under time reversal or spatial inversion.
On the other hand, 3D topological insulators have time-reversal symmetry, which indicates that $S_\theta$ remains unchanged under time reversal.
In other words, the value of $\theta$ must be invariant under the transformation $\theta\rightarrow -\theta$.
It follows that $\theta=\pi$ (mod $2\pi$) in time-reversal invariant topological insulators and $\theta=0$ in normal (topologically trivial) insulators.

Note that $S_\theta$ is a surface term when the value of $\theta$ is constant, i.e., independent of spatial coordinate and time, since we can rewrite the integrand of $S_\theta$ in a total derivative form,
\begin{align}
\epsilon^{\mu\nu\rho\lambda}F_{\mu\nu}F_{\rho\lambda}=4\epsilon^{\mu\nu\rho\lambda}\partial_\mu(A_\nu \partial_\rho A_\lambda),
\label{integrand-total-derivative}
\end{align}
which indicates that the topological magnetoelectric effect in the bulk is a consequence of the surface response to the electric and magnetic fields.
However, as we shall see later, the presence of the $\theta$ term that is dependent of spatial coordinate and/or time results in an electric current generation in the bulk.

Here, let us consider the inverse process of the derivation of the $\theta$ term~(\ref{Theta-term-phenomenological}).
Namely, we derive the surface anomalous Hall current from Eq.~(\ref{Theta-term-phenomenological}).
We have seen in Eq.~(\ref{integrand-total-derivative}) that the integrand of the $\theta$ term is a total derivative when the value of $\theta$ is constant.
For definiteness, let us see what happens at a given surface in the $z$ direction.
Using Eq.~(\ref{integrand-total-derivative}) and integrating out with respect to $z$, the surface term can be obtained from Eq.~(\ref{Theta-term-phenomenological}) as
\begin{align}
S_{\mathrm{surface}}=\int d^3x\, \frac{\theta e^2}{8\pi^2\hbar c}\epsilon^{z\nu\rho\lambda}A_\nu \partial_\rho A_\lambda,
\end{align}
where $d^3x=dtdxdy$.
Recall that, in general, an electric current density $j^\nu$ in the $\nu$ direction can be obtained from the variation of an action with respect to the electromagnetic vector potential $A_\nu$: $j^\nu=\delta S/\delta A_\nu$.
Without loss of generality we may consider the current in the $x$ direction,
\begin{align}
j^x=\frac{\delta S_{\mathrm{surface}}}{\delta A_x}=\frac{\theta e^2}{4\pi^2\hbar c}\epsilon^{zx\rho\lambda} \partial_\rho A_\lambda=\frac{\theta e^2}{4\pi^2\hbar c}E_y,
\label{surface-anomalous-Hall-current}
\end{align}
where $E_y=-\partial_yA^0-\partial_tA^y$ is the electric field in the $y$ direction.
Since $\theta=\pi$ in topological insulators, Eq.~(\ref{surface-anomalous-Hall-current}) clearly shows the surface half-quantized anomalous Hall effect.

More precisely, we should consider an electric current derived directly from the $\theta$ term.
Namely, we should consider the spatial dependence of $\theta$ such that $\theta=0$ in vacuum and $\theta=\pi$ inside the topological insulator.
Notice that the $\theta$ term can be rewritten as
\begin{align}
S_\theta=-\int dt d^3r\, \frac{e^2}{8\pi^2 \hbar}\epsilon^{\mu\nu\rho\lambda}[\partial_\mu \theta(\bm{r},t)] A_\nu\partial_\rho A_\lambda.
\label{Theta-term-Chern-Simons}
\end{align}
Then the electric current density is obtained as
\begin{align}
j^x=\frac{\delta S_\theta}{\delta A_x}=\frac{e^2}{4\pi^2 \hbar}\left[\partial_t\theta(\bm{r},t)B_x-\partial_z\theta(\bm{r},t)E_y\right].
\label{surface-anomalous-Hall-current2}
\end{align}
The magnetic-field induced term is the so-called chiral magnetic effect \cite{Fukushima2008}, which will be mentioned later.
For concreteness, we require that the region $z\le 0$ ($z>0$) be the topological insulator (vacuum).
The $z$ dependence of $\theta(\bm{r},t)$ can be written in terms of the Heaviside step function as $\theta(z)=\pi[1-\Theta(z)]$, since $\theta=\pi$ ($\theta=0$) inside (outside) the topological insulator.
Then, we obtain $\partial_z\theta=-\pi\delta(z)$, which gives rise to the half-quantized Hall conductivity at the topological insulator surface $z=0$.

\subsection{Microscopic derivation of the $\theta$ term}
So far we have derived the topological magnetoelectric effect [Eq.~(\ref{topological-ME-effect})] from a surface property of 3D topological insulators.
In this section, we derive the $\theta$ term microscopically from a low-energy effective model of 3D topological insulators.
There are several ways to derive the $\theta$ term microscopically.
One way is to use the so-called Fujikawa's method \cite{Fujikawa1979,Fujikawa1980}.
Another way is the dimensional reduction from (4+1)-dimensions to (3+1)-dimensions \cite{Qi2008}, which will be briefly mentioned in Sec.~\ref{Sec-Dimensional-Reduction}.
Here, we show the derivation of the $\theta$ term based on Fujikawa's method.

\subsubsection{Effective Hamiltonian for 3D topological insulators}
Let us start from the low-energy continuum model for prototypical 3D topological insulators such as Bi$_2$Se$_3$.
The bulk electronic structure of Bi$_2$Se$_3$ near the Fermi level is described by two $p$-orbitals $P1^+_z$ and $P2^-_z$ with $\pm$ denoting parity.
Defining the basis $[|P1^+_z,\uparrow\rangle,|P1^+_z,\downarrow\rangle,|P2^-_z,\uparrow\rangle,|P2^-_z,\uparrow\rangle]$ and retaining the wave vector $\bm{k}$ up to quadratic order, the low-energy effective Hamiltonian around the $\Gamma$ point is given by \cite{Zhang2009,Liu2010}
\begin{align}
\mathcal{H}_{\rm eff}(\bm{k})&=
\begin{bmatrix}
\mathcal{M}(\bm{k}) & 0 & A_1k_z & A_2k_-\\
0 & \mathcal{M}(\bm{k}) & A_2k_+ & -A_1k_z\\
A_1k_z & A_2k_- & -\mathcal{M}(\bm{k}) & 0\\
A_2k_+ & -A_1k_z & 0 & -\mathcal{M}(\bm{k})
\end{bmatrix}\nonumber \\
&=A_2k_x\alpha_1+A_2k_y\alpha_2+A_1k_z\alpha_3+\mathcal{M}(\bm{k})\alpha_4,
\label{Bi2Se3-effective}
\end{align}
where $k_{\pm}=k_x\pm ik_y$ and $\mathcal{M}(\bm{k})=m_0-B_1k_z^2-B_2k^2_\bot$.
The coefficients for Bi$_2$Se$_3$ estimated by a first-principles calculation read $m_0=0.28$ eV, $A_1 =2.2$ eV$\cdot$\AA, $A_2 =4.1$ eV$\cdot$\AA, $B_1 =10$ eV$\cdot$\AA$^2$, and $B_2=56.6$ eV$\cdot$\AA$^2$ \cite{Zhang2009,Liu2010}.
Here, note that we have introduced a basis in Eq.~(\ref{Bi2Se3-effective}) that is slightly different from that Refs.~\onlinecite{Zhang2009,Liu2010}.
The $4\times 4$ matrices $\alpha_\mu$ are given by the so-called Dirac representation,
\begin{align}
\begin{split}
\alpha_j=
\begin{bmatrix}
0 & \sigma_j\\
\sigma_j & 0
\end{bmatrix},\ \ \ \ \
\alpha_4=
\begin{bmatrix}
1 & 0\\
0 & -1
\end{bmatrix},
\end{split}
\end{align}
where the Clifford algebra $\{\alpha_\mu,\alpha_\nu\}=2\delta_{\mu\nu}\bm{1}$ is satisfied.
The above Hamiltonian is nothing but an anisotropic 3D Dirac Hamiltonian with a momentum-dependent mass.

Before proceeding to the derivation of the $\theta$ term, it is informative to consider the lattice version of Eq.~(\ref{Bi2Se3-effective}).
Here, recall that the $Z_2$ invariant \cite{Kane2005,Fu2007,Fu2007a,Moore2007}, which identifies whether a phase is topologically nontrivial or trivial, is calculated in lattice models.
This means that we cannot directly show that the phase described by the effective Hamiltonian~(\ref{Bi2Se3-effective}) represents a 3D topological insulator.
From this viewpoint, we need to construct a lattice Hamiltonian from the continuum Hamiltonian~(\ref{Bi2Se3-effective}).
The simplest 3D lattice is the cubic lattice.
We replace $k_i$ and $k_i^2$ terms by $k_i\rightarrow\sin k_i$ and $k_i^2\rightarrow 2(1-\cos k_i)$.
Although this replacement is valid only when $k_i\ll 1$, as is shown below, it turns out that this replacement describes the topological insulator phase.
We also simplify the coefficients to obtain the isotropic lattice Hamiltonian
\begin{align}
\mathcal{H}_{\rm eff}(\bm{k})=&\ \hbar v_{\mathrm{F}}(\alpha_1\sin k_x+\alpha_2\sin k_y+\alpha_3\sin k_z)\nonumber\\
&+\left[m_0+r\sum_{i=x,y,z}(1-\cos k_i)\right]\alpha_4,
\label{Wilson-Hamiltonian}
\end{align}
where we have defined $\hbar v_{\mathrm{F}}=A_1=A_2$ and $r=-2B_1=-2B_2$.
As is mentioned below, the Hamiltonian~(\ref{Wilson-Hamiltonian}) is also called the Wilson-Dirac Hamiltonian \cite{Wilson1977,Creutz1994,Sekine2013}, which was originally introduced in lattice quantum chromodynamics.

In cubic lattices, the eight time-reversal invariant momenta $\bm{\Lambda}_\alpha$, which are invariant under $k_i\to -k_i$, are given by $(0,0,0)$, $(\pi/a,0,0)$, $(0,\pi/a,0)$, $(0,0,\pi/a)$, $(\pi/a,\pi/a,0)$, $(\pi/a,0,\pi/a)$, $(0,\pi/a,\pi/a)$, and $(\pi/a,\pi/a,\pi/a)$ where $a$ is the lattice constant.
We can calculate the $Z_2$ invariant of the system as \cite{Fu2007,Fu2007a}
\begin{align}
(-1)^\nu&=\prod_{\alpha=1}^8\mathrm{sgn}\left[m_0+r\sum_{i=x,y,z}(1-\cos \Lambda_\alpha^i)\right]\nonumber\\
&=
\left\{
\begin{aligned}
&-1\ \ \ (0>m_0/r>-2,\ -4>m_0/r>-6)\\
&+1\ \ \ (m_0/r>0,\ -2>m_0/r>-4,\ -6>m_0/r).
\end{aligned}
\right.
\label{Z2-invariant-3D-Dirac}
\end{align}
Indeed, the topological insulator phase with $0>m_0/r>-2$ satisfies the above realistic value for Bi$_2$Se$_3$; $m_0/r\sim -0.1$, where we have assumed the value of the lattice constant as $a=3$ \AA.

It should be noted here that the lattice Dirac Hamiltonian~(\ref{Wilson-Hamiltonian}) is exactly the same as the Hamiltonian of the {\it Wilson fermions}, which was originally introduced in lattice gauge theory to avoid the fermion doubling problem \cite{Wilson1977}.
Namely, we can see that Eq.~(\ref{Wilson-Hamiltonian}) around the $\Gamma$ point $(0,0,0)$ represents the usual (continuum) massive Dirac fermions with the mass $m_0$, while Eq.~(\ref{Wilson-Hamiltonian}) around other momentum points, e.g., $(\pi/a,0,0)$, represent massive Dirac fermions with the mass $m_0+2r$.

\subsubsection{Fujikawa's method \label{Sec-3DTI-FujikawaMethod}}
Now, let us return to the continuum Hamiltonian~(\ref{Bi2Se3-effective}) to obtain the $\theta$ term.
As we have seen in Eq.~(\ref{Z2-invariant-3D-Dirac}), the lattice Hamiltonian~(\ref{Wilson-Hamiltonian}) describes a topological insulator when $0>m_0/r>-2$.
Without loss of generality, we can set $m_0<0$ and $r>0$.
Then, the Hamiltonian~(\ref{Bi2Se3-effective}) with $m_0<0$ and $r>0$, which describes a topological insulator, around the $\Gamma$ point can be simplified by ignoring the terms second-order in $k_i$ as
\begin{align}
\mathcal{H}_{\mathrm{TI}}(\bm{k})=\hbar v_{\mathrm{F}}\bm{k}\cdot\bm{\alpha}+m_0\alpha_4,
\label{Hamiltonian-3DTI}
\end{align}
where $m_0<0$.
Except for the negative mass $m_0$, this is the usual Dirac Hamiltonian.
In the presence of an external electromagnetic vector potential $\bm{A}$, minimal coupling results in $\bm{k}\to\bm{k}+e\bm{A}$, with $e>0$ being the magnitude of the electron charge.
In the presence of an external electromagnetic scalar potential $A_0$, the energy density is modified as $\psi^\dag\mathcal{H}_0\psi\to \psi^\dag(\mathcal{H}_0-eA_0)\psi$.
Using these facts, the action of the system in the presence of an external electromagnetic four-potential $A_\mu=(A_0,-\bm{A})$ is written in the usual relativistic form \cite{Peskin-Schroeder-Book}
\begin{align}
S_{\mathrm{TI}}&=\int dtd^3r\, \psi^\dag\left\{i(\partial_t-ieA_0)-[\mathcal{H}_{\mathrm{TI}}(\bm{k}+e\bm{A})]\right\}\psi \nonumber\\
&=\int dtd^3r\, \bar{\psi}[i\gamma^\mu(\partial_\mu-ieA_\mu)-m_0]\psi,
\label{action-Dirac-fermion}
\end{align}
where $\psi^\dag(\bm{r},t)$ is a fermionic field representing the basis of the Hamiltonian~(\ref{Bi2Se3-effective}) and $\bar{\psi}=\psi^\dag\gamma^0$.
Here, the gamma matrices $\gamma^\mu$ are given by the so-called Dirac representation as
\begin{align}
&\gamma^0=\alpha_4=
\begin{bmatrix}
1 & 0\\
0 & -1
\end{bmatrix},\ \ \ \ \
\gamma^j=\alpha_4\alpha_j=
\begin{bmatrix}
0 & \sigma_j\\
-\sigma_j & 0
\end{bmatrix},\nonumber\\
&\gamma^5=i\gamma^0\gamma^1\gamma^2\gamma^3=
\begin{bmatrix}
0 & 1\\
1 & 0
\end{bmatrix},
\end{align}
which satisfy the relation $\{\gamma^\mu,\gamma^\nu\}=2g^{\mu\nu}$ with $g^{\mu\nu}=\mathrm{diag}(+1,-1,-1,-1)$ being the metric tensor.
It is convenient to study the system in the imaginary time notation, i.e., in Euclidean spacetime.
Namely, we rewrite $t$, $A_0$, and $\gamma^j$ as $t\rightarrow -i\tau$, $A_0\rightarrow iA_0$, and $\gamma^j\rightarrow i\gamma_j$ ($j=1,2,3$).
The Euclidean action of the system is then written as 
\begin{align}
S_{\mathrm{TI}}^{\mathrm{E}}=-iS_{\mathrm{TI}}=\int d\tau d^3r\, \bar{\psi}[\gamma_\mu(\partial_\mu-ieA_\mu)-m_0e^{i\pi\gamma_5}]\psi,
\label{TME-TI-Action}
\end{align}
where we have used the fact that $m_0=-m_0(\cos\pi+i\gamma_5\sin\pi)=-m_0e^{i\pi\gamma_5}$.
Note that $\gamma^0$ and $\gamma^5$ are unchanged ($\gamma_0=\gamma^0$ and $\gamma_5=\gamma^5$), so that the anticommutation relation $\{\gamma_\mu,\gamma_\nu\}=2\delta_{\mu\nu}$ is satisfied.
Note also that, in Euclidean spacetime, we do not distinguish between superscripts and subscripts.

Now, we are in a position to apply Fujikawa's method \cite{Fujikawa1979,Fujikawa1980} to the action~(\ref{TME-TI-Action}).
First let us consider an infinitesimal chiral transformation defined by
\begin{align}
\psi\rightarrow \psi'= e^{-i\pi d\phi\gamma_5/2}\psi,\ \ \ \ \ \ \ \bar{\psi}\rightarrow \bar{\psi}'=\bar{\psi}e^{-i\pi d\phi\gamma_5/2},
\end{align}
where $\phi\in[0,1]$.
Then the partition function $Z$ is transformed as
\begin{align}
Z=\int \mathcal{D}[\psi,\bar{\psi}]\, e^{-S_{\mathrm{TI}}^{\mathrm{E}}[\psi,\bar{\psi}]}\ \ \rightarrow\ \ Z'=\int \mathcal{D}[\psi',\bar{\psi}']\, e^{-S_{\mathrm{TI}}^{'\mathrm{E}}[\psi',\bar{\psi}']}.
\end{align}
The $\theta$ term comes from the Jacobian defined by $\mathcal{D}[\psi',\bar{\psi}']=J\mathcal{D}[\psi,\bar{\psi}]$.
The action (\ref{TME-TI-Action}) is transformed as
\begin{align}
S_{\mathrm{TI}}^{'\mathrm{E}}=\ &\int d\tau d^3r\, \bar{\psi}[\gamma_\mu(\partial_\mu-ieA_\mu)-m_0e^{i\pi(1-d\phi)\gamma_5}]\psi \nonumber\\
&+\frac{i}{2}\pi\int d\tau d^3rd\phi\, \partial_\mu(\bar{\psi}\gamma_\mu\gamma_5\psi).
\end{align}
The Jacobian is written as \cite{Fujikawa1979,Fujikawa1980}
\begin{align}
J=\exp\left[-i\int d\tau d^3rd\phi\, \frac{\pi e^2}{32\pi^2\hbar c}\epsilon^{\mu\nu\rho\lambda}F_{\mu\nu}F_{\rho\lambda}\right].
\end{align}
Here $F_{\mu\nu}=\partial_\mu A_\nu-\partial_\nu A_\mu$ and we have written $\hbar$ and $c$ explicitly.
We repeat this procedure infinite times, i.e., integrate with respect to the variable $\phi$ from $0$ to $1$.
Due to the invariance of the partition function, finally we arrive at the following expression of $S_{\mathrm{TI}}^{\mathrm{E}}$:
\begin{align}
S_{\mathrm{TI}}^{\mathrm{E}}=\ &\int d\tau d^3r\, \bar{\psi}[\gamma_\mu(\partial_\mu-ieA_\mu)-m_0]\psi \nonumber\\
&+i\int d\tau d^3r\, \frac{\pi e^2}{32\pi^2\hbar c}\epsilon^{\mu\nu\rho\lambda}F_{\mu\nu}F_{\rho\lambda},
\end{align}
where we have dropped the irrelevant surface term.
The first term is the action of a topologically trivial insulator, since the mass $-m_0$ is positive.
The second term is the $\theta$ term in imaginary time, and we obtain Eq. (\ref{Theta-term-phenomenological}) by  substituting $\tau=it$.

\subsection{Toward observations of the topological magnetoelectric effect}
\subsubsection{Utilizing topological insulator thin films}
\begin{figure}[!t]
\centering
\includegraphics[width=\columnwidth]{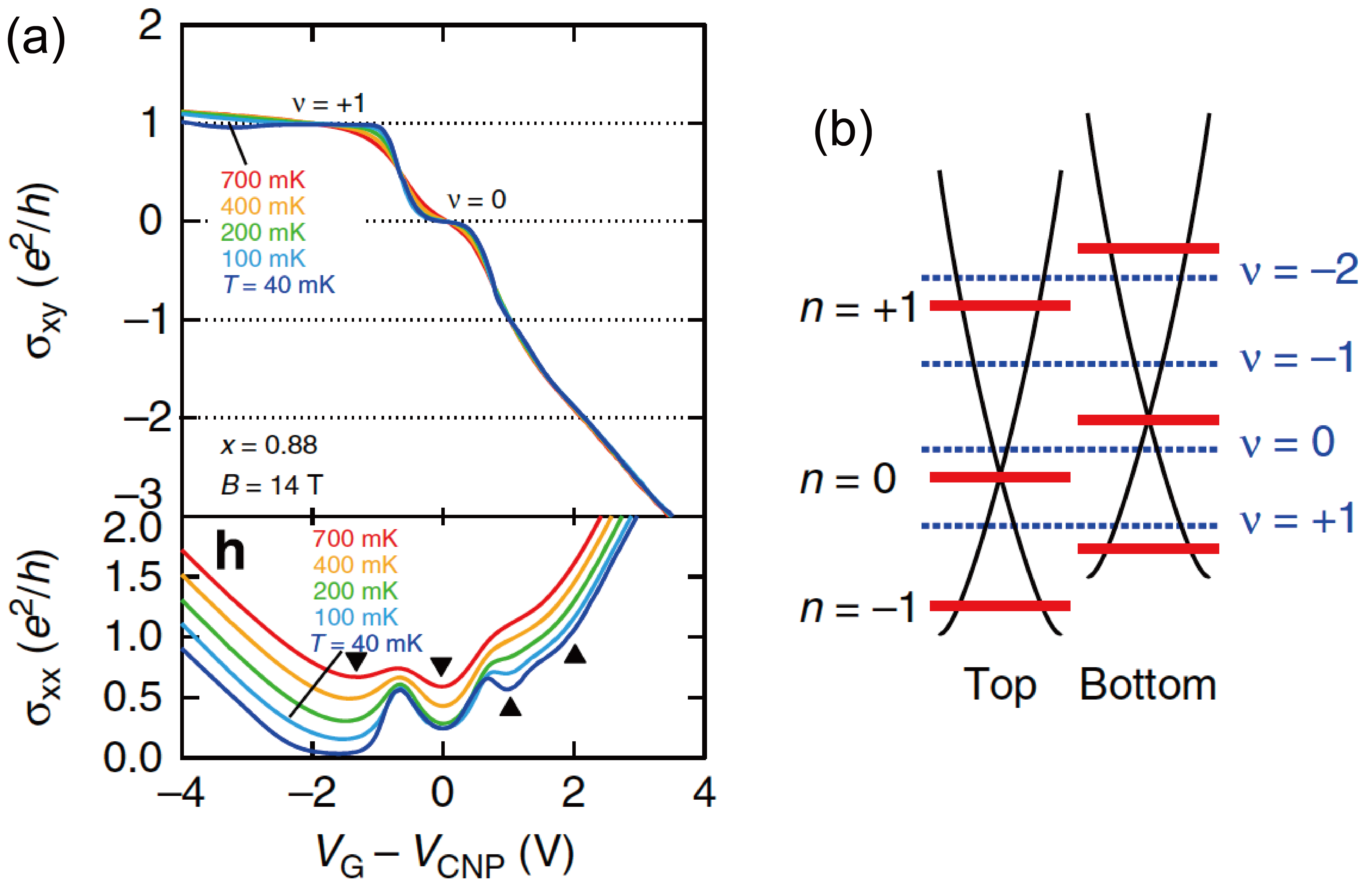}
\caption{(a) Quantum Hall effect in a topological insulator (Bi$_{1-x}$Sb$_x$)$_2$Te$_3$ thin film.
(b) Schematic illustration of the Landau levels of the top and bottom surface states in the presence of an energy difference between the two surfaces.
Adapted from Ref.~\onlinecite{Yoshimi2015}.}
\label{Fig-Yoshimi2015}
\end{figure}
As we have seen in Sec.~\ref{Phenomenological-derivation}, the experimental realization of the topological magnetoelectric effect in topological insulators requires that all the surface Dirac states are gapped by magnetic proximity effect or magnetic doping, resulting in the zero anomalous Hall conductivity of the system.
However, such an experimental setup is rather difficult to to be realized.
As an alternate route to realize the topological magnetoelectric effect, it has been proposed theoretically that the $\nu=0$ quantum Hall state, which attributes to the difference between the Landau levels of the top and bottom surface Dirac states, can be utilized \cite{Morimoto2015,Wang2015}.
The $\nu=0$ quantum Hall state has been experimentally observed in topological insulator (Bi$_{1-x}$Sb$_x$)$_2$Te$_3$ films \cite{Yoshimi2015}, as shown in Fig.~\ref{Fig-Yoshimi2015}(a).
The two-component Dirac fermions in a magnetic field are known to show the quantum Hall effect with the Hall conductivity
\begin{align}
\sigma_{xy}=\left(n+\frac{1}{2}\right)\frac{e^2}{h},
\end{align}
where $n$ is an integer.
Note that, as we have seen in Eq.~(\ref{Surface-QHE}), the $\frac{1}{2}$ contribution arises as a Berry phase effect.
The total Hall conductivity contributed from the top and bottom surfaces of a topological insulator film in a magnetic field is then written as
\begin{align}
\sigma_{xy}=\left(n_{\mathrm{T}}+n_{\mathrm{B}}+1\right)\frac{e^2}{h}\equiv \nu\frac{e^2}{h},
\label{QHE-TI-film}
\end{align}
The $\nu=0$ quantum Hall state is realized when the Landau levels of the top and bottom surface states are $N_{\mathrm{T}}=-N-1$ and $N_{\mathrm{B}}=N$ (and vice versa), where $N$ is an integer \cite{Morimoto2015}.
This state corresponds to $n_{\mathrm{T}}=-N-1$ and $n_{\mathrm{B}}=N$ in Eq.~(\ref{QHE-TI-film}), which can be achieved in the presence of an energy difference between the two surface states, as shown in Fig.~\ref{Fig-Yoshimi2015}(b).
Here, recall that the electron density is given by $n_e=\sigma_{xy}B/e$, with $B$ the magnetic field strength and $e$ the elementary charge.
Using this fact, the charge densities ($\rho=-en_e$) at the top and bottom surfaces are obtained as $\rho_{\mathrm{T}}=(N+\frac{1}{2})Be^2/h$ and $\rho_{\mathrm{B}}=-(N+\frac{1}{2})Be^2/h$, respectively.
We consider the case of $N=0$, which is experimentally relevant \cite{Yoshimi2015}.
The induced electric polarization in a topological insulator film of thickness $d$ reads
\begin{align}
P=\frac{1}{2d}\left[d\rho_{\mathrm{T}}+(-d)\rho_{\mathrm{B}}\right]=\frac{e^2}{2h}B,
\end{align}
which is indeed the topological magnetoelectric effect with the quantized coefficient $\theta=\pi$.
Note that the case of $N\neq 0$, which gives rise to $\theta=(2N+1)\pi$, still describes the topological magnetoelectric effect, since $\theta=\pi$ modulo $2\pi$.

\begin{figure}[!t]
\centering
\includegraphics[width=\columnwidth]{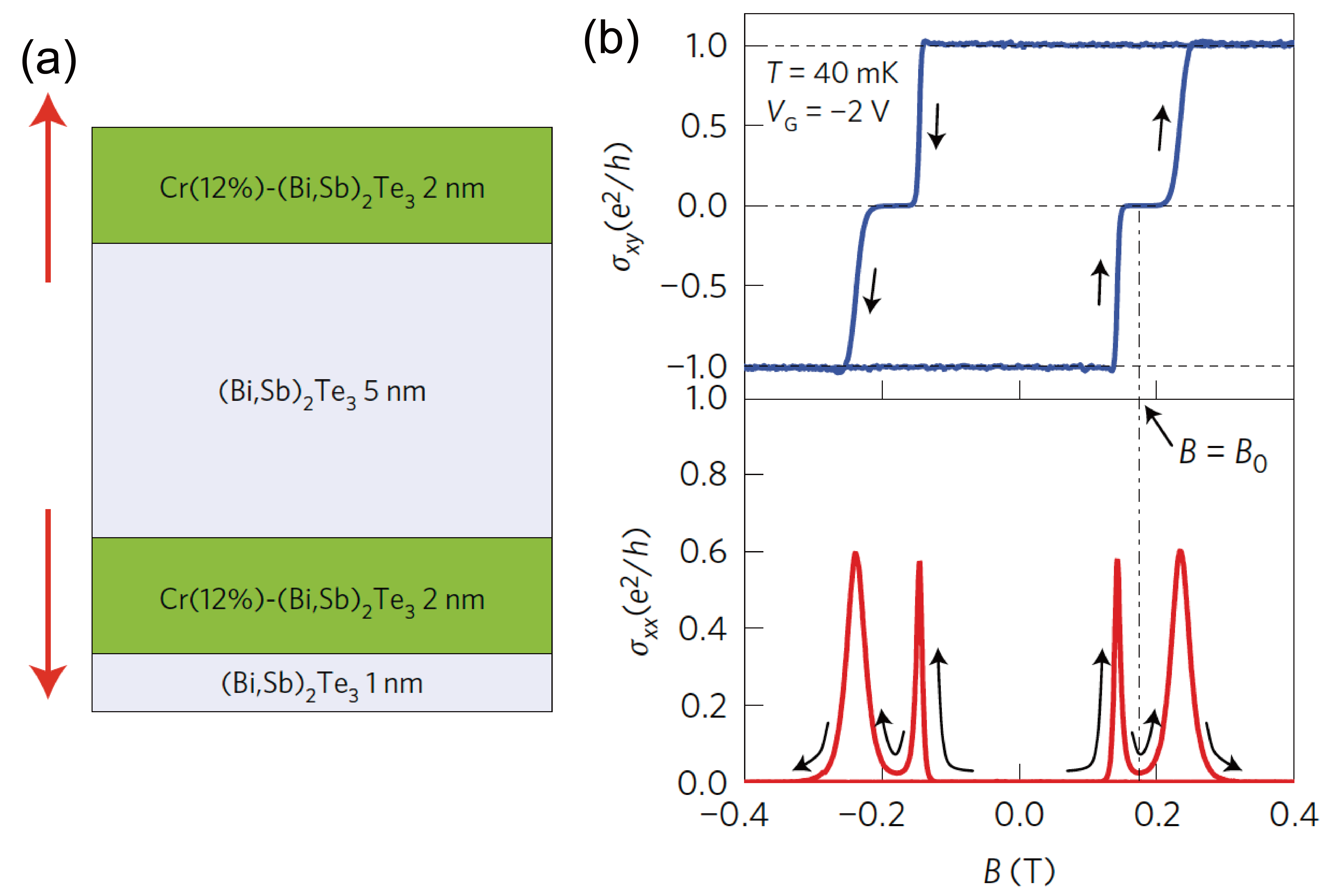}
\caption{(a) Schematic illustration of the magnetic heterostructure.
Red arrows indicate the magnetization directions.
(b) The observed Hall conductivity as a function of an external magnetic field.
Adapted from Ref.~\onlinecite{Mogi2017}.}
\label{Fig-Mogi2017}
\end{figure}
\begin{figure}[!t]
\centering
\includegraphics[width=0.8\columnwidth]{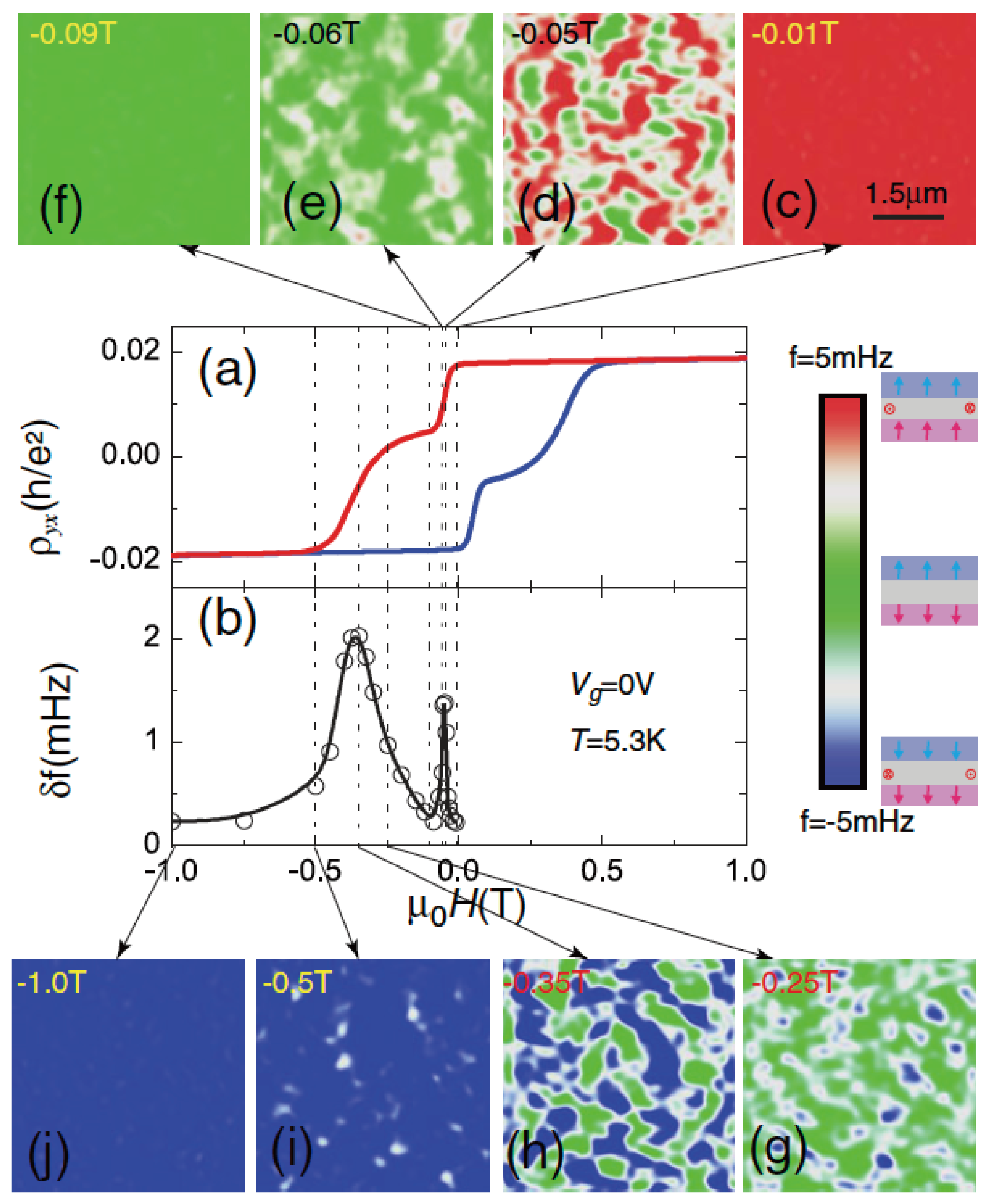}
\caption{Magnetic field dependence of (a) Hall resistivity and (b) magnetic domain contrasts.
(c)-(j) Magnetic force microscopy images of the magnetic domains.
Red and blue represent respectively upward and downward parallel magnetization alignment regions, while green represents antiparallel magnetization alignment regions.
Adapted from Ref.~\onlinecite{Xiao2018}.}
\label{Fig-Xiao2018}
\end{figure}

Another route to realize the topological magnetoelectric effect is a magnetic heterostructure in which the magnetization directions of the top and bottom magnetic insulators are antiparallel \cite{Morimoto2015,Wang2015}.
Several experiments have succeeded in fabricating magnetic heterostructures that exhibits a zero Hall plateau \cite{Mogi2017,Mogi2017a,Xiao2018}.
In Ref.~\onlinecite{Mogi2017}, a magnetic heterostructure consisting of a magnetically doped topological insulator Cr-doped (Bi,Sb)$_2$Te$_3$ and a topological insulator (Bi,Sb)$_2$Te$_3$ was grown by molecular beam epitaxy.
A zero Hall conductivity plateau was observed in this study as shown in Fig.~\ref{Fig-Mogi2017}, implying an axion insulator state.
In Ref.~\onlinecite{Xiao2018}, a magnetic heterostructure of a topological insulator (Bi,Sb)$_2$Te$_3$ sandwiched by two kinds of magnetically doped topological insulators V-doped (Bi,Sb)$_2$Te$_3$ and Cr-doped (Bi,Sb)$_2$Te$_3$ was grown by molecular beam epitaxy.
Importantly, as shown in Fig.~\ref{Fig-Xiao2018}, the antiparallel magnetization alignment of the top and bottom magnetic layers was directly observed by magnetic force microscopy when the system exhibited a zero Hall resistivity plateau.
Note, however, that the above experiments did not make a direct observation of the magnetoelectric effect, i.e., the electric polarization induced by a magnetic field or the magnetization induced by an electric field.

\subsubsection{Faraday and Kerr rotations}
\begin{figure}[!t]
\centering
\includegraphics[width=0.9\columnwidth]{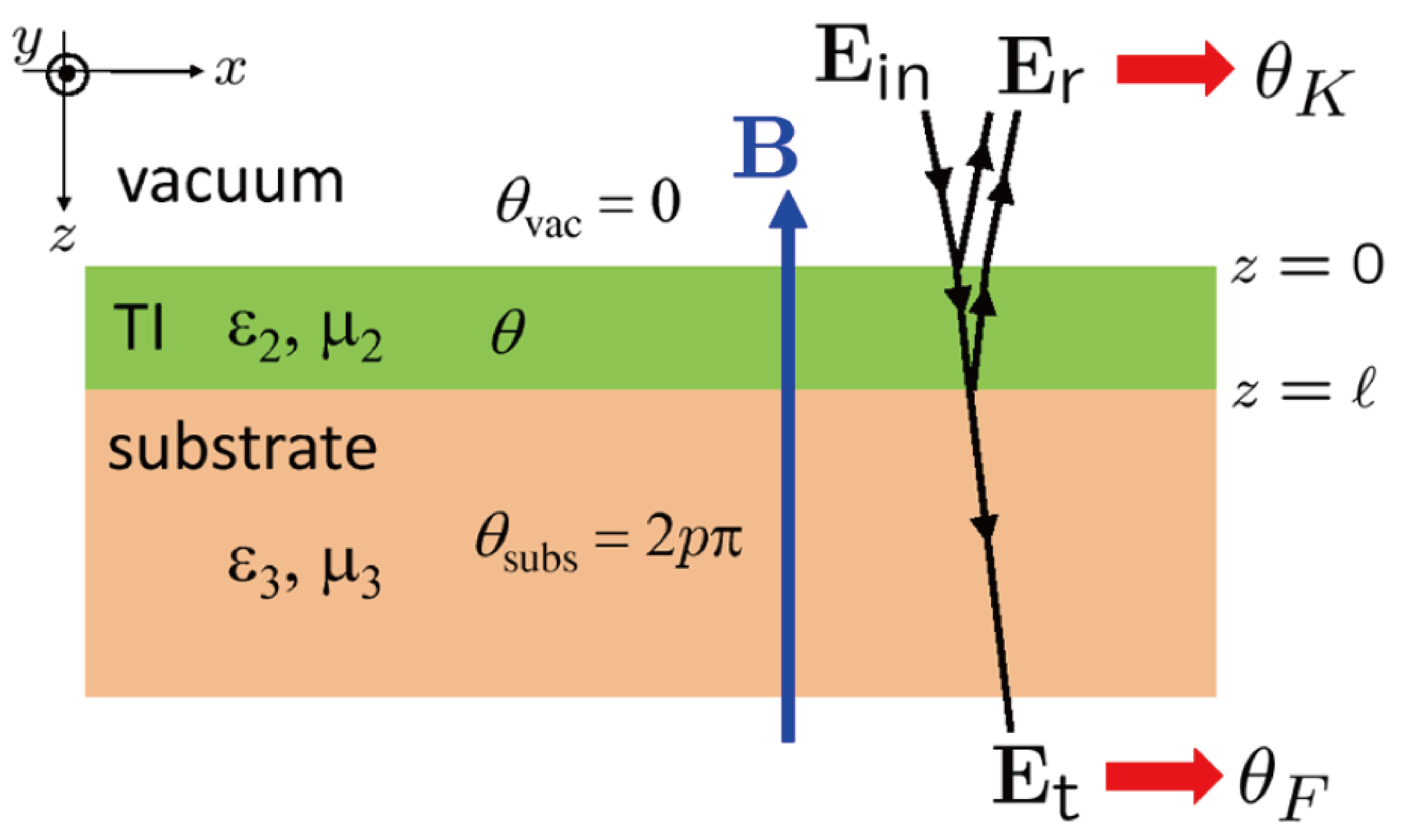}
\caption{Schematic figure of a measurement of the quantized Faraday and Kerr rotations in a topological insulator thin film.
Adapted from Ref.~\onlinecite{Maciejko2010}.}
\label{Fig-Maciejko2010}
\end{figure}
As has been known in particle physics \cite{Sikivie1983,Wilczek1987} before the discovery of 3D topological insulators, the $\theta$ term modifies the Maxwell's equations.
Since the Maxwell's equations describe electromagnetic wave propagation in materials, the presence of the $\theta$ term leads to unusual optical properties such as the quantized Faraday and Kerr rotations in topological insulators \cite{Qi2008,Tse2010,Maciejko2010} which can be viewed as as a consequence of the topological magnetoelectric effect.
To see this, let us start from the total action of an electromagnetic field $A_\mu=(A_0,-\bm{A})$ in the presence of a $\theta$ term is given by
\begin{align}
S=\int dtd^3 r\, \frac{\alpha}{4\pi^2}\theta \bm{E}\cdot\bm{B}-\frac{1}{16\pi}\int dtd^3 r\, F_{\mu\nu}F^{\mu\nu},
\label{action-EM-axion}
\end{align}
where $\alpha=e^2/\hbar c\simeq 1/137$ is the fine-structure constant and $F_{\mu\nu}=\partial_\mu A_\nu-\partial_\nu A_\mu$ is the electromagnetic field tensor.
The electric and magnetic fields are respectively given by $\bm{E}=-\nabla A_0-(1/c)\partial \bm{A}/\partial t$ and $\bm{B}=\nabla\times\bm{A}$.
Note that $\bm{E}\cdot\bm{B}=(1/8)\epsilon^{\mu\nu\rho\lambda}F_{\mu\nu}F_{\rho\lambda}$ and $F_{\mu\nu}F^{\mu\nu}=2(\bm{B}^2/\mu_0-\epsilon_0\bm{E}^2)$.
Here, recall that the classical equation of motion for the field $A_\mu$ is obtained from the Euler-Lagrange equation:
\begin{align}
\frac{\delta S}{\delta A_\mu}=\frac{\partial\mathcal{L}}{\partial A_\mu}-\partial_\nu\left(\frac{\partial\mathcal{L}}{\partial(\partial_\nu A_\mu)}\right)=0,
\label{EulerLagrange-EM}
\end{align}
where $\mathcal{L}$ is the Lagrangian density of the system.
From Eqs.~(\ref{action-EM-axion}) and (\ref{EulerLagrange-EM}), one finds that the Maxwell's equations are modified in the presence of a $\theta$ term \cite{Sikivie1983,Wilczek1987,Qi2008}
\begin{align}
\nabla\cdot\bm{E}&=4\pi\rho-2\alpha\nabla\biggl(\frac{\theta}{2\pi}\biggr)\cdot\bm{B}, \nonumber\\
\nabla\times\bm{E}&=-\frac{1}{c}\frac{\partial \bm{B}}{\partial t}, \nonumber\\
\nabla\cdot\bm{B}&=0, \nonumber\\
\nabla\times\bm{B}&=\frac{4\pi}{c}\bm{J}+\frac{1}{c}\frac{\partial \bm{E}}{\partial t}+\frac{2\alpha}{c}\left[\frac{\partial}{\partial t}\biggl(\frac{\theta}{2\pi}\biggr)\bm{B}+c\nabla\biggl(\frac{\theta}{2\pi}\biggr)\times\bm{E}\right].
\label{modified-Maxwell's-Eq}
\end{align}
The $\nabla\theta$ terms in Eq.~(\ref{modified-Maxwell's-Eq}) play roles when there is a boundary, e.g., gives rise to the surface Hall current as we have seen in Eq.~(\ref{surface-anomalous-Hall-current2}).

\begin{figure}[!t]
\centering
\includegraphics[width=\columnwidth]{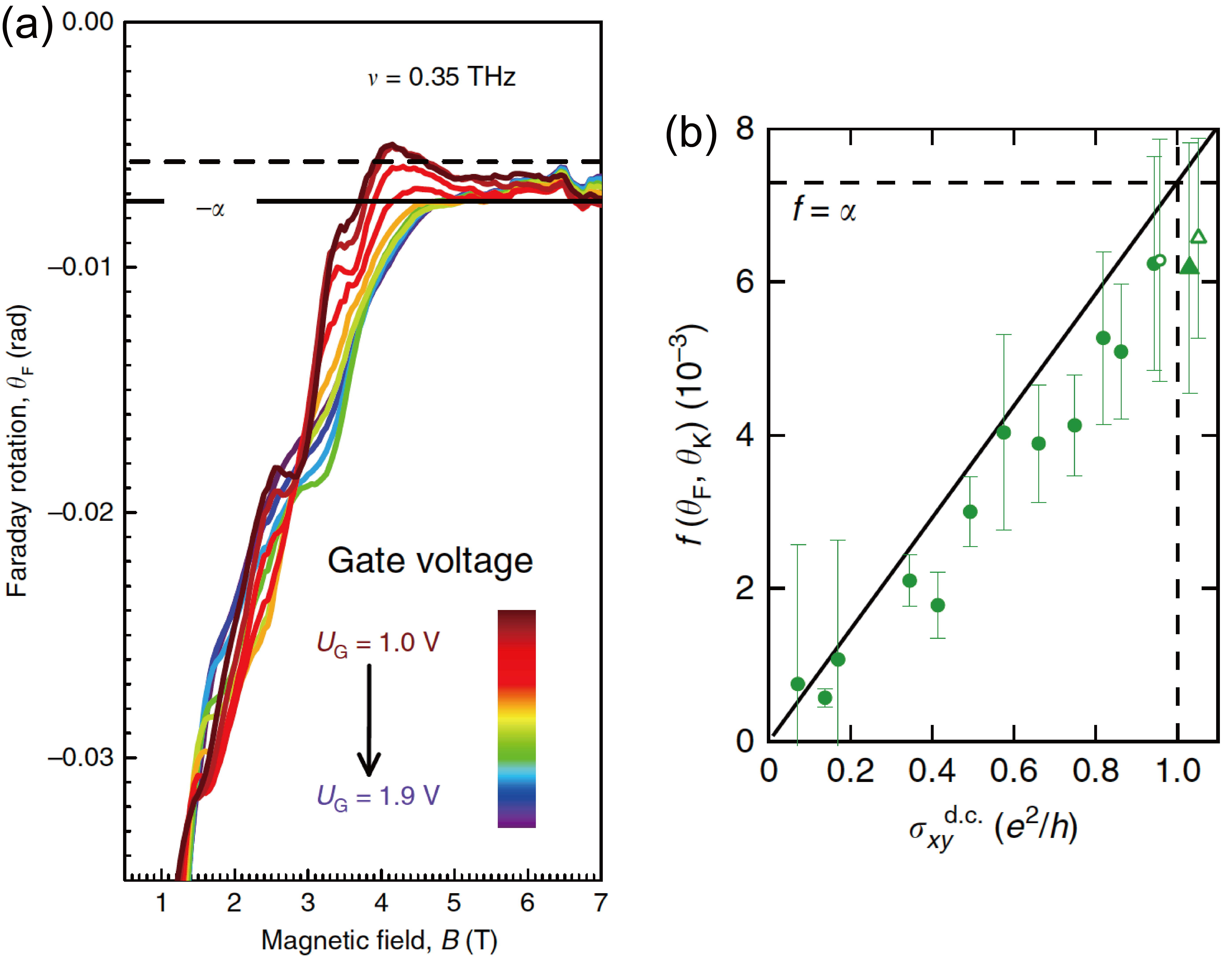}
\caption{(a) Magnetic field dependence of the Faraday rotation angle.
(b) Evolution of the scaling function $f(\theta_{\mathrm{F}}, \theta_{\mathrm{K}})=\frac{\cot\theta_{\mathrm{F}}-\cot\theta_{\mathrm{K}}}{\cot^2\theta_{\mathrm{F}}-2\cot\theta_{\mathrm{F}}\cot\theta_{\mathrm{K}}-1}$ as a function of dc Hall conductance towards the universal relationship $f(\theta_{\mathrm{F}}, \theta_{\mathrm{K}})=\alpha$.
(a): Adapted from Ref.~\onlinecite{Dziom2017}.
(b): Adapted from Ref.~\onlinecite{Okada2016}.}
\label{Fig-Okada2016-Dziom2017}
\end{figure}
The modified Maxwell's equations~(\ref{modified-Maxwell's-Eq}) can be solved under the boundary conditions (see Fig.~\ref{Fig-Maciejko2010}).
It is found that the Faraday and Kerr rotation angles are independent of the material (i.e., topological insulator thin film) parameters such as the dielectric constant and thickness \cite{Tse2010,Maciejko2010}.
Especially, in the quantized limit the Faraday and Kerr rotation angles are given respectively by \cite{Tse2010,Maciejko2010}
\begin{align}
\theta_{\mathrm{F}}=\tan^{-1}(\alpha)\simeq\alpha,\ \ \ \ \ \ \ \ \theta_{\mathrm{K}}=\tan^{-1}(1/\alpha)\simeq\frac{\pi}{2}.
\end{align}
These quantized angles have been experimentally observed in the anomalous Hall state \cite{Okada2016} and the quantum Hall state [Fig.~\ref{Fig-Okada2016-Dziom2017}(a)] \cite{Wu2016,Dziom2017}.
Also, as predicted in Ref.~\onlinecite{Maciejko2010}, a universal relationship in units of the fine-structure constant $\alpha$ between the Faraday and Kerr rotation angles has been observed [Fig.~\ref{Fig-Okada2016-Dziom2017}(b)] \cite{Okada2016,Wu2016}.

\section{Axion insulators \label{Sec-Axion-Insulators}}
In Sec.~\ref{Sec-Quantized-Magnetoelectric-Effect} we have seen that the topological magnetoelectric effect with the quantized coefficient $\theta=\pi$ (mod $2\pi$) occurs in 3D time-reversal invariant topological insulators.
In general, the value of $\theta$ is no longer quantized and becomes arbitrary in systems with broken time-reversal symmetry.
However, in a class of 3D antiferromagnetic insulators, an ``effective'' time-reversal symmetry represented by a combination of time-reversal and a lattice translation is present, leading to the topological magnetoelectric effect with the quantized coefficient $\theta=\pi$ (mod $2\pi$).
In this section, we review theoretical and experimental studies on such antiferromagnetic topological insulators which are also called the {\it axion insulators}.
Starting from the basics of the antiferromagnetic topological insulators, we focus on the MnBi$_2$Te$_4$ family of materials which are layered van der Waals compounds and have recently been experimentally realized.

\subsection{Quantized magnetoelectric effect in antiferromagnetic topological insulators \label{AF-topolgical-insulators}}
Following Ref.~\onlinecite{Mong2010}, we consider a class of insulators in which time-reversal symmetry is broken but the combined symmetry of time reversal and a lattice translation is preserved.
We note here that the presence or absence of inversion symmetry does not affect their topological classification, although the presence of inversion symmetry greatly simplifies the evaluation of their topological invariants as in the case of time-reversal invariant topological insulators \cite{Fu2007}.
Let us start from some general arguments on symmetry operations.
The time-reversal operator $\Theta$ for spin-1/2 systems is generically given by $\Theta=i\sigma_y K$ with $\Theta^2=-1$, where $\sigma_i$ are Pauli matrices and $K$ is complex conjugation operator.
In the presence of time-reversal symmetry, the Bloch Hamiltonian of a system $\mathcal{H}(\bm{k})$ satisfies
\begin{align}
\Theta\mathcal{H}(\bm{k})\Theta^{-1}&=\mathcal{H}(-\bm{k}).
\label{TRS-Bloch-Hamiltonian}
\end{align}
Recall that momentum is the generator of lattice translation.
An operator that denotes a translation by a vector $\bm{x}$ is given by $T(\bm{x})=e^{-i\bm{k}\cdot\bm{x}}$.
Then, the translation operator that moves a lattice by half a unit cell in the $\bm{a}_3$ direction is written as
\begin{align}
T_{1/2}=e^{-(i/2)\bm{k}\cdot\bm{a}_3}
\begin{bmatrix}
0 && \bm{1}\\
\bm{1} && 0
\end{bmatrix},
\end{align}
where $\bm{a}_3$ is a primitive translation vector and $\bm{1}$ is an identity operator that acts on the half of the unit cell \cite{Mong2010}.
One can see that $T_{1/2}^2$ gives a translation by $\bm{a}_3$ because $T_{1/2}^2=e^{-i\bm{k}\cdot\bm{a}_3}$.
\begin{figure}[!t]
\centering
\includegraphics[width=0.9\columnwidth]{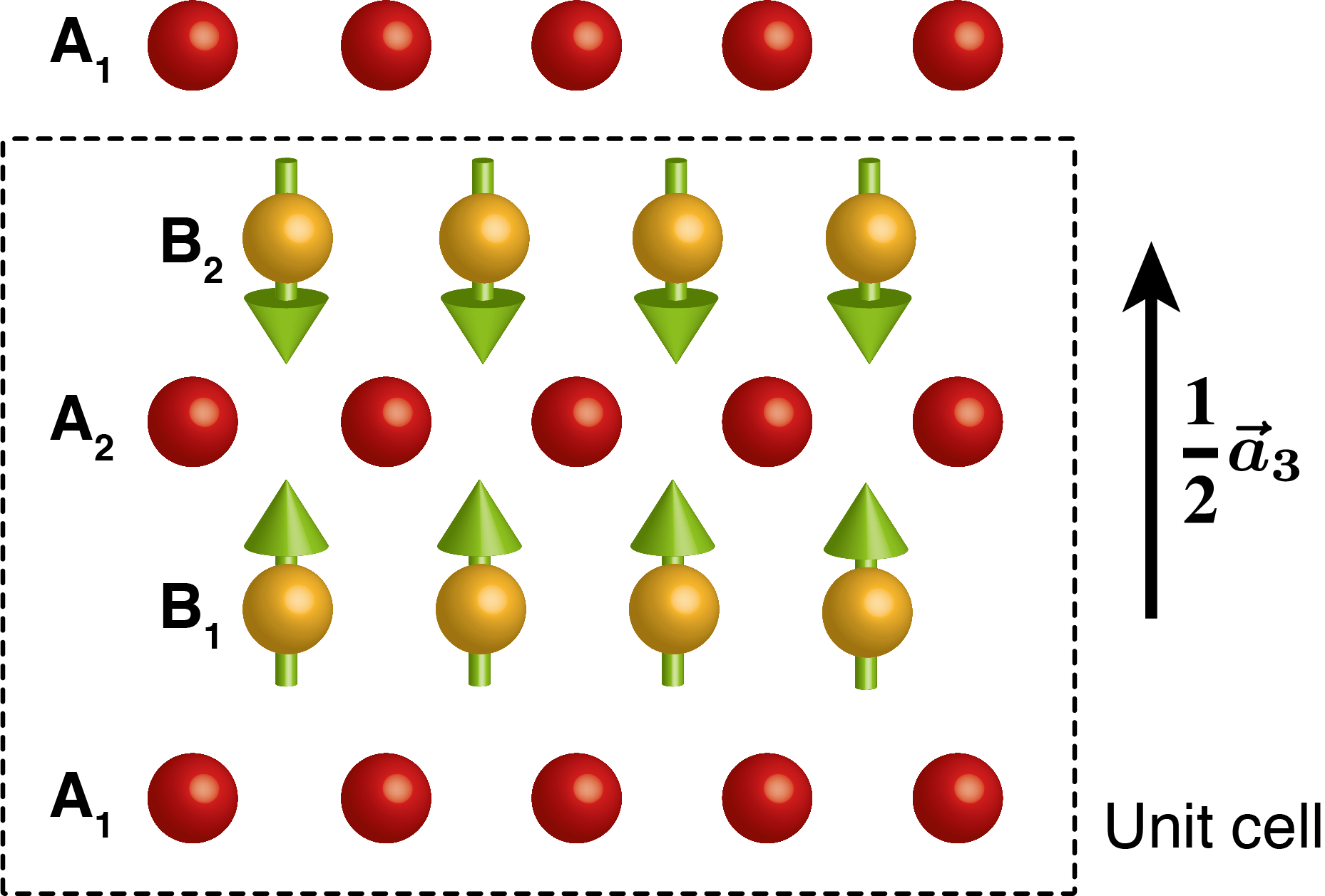}
\caption{Schematic illustration of an antiferromagnetic topological insulator protected by the $S=\Theta T_{1/2}$ symmetry.}
\label{Fig-AFI-TI}
\end{figure}
Now, we consider the combination of $\Theta$ and $T_{1/2}$ defined by $S=\Theta T_{1/2}$.
It follows that $S^2=-e^{-i\bm{k}\cdot\bm{a}_3}$, which means that the operator $S$ is antiunitary like $\Theta$.
Here, we have used the fact that $\Theta$ and $T_{1/2}$ are commute.
Note, however, that $S^2=-1$ only on the Brillouin-zone plane satisfying $\bm{k}\cdot\bm{a}_3=0$, while $\Theta^2=-1$.
When a system is invariant under the operation $S$, the Bloch Hamiltonian $\mathcal{H}(\bm{k})$ satisfies
\begin{align}
S\mathcal{H}(\bm{k})S^{-1}&=\mathcal{H}(-\bm{k}),
\end{align}
which has the same property as time-reversal symmetry in Eq.~(\ref{TRS-Bloch-Hamiltonian}).
Therefore, the $\mathbb{Z}_2$ topological classification can also be applied in systems with the $S$ symmetry \cite{Mong2010,Fang2013}.
Figure~\ref{Fig-AFI-TI} shows a schematic illustration of an antiferromagnetic topological insulator protected by the $S=\Theta T_{1/2}$ symmetry.
In this simple model, the unit cell consists of nonmagnetic equivalent $A_1$ and $A_2$ atomic layers, and antiferromagnetically ordered $B_1$ and $B_2$ atomic layers.
The half-uni-cell translation $ T_{1/2}$ moves the $B_1$ layer to the $B_2$ layer, and time reversal $\Theta$ changes a spin-up state into a spin-down state.
Therefore, the system is obviously invariant under the $S=\Theta T_{1/2}$ transformation.

Next, let us consider the resulting surface states.
Since $S^2=-1$ on the Brillouin-zone plane satisfying $\bm{k}\cdot\bm{a}_3=0$, the 2D subsystem on the $(k_1, k_2)$ plane is regarded as a quantum spin Hall system with time-reversal symmetry.
This means that the $k_1$ or $k_2$ dependence of the surface spectra must be gapless, because the $\bm{k}\cdot\bm{a}_3=0$ line of the surface states is the boundary of 2D subsystem (the $\bm{k}\cdot\bm{a}_3=0$ plane) in the bulk Brillouin zone.
In other words, at the surfaces that are parallel to $\bm{a}_3$, which preserve the $S$ symmetry, there exist an odd number of gapless surface states (as in the case of a strong time-reversal invariant topological insulator).
On the other hand, at the surfaces that are perpendicular to $\bm{a}_3$, which break the $S$ symmetry, such a topological protection of the surface states no longer exists, and the surface states can have gapped spectra.

As we have seen above, the presence of $S$ symmetry results in a realization of a new 3D topological insulator.
This implies that such topological insulators exhibit a quantized magnetoelectric effect described by a $\theta$ term, as in the case of time-reversal invariant 3D topological insulators.
To see this, recall that the magnetoelectric effect resulting from a $\theta$ term is expressed as 
$\bm{P}=\theta e^2/(4\pi^2\hbar c)\bm{B}$, and $\bm{M}=\theta e^2/(4\pi^2\hbar c)\bm{E}$,
where $\bm{P}$ and $\bm{M}$ are the electric polarization and the magnetization, respectively.
Under time reversal $\Theta$, the coefficient $\theta$ changes sign $\theta\to -\theta$, because $\bm{P}\to\bm{P}$ and $\bm{E}\to\bm{E}$ while $\bm{M}\to -\bm{M}$ and $\bm{B}\to -\bm{B}$.
On the other hand, the lattice translation $T_{1/2}$ does not affect $\theta$ \cite{Mong2010}.
Combining these, the $S$ operation implies the transformation such that $\theta\to -\theta+2\pi n$ with $n$ being an integer.
Then, it follows that $\theta=0$ or $\theta=\pi$ modulo $2\pi$.

\subsection{MnBi$_2$Te$_4$}
\subsubsection{Electronic structure of MnBi$_2$Te$_4$ bulk crystals}
\begin{figure}[!t]
\centering
\includegraphics[width=0.7\columnwidth]{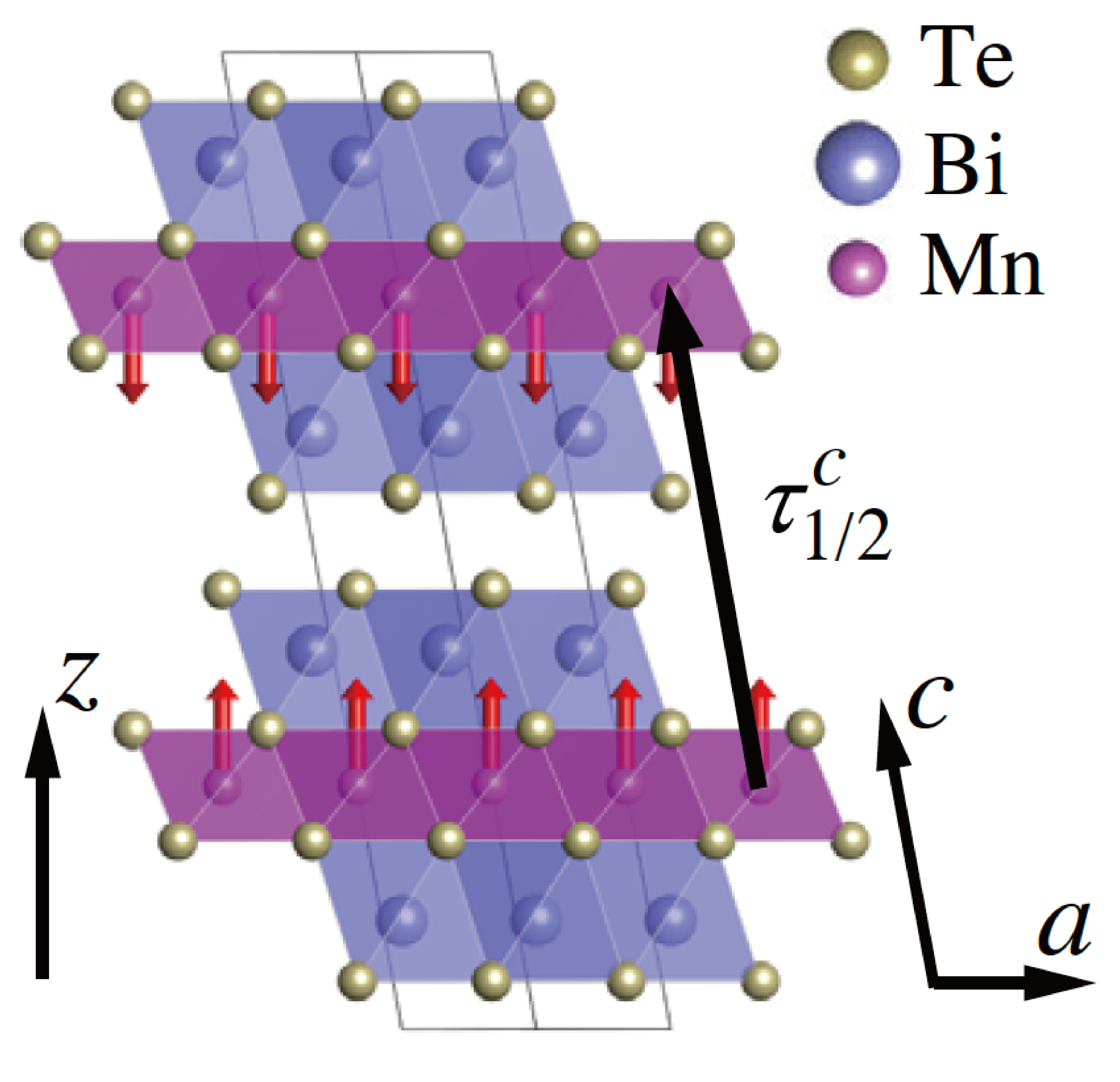}
\caption{Crystal and magnetic structure of the antiferromagnetic topological insulator state in MnBi$_2$Te$_4$.
The unit cell consists of two septuple layers.
$\tau_{1/2}^c$ is the half-cell translation vector along the $c$ axis which connects nearest spin-up and spin-down Mn atomic layers.
Adapted from Ref.~\onlinecite{Hao2019}.}
\label{Fig-Hao2019}
\end{figure}
With the knowledge of antiferromagnetic topological insulators with the $S$ symmetry, here we review recent experimental realizations of the antiferromagnetic topological insulator state in MnBi$_2$Te$_4$ \cite{Otrokov2019,Zhang2019,Li2019,Lee2019,Chen2019a,Vidal2019,Hao2019,Li2019PRX,Chen2019,Gong2019,Yan2019}.
The crystal structure of MnBi$_2$Te$_4$ is shown in Fig.~\ref{Fig-Hao2019}.
The septuple layer consisting of Te-Bi-Te-Mn-Te-Bi-Te is stacked along the [0001] direction by van der Waals forces.
A theoretical calculation of the exchange coupling constants between Mn atoms shows that the intralayer coupling in each Mn layer is ferromagnetic, while the interlayer coupling between neighboring Mn layers is antiferromagnetic \cite{Otrokov2019}.
The magnetic ground state is thus considered to be antiferromagnetic with the N\'{e}el vector pointing the out-of-plane direction (i.e., the $z$ direction), which is called A-type AFM-$z$.
The N\'{e}el temperature is reported to be about $25\ \mathrm{K}$ \cite{Otrokov2019,Lee2019,Chen2019,Yan2019}.
The unit cell of the antiferromagnetic insulator state consists of two septuple layers (Fig.~\ref{Fig-Hao2019}), where $\tau_{1/2}^c$ is the half-cell translation vector along the $c$ axis which connects nearest spin-up and spin-down Mn atomic layers.
It can be easily seen that this interlayer antiferromagnetism between the Mn atonic layers preserves the $S=\Theta \tau_{1/2}^c$ symmetry, indicating that the system is a topological antiferromagnetic insulator which we have discussed in the previous section.
Interestingly, the bulk bandgap is estimated to be about 0.2 eV \cite{Otrokov2019,Zhang2019}, which is comparable to that of the time-reversal invariant topological insulator Bi$_2$Se$_3$.
\begin{figure}[!t]
\centering
\includegraphics[width=\columnwidth]{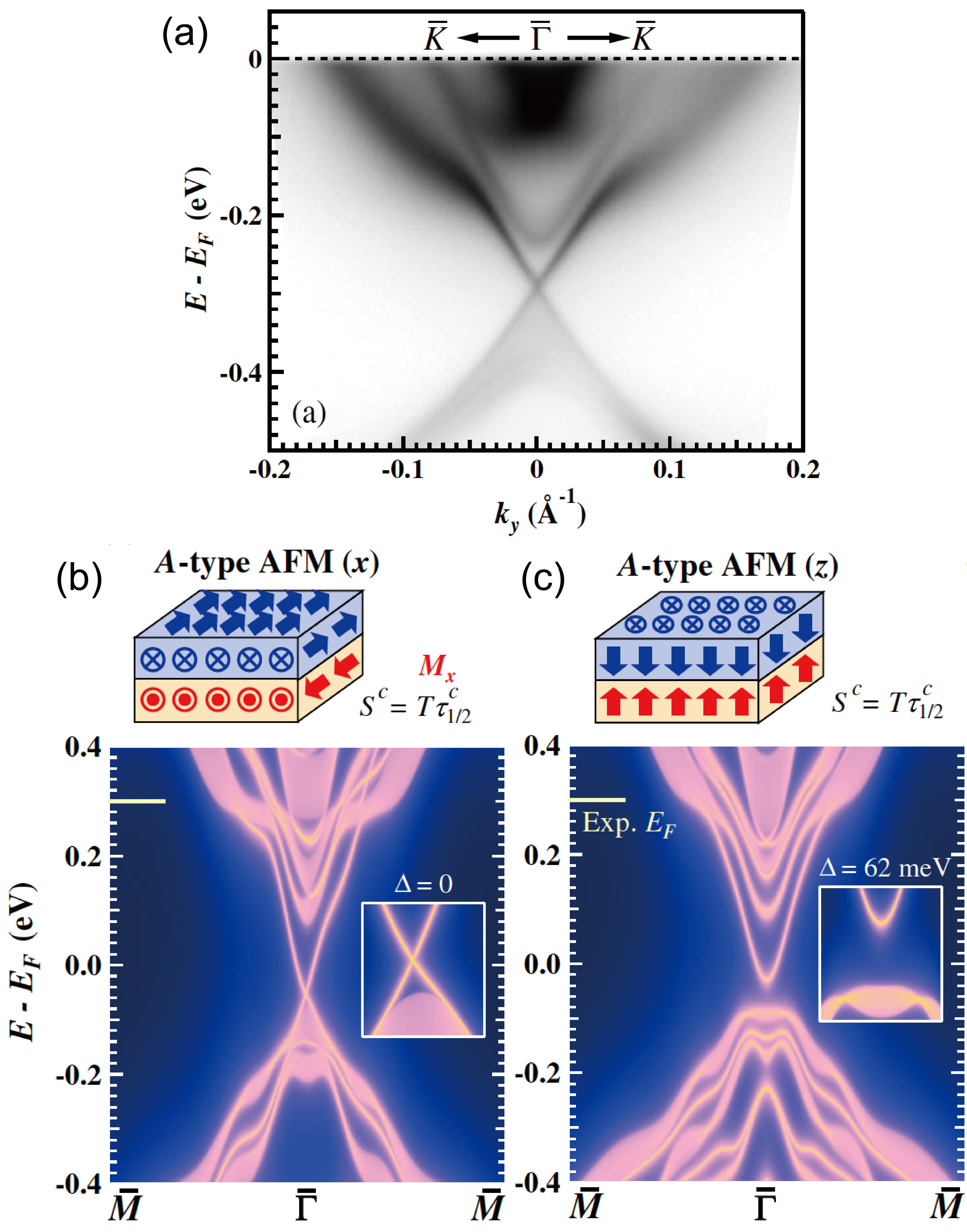}
\caption{(a) Bulk and surface spectra of MnBi$_2$Te$_4$ obtained by an ARPES measurement.
Bulk and surface spectra of MnBi$_2$Te$_4$ obtained by a first-principles calculation which assumes (b) A-type AFM with the magnetic moments along the $x$ axis and (c) A-type AFM with the magnetic moments along the $z$ axis.
Adapted from Ref.~\onlinecite{Hao2019}.}
\label{Fig-Hao2019-2}
\end{figure}
\begin{figure*}[!t]
\centering
\includegraphics[width=1.4\columnwidth]{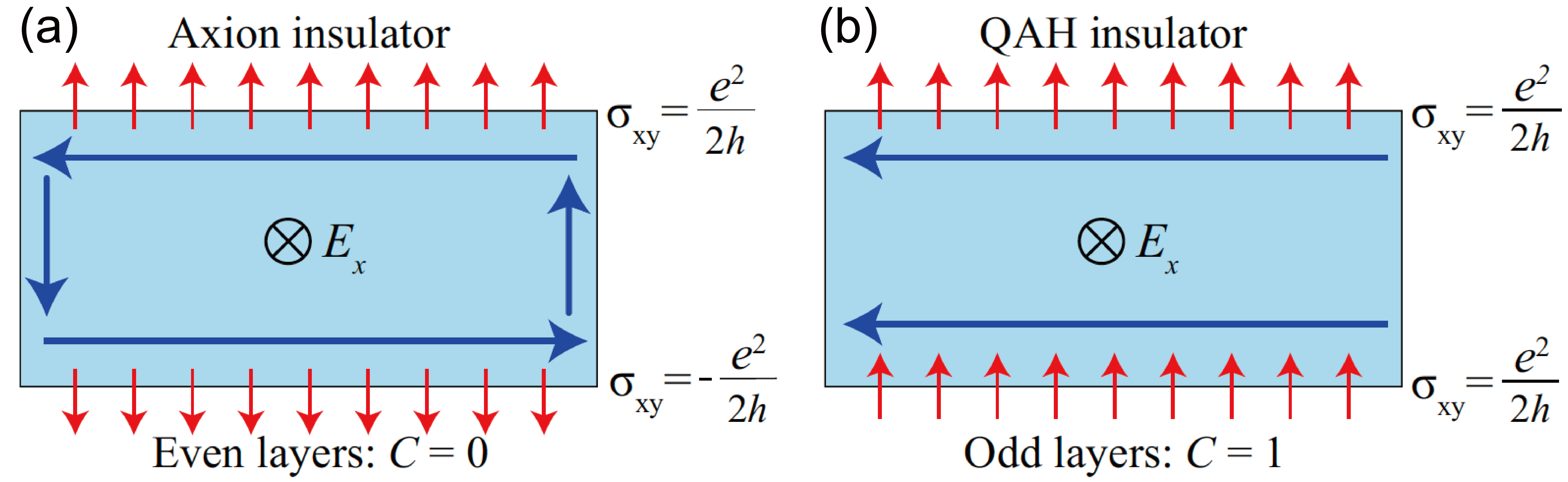}
\caption{Schematic illustration of (a) an axion insulator state realized in an even-septuple-layer MnBi$_2$Te$_4$ film and (b) a quantum anomalous Hall insulator state realized in an odd-septuple-layer MnBi$_2$Te$_4$ film.
In even-septuple-layer (odd-septuple-layer) films, the anomalous Hall conductivities of the top and bottom surfaces are opposite (the same) to each other, resulting in the total anomalous Hall conductivity $\sigma_{xy}=0$ ($\sigma_{xy}=\pm e^2/h$), or equivalently, the Chern number $C=0$ ($C=\pm1$).
Adapted from Ref.~\onlinecite{Li2019}.}
\label{Fig-Li2019}
\end{figure*}
The A-type AFM-$z$ state is invariant under spatial inversion $P_1$ with the inversion center located at the Mn atomic layer in each septuple layers.
Importantly, $P_2\Theta$ symmetry, the combination of spatial inversion $P_2$ with the inversion center located between two septuple layers and time reversal $\Theta$, is also preserved.
The presence of $P_2\Theta$ symmetry leads to doubly degenerate bands even in the absence of time-reversal symmetry \cite{Tang2016,Li2019,Li2019PRB}.
Here, following Ref.~\onlinecite{Zhang2019,Wang2020}, we derive the low-energy effective Hamiltonian of the A-type AFM-$z$ state.
$P_2\Theta$ symmetry requires that
\begin{align}
(P_2\Theta)\mathcal{H}(\bm{k})(P_2\Theta)^{-1}=\mathcal{H}(\bm{k}),
\end{align}
since momentum $\bm{k}$ changes sign under both $P_2$ and $\Theta$.
As in the case of Bi$_2$Se$_3$ [Eq.~(\ref{Bi2Se3-effective})], the low-energy effective Hamiltonian of the nonmagnetic state of MnBi$_2$Te$_4$ around the $\Gamma$ point is written in the basis of $[|P1^+_z,\uparrow\rangle,|P1^+_z,\downarrow\rangle,|P2^-_z,\uparrow\rangle,|P2^-_z,\uparrow\rangle]$, where the states $|P1^+_z,\uparrow\downarrow\rangle$ and $|P2^-_z,\uparrow\downarrow\rangle$ come from the $p_z$ orbitals of Bi and Te, respectively \cite{Zhang2019}.
In this basis, $P_2=\tau_z\otimes\bm{1}$ and $\Theta=\bm{1}\otimes i\sigma_y K$, where $\tau_i$ and $\sigma_i$ act on the orbital and spin spaces, respectively, and $K$ is complex conjugation operator.
$P_2\Theta$ symmetry constrains the possible form of the $4\times 4$ Bloch Hamiltonian $\mathcal{H}(\bm{k})=\sum_{i,j}d_{ij}(\bm{k})\tau_i\otimes\sigma_j$.
It follows that the following five matrices and the identity matrix are allowed by $P_2\Theta$ symmetry:
\begin{align}
\tau_x\otimes\sigma_x,\ \ \tau_x\otimes\sigma_y,\ \ \tau_x\otimes\sigma_z,\ \ \tau_y\otimes\bm{1},\ \ \tau_z\otimes\bm{1},
\end{align}
due to the property $(P_2\Theta)(\tau_i\otimes\sigma_j)(P_2\Theta)^{-1}=\tau_i\otimes\sigma_j$.
Note that these five matrices anticommute with each other, leading to doubly degenerate energy eigenvalues.
Using these five matrices, the low-energy effective Hamiltonian around the $\Gamma$ point is written as \cite{Zhang2019,Wang2020}
\begin{align}
\mathcal{H}(\bm{k})=\tau_x(A_2k_y\sigma_x-A_2k_x\sigma_y+m_5\sigma_z)+A_1k_z\tau_y+M(\bm{k})\tau_z,
\label{effective-Hamiltonian-MnBi2Te4}
\end{align}
where $M(\bm{k})=M+B_1k_z^2+B_2(k_x^2+k_y^2)$.
The mass $m_5$ is induced by the antiferromagnetic order.
One can see that the Hamiltonian~(\ref{effective-Hamiltonian-MnBi2Te4}) is invariant under both $P_2$ and $\Theta$ when $m_5=0$.
Indeed, the surface states of the lattice model constructed from Eq.~(\ref{effective-Hamiltonian-MnBi2Te4}) in a slab geometry in the $z$ direction exhibit the half-quantized anomalous Hall conductivity $\sigma_{xy}=\pm \mathrm{sgn}(m_5)e^2/2h$, implying the axion insulator state \cite{Wang2020}.

The surface states of antiferromagnetic MnBi$_2$Te$_4$ are somewhat complicated.
Theoretical studies have predicted that the (0001) surface state (i.e., at the surface perpendicular to the $z$ axis) which breaks the $S$ symmetry of the A-type AFM-$z$ state is gapped \cite{Otrokov2019,Zhang2019}, as indicated by the property of antiferromagnetic topological insulators (see Sec.~\ref{AF-topolgical-insulators}).
The first experimental study reported that the (0001) surface state is gapped \cite{Otrokov2019}.
However, subsequent studies reported that it is gapless \cite{Hao2019,Li2019PRX,Chen2019,Swatek2020,Nevola2020}.
Figure~\ref{Fig-Hao2019-2}(a) shows an ARPES measurement of the bulk and surface states, in which the surface state is clearly gapless Dirac cone at the (0001) surface.
Among possible spin configurations that are allowed by symmetry, Ref.~\onlinecite{Hao2019} proposed that the gapless surface state is protected by the mirror symmetry $M_x$, while the $S$ symmetry is broken at the surface.
(Note that the mirror symmetry $M_x$ is broken in the A-type AFM-$z$ state.)
In other words, A-type AFM with the magnetic moments along the $x$ axis (i.e., in-plane direction), whose bulk and surface spectra obtained by a first-principles calculation is shown in Fig.~\ref{Fig-Hao2019-2}(b), might be realized in MnBi$_2$Te$_4$ instead of the A-type AFM-$z$ shown in Fig.~\ref{Fig-Hao2019-2}(c).
These observations of the gapless surface states imply the occurrence of a surface-mediated spin reconstruction.

As pointed in Ref.~\onlinecite{Zhang2019}, it should be noted here that the antiferromagnetic order in MnBi$_2$Te$_4$ is essentially different from such an antiferromagnetic order in Fe-doped Bi$_2$Se$_3$ which has been proposed to realize a dynamical axion field \cite{Li2010}.
In the latter case, time-reversal $\Theta$ and inversion symmetries are both broken, allowing the deviation of the value of $\theta$ from $\pi$.
The antiferromagnetic fluctuation contributes to the dynamical axion field at linear order in the N\'{e}el field.
In contrast, in MnBi$_2$Te$_4$ an effective time-reversal $S$ symmetry and inversion symmetry are both preserved, keeping the quantization $\theta=\pi$ and making no contribution to the dynamical axion field at linear order in the N\'{e}el field.

\subsubsection{Transport properties of MnBi$_2$Te$_4$ thin films}
Due to the intralayer ferromagnetism and interlayer antiferromagnetism of the Mn layers, the layered van der Waals crystal MnBi$_2$Te$_4$ exhibit interesting properties in its few-layer thin films.
In even-septuple-layer films, $P_2$ and $\Theta$ symmetries are both broken, but $P_2\Theta$ symmetry is preserved \cite{Li2019}.
As we have seen above, the presence of $P_2\Theta$ symmetry leads to doubly degenerate bands.
On the other hand, in odd-septuple-layer films, $P_1$ symmetry is preserved, but $\Theta$ and $P_1\Theta$ symmetries are both broken, leading to spin-split bands \cite{Li2019}.
Consequently, the Chern number is zero in even-septuple-layer films as required by the $P_2\Theta$ symmetry, while the Chern number in odd-septuple-layer films can be nonzero.
Indeed, first-principles calculations show that there exist gapless chiral edge states in odd-septuple-layer films, whereas there do not in even-septuple-layer films \cite{Otrokov2019PRL,Li2019}.
It should be noted that the zero-Chern-number state with $\sigma_{xy}=0$ is realized by the combination of half-quantized anomalous Hall states with opposite conductivities $\sigma_{xy}=\pm e^2/2h$ at the top and bottom surfaces, as shown in Fig.~\ref{Fig-Li2019}(a).
In other words, this state is an axion insulator exhibiting a topological magnetoelectric effect with the quantized coefficient $\theta=\pi$ (see Sec.~\ref{Phenomenological-derivation} for a phenomenological derivation of the topological magnetoelectric effect).
In contrast, even-septuple-layer films have the quantized anomalous Hall conductivity $\sigma_{xy}=\pm e^2/h$ which results from the half-quantized anomalous Hall conductivity $\sigma_{xy}=\pm e^2/2h$ of the same sign at the top and bottom surfaces, giving rise to the Chern number $C=\pm 1$ as shown in Fig.~\ref{Fig-Li2019}(b).

\begin{figure}[!t]
\centering
\includegraphics[width=\columnwidth]{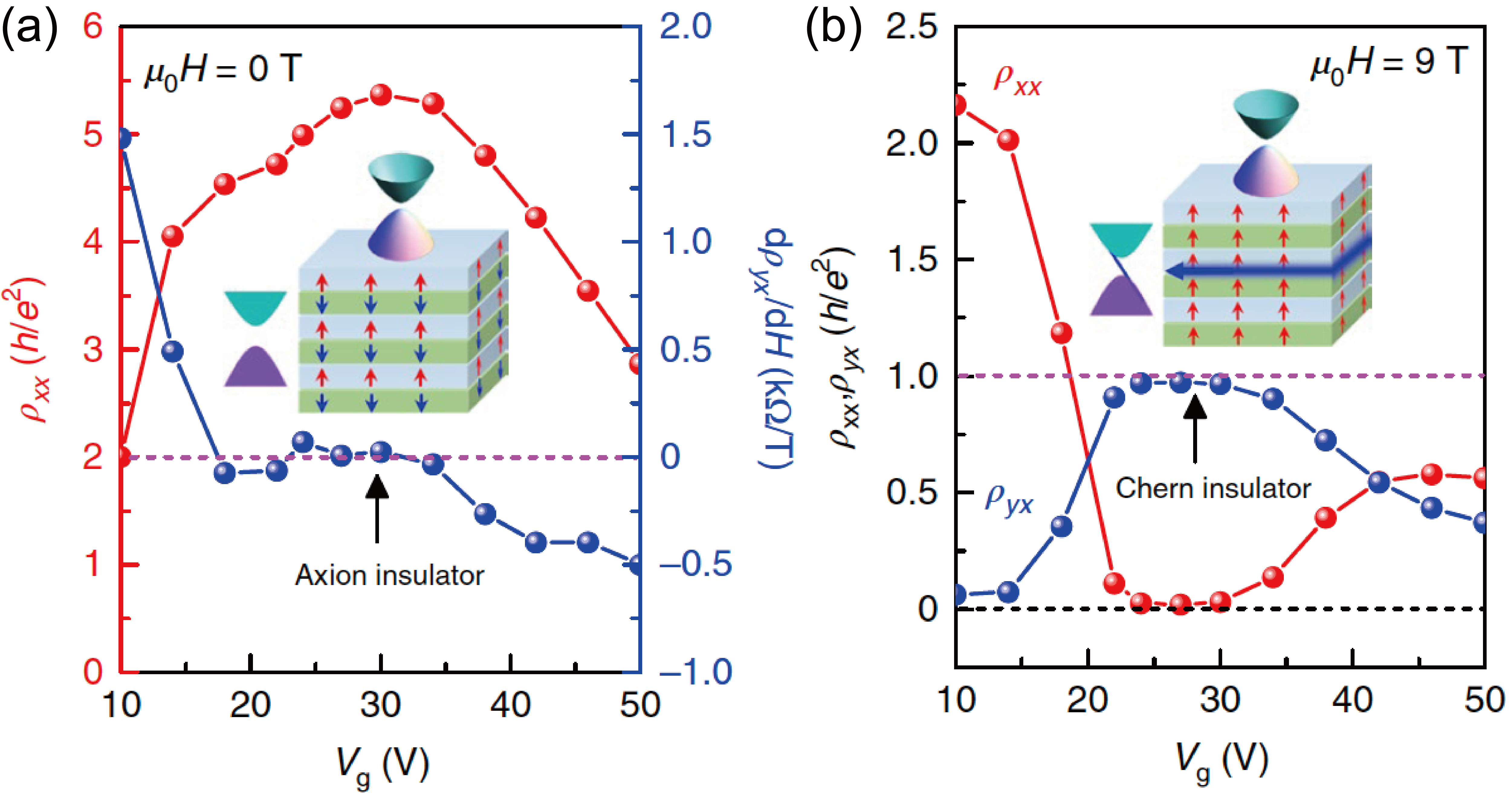}
\caption{Resistivity measurement in a six-septuple-layer MnBi$_2$Te$_4$ film, showing (a) an axion insulator behavior with a zero Hall plateau at zero magnetic field and (b) a Chern insulator behavior with the quantized Hall resistivity $h/e^2$ in a magnetic field of $9\ \mathrm{T}$.
Adapted from Ref.~\onlinecite{Liu2020}.}
\label{Fig-Liu2020}
\end{figure}
\begin{figure}[!t]
\centering
\includegraphics[width=\columnwidth]{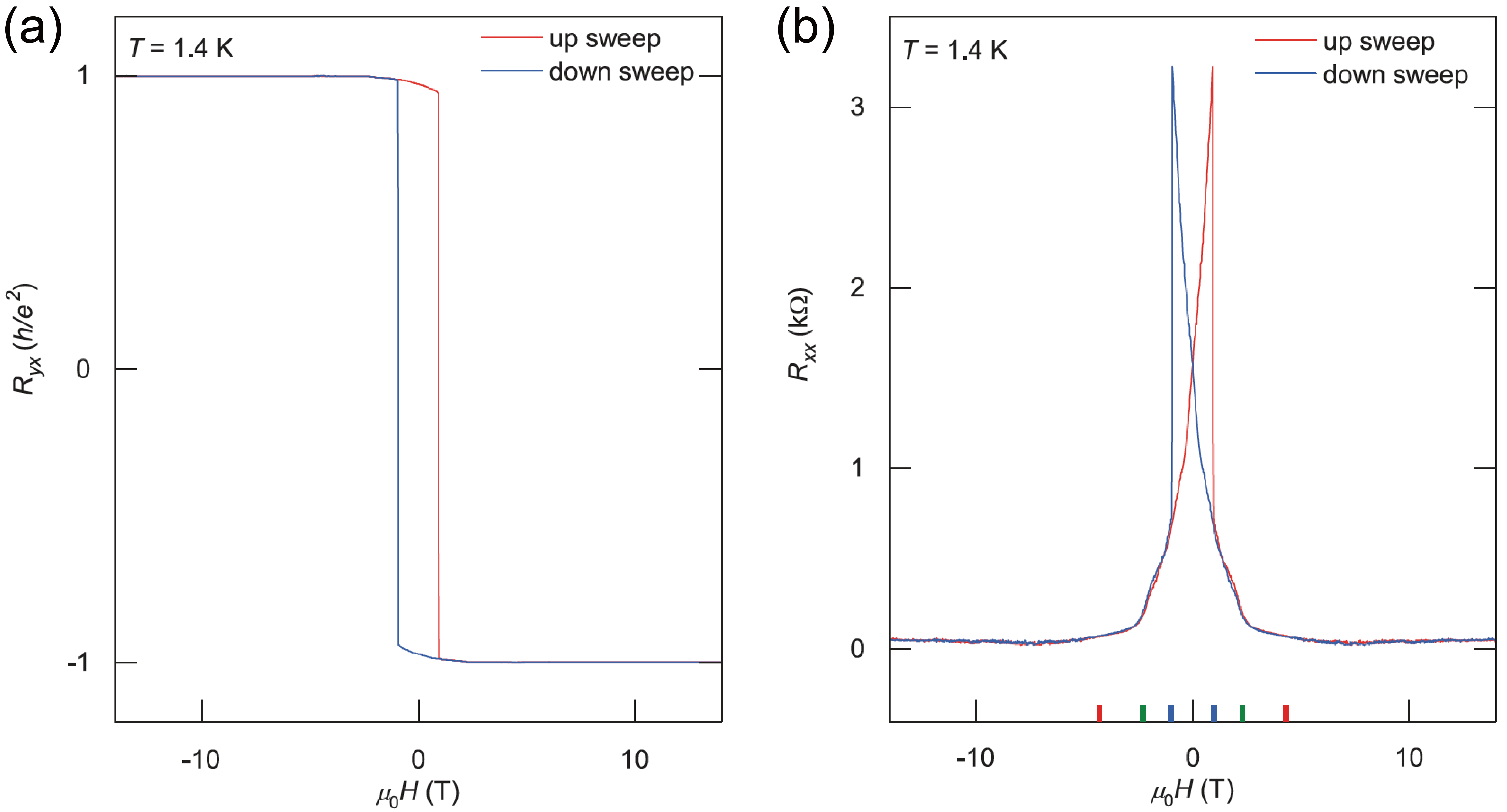}
\caption{Resistivity measurement in a five-septuple-layer MnBi$_2$Te$_4$ film, showing a quantum anomalous Hall effect with the quantized transverse resistivity $h/e^2$ at zero magnetic field.
Adapted from Ref.~\onlinecite{Deng2020}.}
\label{Fig-Deng2020}
\end{figure}
Experimental observations that are consistent with theoretical predictions have been made.
Figure~\ref{Fig-Liu2020} shows the resistivity measurement in a six-septuple-layer MnBi$_2$Te$_4$ film \cite{Liu2020}, in which an axion insulator behavior with a zero Hall plateau at zero magnetic field and a Chern insulator behavior with the quantized Hall resistivity $h/e^2$ in a strong magnetic field were clearly observed.
Also, the change in the Chern number between $C=\pm1$ was observed in response to the change in the magnetic field direction.
Figure~\ref{Fig-Deng2020} shows the resistivity measurement in a five-septuple-layer MnBi$_2$Te$_4$ film \cite{Deng2020}, in which a quantum anomalous Hall effect with the quantized Hall resistivity $h/e^2$ was clearly observed.

\subsection{MnBi$_2$Te$_4$ family of materials}
\begin{figure}[!t]
\centering
\includegraphics[width=0.85\columnwidth]{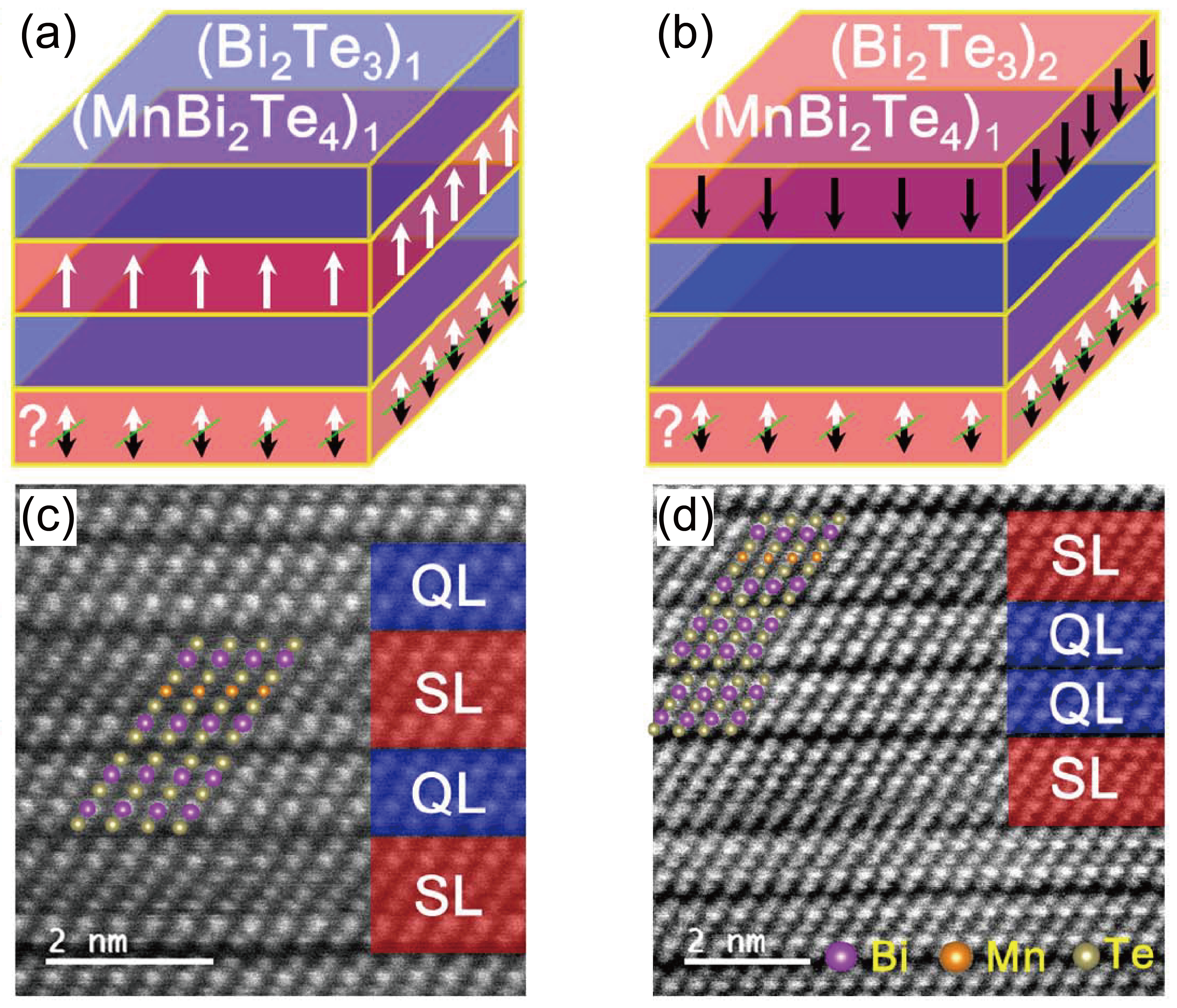}
\caption{Schematic illustrations of (a) MnBi$_4$Te$_7$ and (b) MnBi$_6$Te$_{10}$.
STEM images of (c) MnBi$_4$Te$_7$ and (d) MnBi$_6$Te$_{10}$, showing layered heterostrucrutures.
Here, QL and SL indicate a quintuple layer of Bi$_2$Te$_3$ and a septuple layer of MnBi$_2$Te$_4$, respectively.
Adapted from Ref.~\onlinecite{Wu2019}.}
\label{Fig-Wu2019}
\end{figure}
Taking advantage of the nature of van der Waals materials, the layered van der Waals heterostructures of (MnBi$_2$Te$_4$)$_m$(Bi$_2$Te$_3$)$_n$ can be synthesized.
Here, it is well known that Bi$_2$Te$_3$ is a time-reversal invariant topological insulator \cite{Zhang2009}.
So far, MnBi$_4$Te$_7$ ($m=n=1$) \cite{Wu2019,Vidal2019a,Hu2020,Wu2020PRX,Yan2020} and MnBi$_6$Te$_{10}$ ($m=1$ and $n=2$) \cite{Wu2019,Jo2020,Yan2020} have been experimentally realized.
Figure~\ref{Fig-Wu2019} shows schematic illustrations of MnBi$_4$Te$_7$ and MnBi$_6$Te$_{10}$ and their STEM images.
In MnBi$_4$Te$_7$, a quintuple layer of Bi$_2$Te$_3$ and a septuple layer of MnBi$_2$Te$_4$ stack alternately.
In MnBi$_6$Te$_{10}$, two quintuple layers of Bi$_2$Te$_3$ are sandwiched by septuple layers of MnBi$_2$Te$_4$.
As in the case of MnBi$_2$Te$_4$, interlayer antiferromagnetism (between Mn layers) develops with a N\'{e}el temperature $T_{\mathrm{N}}=13\ \mathrm{K}$ in MnBi$_4$Te$_7$ \cite{Wu2019,Vidal2019a,Yan2020} and $T_{\mathrm{N}}=11\ \mathrm{K}$ in MnBi$_6$Te$_{10}$ \cite{Yan2020}, and this antiferromagnetic insulator state is protected by the $S=\Theta T_{1/2}$ symmetry, which indicates that MnBi$_4$Te$_7$ and MnBi$_6$Te$_{10}$ are also antiferromagnetic topological insulators.

\begin{figure}[!t]
\centering
\includegraphics[width=0.85\columnwidth]{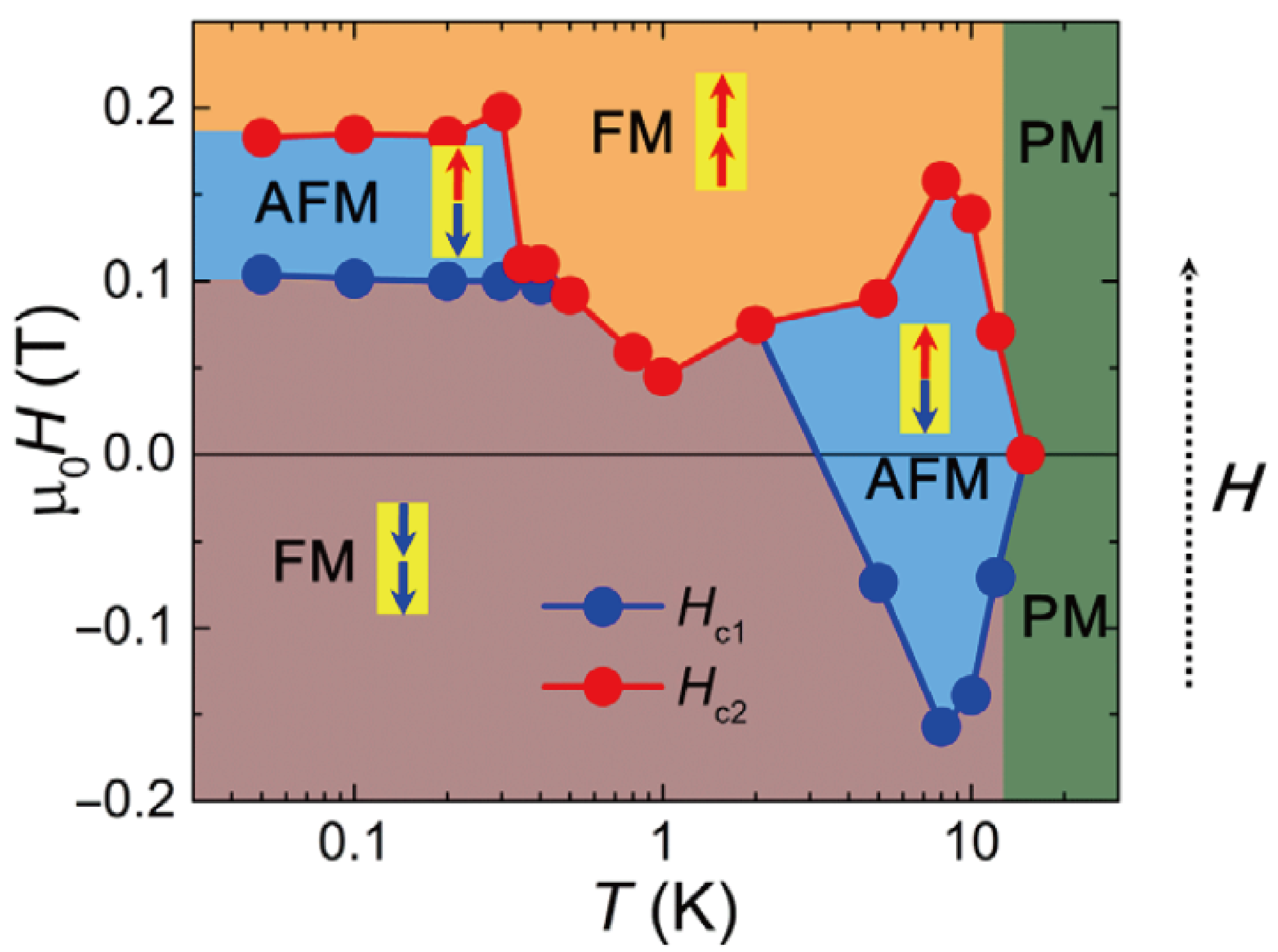}
\caption{Magnetic phase diagram of MnBi$_4$Te$_7$ as functions of temperature and out-of-plane magnetic field, showing a complex competition between antiferromagnetism (AFM) and ferromagnetism (FM).
Adapted from Ref.~\onlinecite{Wu2019}.}
\label{Fig-Wu2019-2}
\end{figure}

It was reported that, due to the gradual weakening of the antiferromagnetic exchange coupling associated with the  increasing separation distance between Mn layers, a competition between antiferromagnetism and ferromagnetism occurs at low temperature $\approx 5\ \mathrm{K}$ \cite{Wu2019,Vidal2019a}.
A magnetic phase diagram of MnBi$_4$Te$_7$ is shown in Fig.~\ref{Fig-Wu2019-2}.
Also, two distinct types of topological surface states are realized depending on the Bi$_2$Te$_3$ quintuple-layer termination or the MnBi$_2$Te$_4$ septuple-layer termination \cite{Hu2020,Wu2020PRX}.
ARPES studies showed that the Bi$_2$Te$_3$ quintuple-layer termination gives rise to gapped surface states, while the MnBi$_2$Te$_4$ septuple-layer termination gives rise to gapless surface states \cite{Hu2020,Wu2020PRX}.
Note that these terminations break the $S$ symmetry, which implies in principle gapped surface states (see Sec.~\ref{AF-topolgical-insulators}).
It is suggested that the gap opening in the Bi$_2$Te$_3$ quintuple-layer termination can be explained by the magnetic proximity effect from the MnBi$_2$Te$_4$ septuple layer beneath, and that the gaplessness in MnBi$_2$Te$_4$ septuple-layer termination can be explained by the restoration of time-reversal symmetry at the septuple-layer surface due to disordered spin \cite{Wu2020PRX}.
On the other hand, an ARPES study of MnBi$_6$Te$_{10}$ observed a gapped Dirac surface state in the MnBi$_2$Te$_4$ septuple-layer termination \cite{Jo2020}.

Since the bulk crystals of MnBi$_4$Te$_7$ and MnBi$_6$Te$_{10}$ are realized by van der Waals forces, various heterostructures in the 2D limit, which are made from the building blocks of the MnBi$_2$Te$_4$ septuple layer and the Bi$_2$Te$_3$ quintuple layer, can be obtained by exfoliation.
A theoretical calculation shows that such 2D heterostructures exhibit the quantum spin-Hall effect without time-reversal symmetry and quantum anomalous Hall effect \cite{Sun2019}.
Theoretically, it is suggested that (MnBi$_2$Te$_4$)(Bi$_2$Te$_3$)$_n$ is a higher-order topological insulator hosting surface states with a M\"{o}bius twist \cite{Zhang2020PRL}.
In contrast to MnBi$_2$Te$_4$ in which the value of $\theta$ is quantized to be $\pi$, it is suggested that the antiferromagnetic insulator phases of Mn$_2$Bi$_6$Te$_{11}$ (with $m=2$ and $n=1$) \cite{Wang2020} and Mn$_2$Bi$_2$Te$_5$ \cite{Zhang2020CPL} in which the $S$ symmetry is absent, break both time-reversal and inversion symmetries, realizing a dynamical axion field.

\subsection{EuIn$_2$As$_2$ and EuSn$_2$As$_2$}
\begin{figure}[!t]
\centering
\includegraphics[width=0.8\columnwidth]{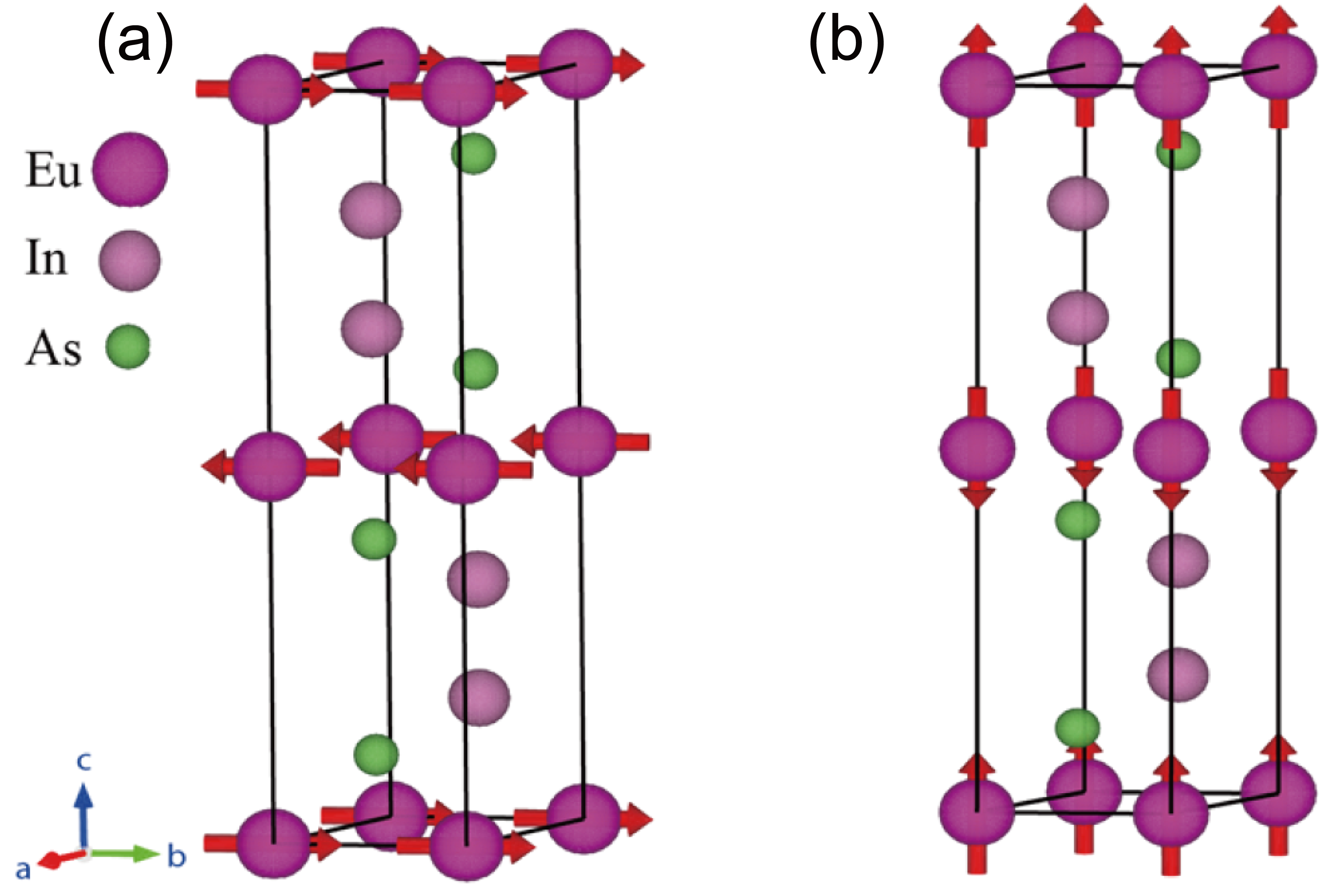}
\caption{Crystal and magnetic structure of EuIn$_2$As$_2$.
There are two metastable magnetic structures where the magnetic moments align parallel to (a) the $b$ axis and (b) the $c$ axis.
Adapted from Ref.~\onlinecite{Xu2019}.}
\label{Fig-Xu2019}
\end{figure}

EuIn$_2$As$_2$ and EuSn$_2$As$_2$ have also been considered a candidate class of materials for antiferromagnetic topological insulators with inversion symmetry \cite{Xu2019}.
Different from MnBi$_2$Te$_4$ which is a layered van der Waals material, EuIn$_2$As$_2$ has a three-dimensional crystal structure as shown in Fig.~\ref{Fig-Xu2019}.
EuSn$_2$As$_2$ has a very similar crystal and magnetic structure to EuIn$_2$As$_2$.
Two metastable magnetic structures with the magnetic moments parallel to the $b$ axis (AFM$\parallel b$) and the $c$ axis (AFM$\parallel c$) have been known in EuIn$_2$As$_2$ and EuSn$_2$As$_2$ \cite{Goforth2008,Zhang2020PRB}.
As in the case of MnBi$_2$Te$_4$, the antiferromagnetic insulator phases of EuIn$_2$As$_2$ and EuSn$_2$As$_2$ are protected by the $S=\Theta T_{1/2}$ symmetry, with the half-unit-cell translation vector connecting four Eu atoms along the $c$ axis.
Indeed, ARPES measurements in EuIn$_2$As$_2$ \cite{Sato2020} and EuSn$_2$As$_2$ \cite{Li2019PRX} suggests that they are antiferromagnetic topological insulators.
Theoretically, it is suggested that antiferromagnetic EuIn$_2$As$_2$ (both AFM$\parallel b$ and AFM$\parallel c$) is  at the same time a higher-order topological insulator with gapless chiral hinge states lying within the gapped surface states \cite{Xu2019}.
%

\section{Expressions for $\theta$ in insulators \label{Sec-Expressions-for-theta}}
We have seen in Sec.~\ref{Sec-Quantized-Magnetoelectric-Effect} that time-reversal symmetry and inversion symmetry impose the constraint on the coefficient $\theta$ of the topological magnetoelectric effect such that $\theta=\pi$ in 3D topological insulators and $\theta=0$ in 3D normal insulators.
In this section, first we derive a generic expression for $\theta$ which is given in terms of the Bloch-state wave function.
Then, we show explicitly that the value of $\theta$ can be arbitrary in a class of antiferromagnetic insulators with broken time-reversal and inversion symmetries, taking a microscopic tight-binding model called the Fu-Kane-Mele-Hubbard model as an example.

\subsection{General expression for $\theta$ from the dimensional reduction \label{Sec-Dimensional-Reduction}}
It is known that the chiral anomaly in (1+1) dimensions can be derived from the dimensional reduction from (2+1)D Chern-Simons action.
A similar way of deriving the effective action of (3+1)D time-reversal invariant topological insulators from the dimensional reduction from (4+1)D Chern-Simons action was considered in Ref~\onlinecite{Qi2008}.
To see this, let $k_w$ be the momentum in the fourth dimension and $(k_x, k_y, k_z)$ be the momentum in 3D spatial dimensions.
The second Chern number in 4D momentum space $(k_x, k_y, k_z, k_w)$ is given by \cite{Niemi1983,Golterman1993,Qi2008}
\begin{align}
\nu^{(2)}=\frac{1}{32\pi^2}\int d^4k\, \varepsilon^{ijkl}\mathrm{tr}\left[f_{ij}f_{kl}\right],
\label{Second-Chern-number}
\end{align}
where
\begin{align}
f_{ij}&=\partial_i\mathcal{A}_j-\partial_j\mathcal{A}_i-i[\mathcal{A}_i,\mathcal{A}_j], \nonumber\\
\mathcal{A}^{\alpha\beta}_j&=i\langle u_\alpha|\partial_{k_j}|u_\beta\rangle.
\label{expression-for-f_ij}
\end{align}
Here, $|u_\alpha\rangle$ is the periodic part of the Bloch wave function of the occupied band $\alpha$.
By substituting the explicit expression for $f_{ij}$~(\ref{expression-for-f_ij}) into Eq.~(\ref{Second-Chern-number}), we obtain
\begin{align}
\nu^{(2)}&=\frac{1}{8\pi^2}\int d^4k\, \frac{\partial}{\partial k_w}\left\{\varepsilon^{4jkl}\mathrm{tr}\left[\mathcal{A}_j\partial_k\mathcal{A}_l-\frac{2}{3}i\mathcal{A}_j\mathcal{A}_k\mathcal{A}_l\right]\right\}\nonumber\\
&\equiv \int dk_w\, \frac{\partial P_3(k_w)}{\partial k_w},
\label{Second-Chern-number2}
\end{align}
where $j,k,l=1,2,3$ indicate the 3D spatial direction.
Here, note that $\varepsilon^{4jkl}=-\varepsilon^{jkl4}\equiv-\varepsilon^{jkl}$ due to the convention $\varepsilon^{1234}=1$.
On the other hand, the corresponding topological action in (4+1) dimension $(x,y,z,w)$ is given by
\begin{align}
S&=\frac{\nu^{(2)}}{24\pi^2}\int dtd^4x\, \varepsilon^{\mu\nu\rho\sigma\tau}A_{\mu}\partial_\nu A_{\rho}\partial_\sigma A_\tau,
\end{align}
which can be rewritten as
\begin{align}
S&=\frac{\nu^{(2)}}{8\pi^2}\int dtd^3xdw\, \varepsilon^{4\nu\rho\sigma\tau}A_4\partial_\nu A_{\rho}\partial_\sigma A_\tau\nonumber\\
&=\frac{1}{32\pi^2}\int dtd^3x\, \theta(\bm{r},t)\varepsilon^{\nu\rho\sigma\tau}F_{\nu\rho}F_{\sigma\tau},
\end{align}
where we have used the identity $\varepsilon^{4\nu\rho\sigma\tau}=\varepsilon^{\nu\rho\sigma\tau}$, and defined
$\theta(\bm{r},t)\equiv {\nu^{(2)}}\phi$.
Here, $\phi=\oint dw\ A_4(\bm{r},w,t)$ can be regarded as the flux due to the extra dimension.
In analogy with the (1+1)D case in which the first Chern number is given by $\nu^{(1)}=\int d\phi \partial P/\partial \phi$ with $P$ the electric polarization, Eq.~(\ref{Second-Chern-number2}) indicates a relation between the generalized polarization $P_3$ and the Chern number $\nu^{(2)}$.
Then, it follows that
$P_3={\nu^{(2)}}\phi/2\pi$.
Finally, we arrive at a general expression for $\theta$ \cite{Qi2008,Essin2009}
\begin{align}
\theta=-\frac{1}{4\pi}\int_{\rm BZ}d^3k\, \epsilon^{ijk}\mathrm{tr}\left[\mathcal{A}_i\partial_j\mathcal{A}_k-\frac{2}{3}i\mathcal{A}_i\mathcal{A}_j\mathcal{A}_k\right],
\label{exact-theta1}
\end{align}
where $i,j,k=1,2,3$, $d^3k=dk_xdk_ydk_z$, and the integration is done over the Brillouin zone of the system.
Equation~(\ref{exact-theta1}) can be derived more rigorously and microscopically, starting from a generic Bloch Hamiltonian and its wave function \cite{Malashevich2010,Essin2010}.

\begin{figure}[!t]
\centering
\includegraphics[width=0.85\columnwidth]{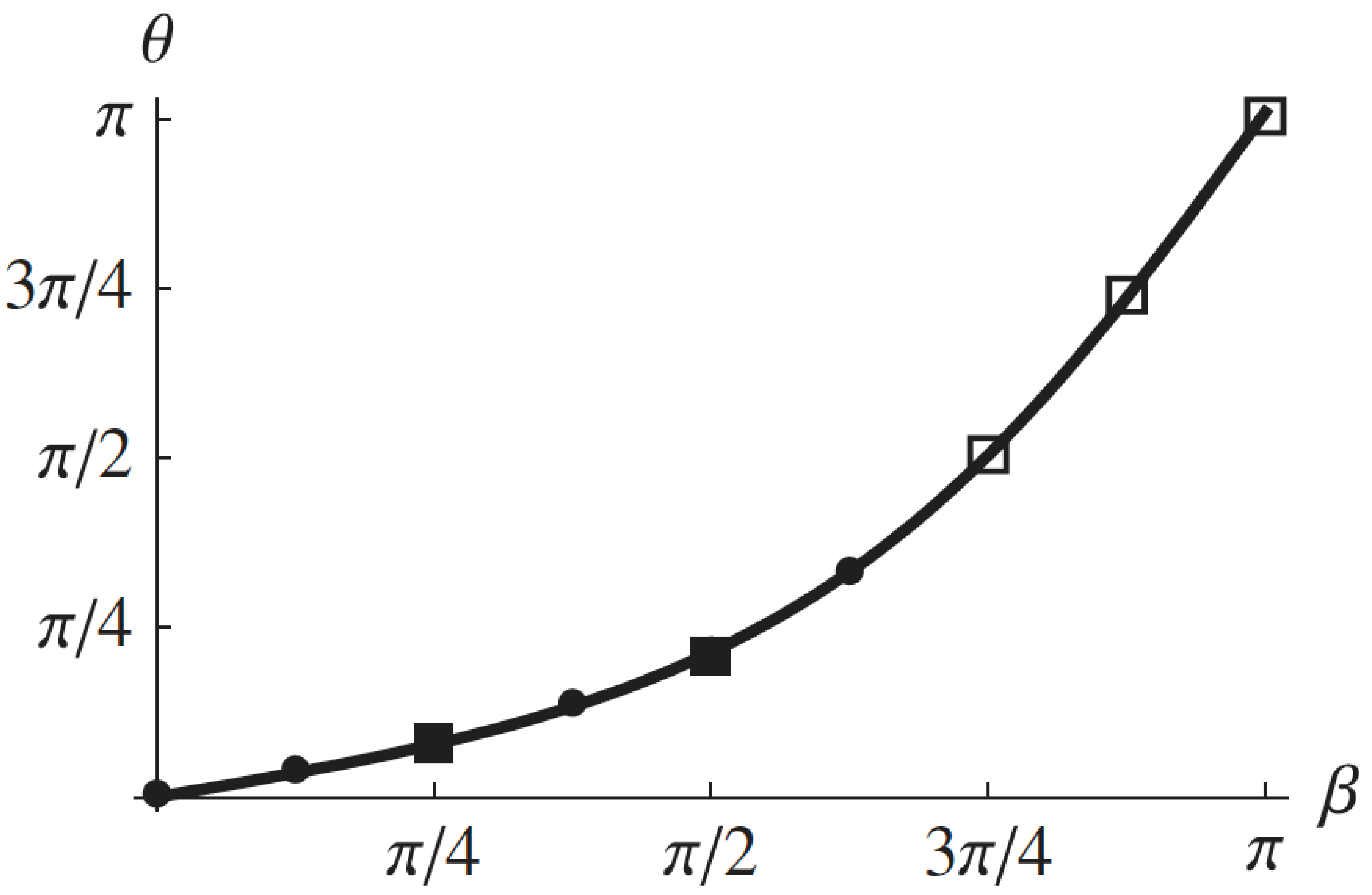}
\caption{Numerically obtained value of $\theta$ in the Fu-Kane-Mele model on a diamond lattice.
Here, $\beta=\tan^{-1}(|\bm{h}|/\delta t_1)$ with $\bm{h}(=U\bm{n})$ being a staggered Zeeman field in the [111] direction of the diamond lattice, and $\delta t_1$ being the hopping strength anisotropy due to the lattice distortion in  the [111] direction.
When $\beta=\pi$ ($\beta=0$), the system is a topological (normal) insulator.
Adapted from Ref.~\onlinecite{Essin2009}.}
\label{Fig-Essin2009}
\end{figure}
Figure~\ref{Fig-Essin2009} shows a numerically calculated value of $\theta$ using Eq.~(\ref{exact-theta1}) and other equivalent expressions for $\theta$ in the Fu-Kane-Mele model on a diamond lattice with a staggered Zeeman field that breaks both time-reversal and inversion symmetries \cite{Essin2009}.
One can see that the value of $\theta$ is no longer quantized once time-reversal symmetry is broken and varies continuously between $\theta=0$ corresponding to the case of a normal insulator and $\theta=\pi$ corresponding to the case of a topological insulator.

\subsection{Expression for $\theta$ in topological magnetic insulators}
A generic expression for $\theta$ [Eq.~(\ref{exact-theta1})] is applicable to arbitrary band structure.
However, some techniques (such as choosing a gauge for the Berry connection $\mathcal{A}$) are required to calculate numerically.
On the other hand, it has been shown that there exists an explicit expression for $\theta$ that can be calculated easily from the Bloch Hamiltonian of a certain class of insulators with broken time-reversal and inversion symmetries \cite{Li2010}, which calculation does not rely on a specific choice of gauge.
Here, we consider a generic $4\times 4$ Bloch Hamiltonian of the form
\begin{align}
\mathcal{H}(\bm{k})=\sum_{i=1}^5R_i(\bm{k})\alpha_i,
\label{Generic-Hamiltonian-alpha_5}
\end{align}
with matrices $\alpha_i$ satisfying the Clifford algebra $\{\alpha_i,\alpha_j \}=2\delta_{ij}\bm{1}$.
Here, the matrix $\alpha_4$ is invariant under both time reversal and spatial inversion.
Specifically, it has been known that the antiferromagnetic insulator phases of 3D correlated systems with spin-orbit coupling, such as Bi$_2$Se$_3$ doped with magnetic impurities such as Fe \cite{Li2010} and $5d$ transition-metal oxides with the corundum structure \cite{Wang2011}, can be described by Eq.~(\ref{Generic-Hamiltonian-alpha_5}).
More recently, it has been suggested that van der Waals layered antiferromagnets such as Mn$_2$Bi$_6$Te$_{11}$ \cite{Wang2020} and Mn$_2$Bi$_2$Te$_5$ \cite{Zhang2020CPL} can also be described by Eq.~(\ref{Generic-Hamiltonian-alpha_5}).
In such systems, we can calculate the value of $\theta$ using the following expression \cite{Li2010,Wang2011}:
\begin{align}
\theta&=\frac{1}{4\pi}\int_{\rm BZ} d^3 k\, \frac{2|R|+R_4}{(|R|+R_4)^2|R|^3}\epsilon^{ijkl}R_i\frac{\partial R_j}{\partial k_x}\frac{\partial R_k}{\partial k_y}\frac{\partial R_l}{\partial k_z},
\label{exact-theta2}
\end{align}
where $i,j,k,l=1,2,3,5$, $|R|=\sqrt{\sum_{i=1}^5R_i^2}$, and the integration is done over the Brillouin zone.

\subsubsection{Four-band Dirac model}
Let us derive a simpler expression for $\theta$ in systems whose effective continuum Hamiltonian is given by a massive Dirac Hamiltonian.
We particularly consider a generic Dirac Hamiltonian with a symmetry-breaking mass term of the form
\begin{align}
\mathcal{H}(\bm{q})=q_x\alpha_1+q_y\alpha_2+q_z\alpha_3+m_0\alpha_4+m_5\alpha_5,
\label{Effective-Hamiltonian-alpha_5}
\end{align}
which can be derived by expanding Eq.~(\ref{Generic-Hamiltonian-alpha_5}) around some momentum points $X$ and retaining only the terms linear in $\bm{q}=\bm{k}-X$.
Here, the matrix $\alpha_4$ is invariant under both time reversal and spatial inversion and the matrix $\alpha_5=\alpha_1\alpha_2\alpha_3\alpha_4$ breaks both time-reversal and inversion symmetries.
In other words, the system has both time-reversal and inversion symmetries when $m_5=0$.
For concreteness, we require that the system be a time-reversal invariant topological insulator when $m_0<0$, as we have considered in Eq.~(\ref{Hamiltonian-3DTI}).
The action of the system in the presence of an external electromagnetic potential $A_\mu$ is given by [see also Eq.~(\ref{action-Dirac-fermion})]
\begin{align}
S=\int dtd^3 r\, \bar{\psi}(\bm{r},t)\left[i\gamma^\mu (\partial_\mu-ieA_\mu)-m' e^{i\theta\gamma^5}\right]\psi(\bm{r},t),
\label{Effective-action-alpha_5}
\end{align}
where $t$ is real time, $\psi(\bm{r},t)$ is a four-component spinor, $\bar{\psi}=\psi^\dagger\gamma^0$, $m'=\sqrt{(m_0)^2+(m_5)^2}$, $\cos\theta=m_0/m'$, $\sin\theta=-m_5/m'$, and we have used the fact that $\alpha_4=\gamma^0$, $\alpha_5=-i\gamma^0\gamma^5$ and $\alpha_j=\gamma^0\gamma^j$ ($j=1,2,3$).
Here, the gamma matrices satisfy the identities $\{\gamma^\mu,\gamma^5\}=0$ and $\{\gamma^\mu,\gamma^\nu\}=2g^{\mu\nu}$ with $g^{\mu\nu}=\mathrm{diag}(1,-1,-1,-1)$ ($\mu,\nu=0,1,2,3$).
One can see that the action~(\ref{Effective-action-alpha_5}) is identical to Eq.~(\ref{TME-TI-Action}), except for the generic value of $\theta$ in the exponent.
By applying Fujikawa's method to the action (\ref{Effective-action-alpha_5}), the $\theta$ term is obtained as \cite{Sekine2014,Sekine2016}
\begin{align}
S_\theta&=\int dtd^3 r\, \frac{e^2}{2\pi h}\theta \bm{E}\cdot\bm{B},
\label{S_theta_realtime2}
\end{align}
where
\begin{align}
\theta=\frac{\pi}{2}[1-\mathrm{sgn}(m_0)]-\tan^{-1}\left(\frac{m_5}{m_0}\right).
\label{Expression-theta-m_5-mass}
\end{align}
Here, the first term in Eq.~(\ref{Expression-theta-m_5-mass}) is $0$ or $\pi$, which describes whether the system is topologically trivial or nontrivial.
The second term in Eq.~(\ref{Expression-theta-m_5-mass}) describes the deviation from the quantized value due to the $m_5$ mass.
Note that $\tan^{-1}(m_5/m_0)\approx m_5/m_0$, i.e., the deviation is proportional to $m_5$ when $m_5\ll m_0$.

\subsubsection{Fu-Kane-Mele-Hubbard model on a diamond lattice \label{Sec-FKMH-model}}
In Eq.~(\ref{Expression-theta-m_5-mass}) we have seen that the $m_5$ mass term which breaks both time-reversal and inversion symmetries generates a deviation of the value of $\theta$ from the quantized value $\pi$ or $0$.
Here, following Ref.~\onlinecite{Sekine2014}, we discuss a microscopic origin of this $m_5$ mass term and derive an expression for $\theta$ of the form of Eq.~(\ref{Expression-theta-m_5-mass}) in a 3D correlated system with spin-orbit coupling.
To this end, we start with the Fu-Kane-Mele-Hubbard (FKMH) model on a diamond lattice, whose tight-binding Hamiltonian is given by \cite{Fu2007,Fu2007a,Sekine2014,Sekine2016}
\begin{align}
H=&\ \sum_{\langle i,j\rangle,\sigma}t_{ij}c^\dag_{i\sigma}c_{j\sigma}+i\frac{4\lambda}{a^2}\sum_{\langle\langle i,j\rangle\rangle}c^\dag_{i}\bm{\sigma}\cdot(\bm{d}^1_{ij}\times\bm{d}^2_{ij})c_{j}\nonumber\\
&+U\sum_i n_{i\uparrow}n_{i\downarrow},
\label{FKMH-Hamiltonian}
\end{align}
where $c^\dag_{i\sigma}$ is an electron creation operator at a site $i$ with spin $\sigma(=\uparrow,\downarrow)$, $n_{i\sigma}=c^\dag_{i\sigma}c_{i\sigma}$, and $a$ is the lattice constant of the fcc lattice.
$\bm{d}^1_{ij}$ and $\bm{d}^2_{ij}$ are the two vectors which connect two sites $i$ and $j$ on the same sublattice.
$\bm{\sigma}=(\sigma_x,\sigma_y,\sigma_z)$ are the Pauli matrices for the spin degree of freedom.
The first through third terms in Eq.~(\ref{FKMH-Hamiltonian}) represent the nearest-neighbor hopping, the next-nearest-neighbor spin-orbit coupling, and the on-site repulsive electron-electron interactions, respectively.
\begin{figure}[!t]
\centering
\includegraphics[width=\columnwidth]{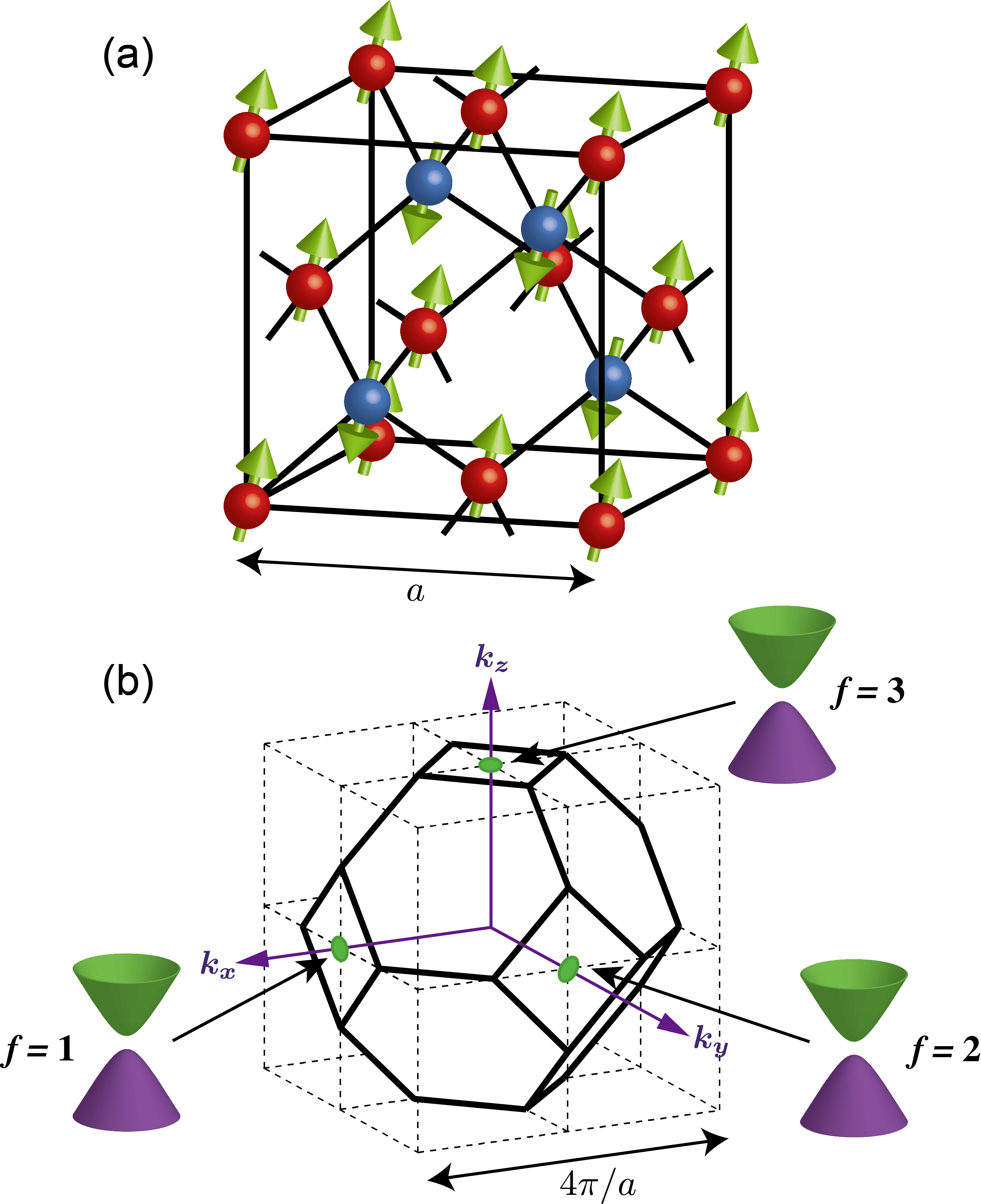}
\caption{(a) Schematic illustration of the antiferromagnetic order between the two sublattices (denoted by red and blue) in the FKMH model.
(b) The first Brillouin zone of an fcc lattice.
Around the $X_r$ points with $r=x,y,z$ (represented by green circles), massive Dirac Hamiltonians are derived.}
\label{Fig-diamondlattice}
\end{figure}

In the mean-field approximation, the interaction term is decomposed as
$U\sum_i n_{i\uparrow}n_{i\downarrow}\approx  U\sum_i \left[ \langle n_{i\downarrow}\rangle n_{i\uparrow}+\langle n_{i\uparrow}\rangle n_{i\downarrow}-\langle n_{i\uparrow}\rangle\langle n_{i\downarrow}\rangle
-\langle c^\dagger_{i\uparrow}c_{i\downarrow}\rangle c^\dagger_{i\downarrow}c_{i\uparrow}-\langle c^\dagger_{i\downarrow}c_{i\uparrow}\rangle\right.$
$\left.\times c^\dagger_{i\uparrow}c_{i\downarrow}+\langle c^\dagger_{i\uparrow}c_{i\downarrow}\rangle \langle c^\dagger_{i\downarrow}c_{i\uparrow}\rangle \right]$.
The spin-orbit coupling breaks spin SU(2) symmetry and therefore the directions of the spins are coupled to the lattice structure.
Hence, we should parametrize the antiferromagnetic ordering between the two sublattices $A$ and $B$ [see Fig.~\ref{Fig-diamondlattice}(a)] in terms of the spherical coordinate $(n,\theta,\varphi)$:
\begin{align}
\langle \bm{S}_{i'A}\rangle=-\langle \bm{S}_{i'B}\rangle&=(n\sin\theta\cos\varphi, n\sin\theta\sin\varphi, n\cos\theta)\nonumber\\
&\equiv n_1\bm{e}_x+n_2\bm{e}_y+n_3\bm{e}_z\ (\equiv \bm{n}),
\label{AF-order}
\end{align}
where $\langle \bm{S}_{i'\mu}\rangle=\frac{1}{2}\langle c^\dagger_{i'\mu\alpha}\bm{\sigma}_{\alpha\beta}c_{i'\mu\beta}\rangle$ $(\mu=A,B)$ with $i'$ denoting the $i'$-th unit cell.
It is convenient to express the mean-field Hamiltonian in terms of the 4$\times$4 $\alpha$ matrices that anticommute with each other. 
We can define the basis $c_{\bm{k}}\equiv[c_{\bm{k}A\uparrow},c_{\bm{k}A\downarrow},c_{\bm{k}B\uparrow},c_{\bm{k}B\downarrow}]^T$ with the wave vector $\bm{k}$ in the first Brillouin zone of the fcc lattice [see Fig. \ref{Fig-diamondlattice}(b)].
Then, the single-particle Hamiltonian $\mathcal{H}_{\mathrm{MF}}(\bm{k})$ [$H_{\mathrm{MF}}\equiv\sum_{\bm{k}}c^\dag_{\bm{k}}\mathcal{H}_{\mathrm{MF}}(\bm{k})c_{\bm{k}}$] is written in the form of Eq.~(\ref{Generic-Hamiltonian-alpha_5}) \cite{Fu2007,Fu2007a}, where the alpha matrices $\alpha_i$ are given by the so-called chiral representation: 
\begin{align}
\alpha_j=
\begin{bmatrix}
\sigma_j & 0\\
0 & -\sigma_j
\end{bmatrix},\ \ \ 
\alpha_4=
\begin{bmatrix}
0 & 1\\
1 & 0
\end{bmatrix},\ \ \ 
\alpha_5=
\begin{bmatrix}
0 & -i\\
i & 0
\end{bmatrix},
\end{align}
which satisfies $\{\alpha_i,\alpha_j\}=2\delta_{ij}\bm{1}$ with $\alpha_5=\alpha_1\alpha_2\alpha_3\alpha_4$.
In the present basis, the time-reversal operator and spatial inversion (parity) operator are given by $\mathcal{T}=\bm{1}\otimes(-i\sigma_2)\mathcal{K}$ ($\mathcal{K}$ is the complex conjugation operator) and $\mathcal{P}=\tau_1\otimes\bm{1}$, respectively.
We have introduced the hopping strength anisotropy $\delta t_1$ due to the lattice distortion along the [111] direction.
Namely, we have set such that $t_{ij}=t+\delta t_1$ for the [111] direction, and $t_{ij}=t$ for the other three directions.
When $\delta t_1=0$, the system is a semimetal, i.e., the energy bands touch at the three points $X^r=2\pi (\delta_{rx},\delta_{ry},\delta_{rz})$ ($r=x,y,z$)
with $\delta_{xx}=\delta_{yy}=\delta_{zz}=1$ (and otherwise zero) indicating a Kronecker delta.
Finite $\delta t_1$ opens a gap of $2|\delta t_1|$ at the $X^r$ points.

It is notable that, in the ground state characterized by the antiferromagnetic order parameter (\ref{AF-order}), the Dirac Hamiltonians around the $X^r$ points acquire another mass induced by $\alpha_5$ which breaks both time-reversal and inversion symmetries.
In the strongly spin-orbit coupled case when the condition $Un_f\ll 2\lambda$ ($f=1,2,3$) is satisfied, we can derive the Dirac Hamiltonians around the $\tilde{X}^r$ points which are slightly deviated from the $X^r$ points \cite{Sekine2014}:
\begin{align}
\mathcal{H}_{\mathrm{MF}}(\tilde{X}^r+\bm{q})&=q_x\alpha_1+q_y\alpha_2+q_z\alpha_3+\delta t_1\alpha_4+Un_f\alpha_5.
\label{Effective-Hamiltonian-FKMH}
\end{align}
Here, the subscript $f$ can be regarded as the ``flavor'' of Dirac fermions.
This Hamiltonian~(\ref{Effective-Hamiltonian-FKMH}) has the same form as Eq.~(\ref{Effective-Hamiltonian-alpha_5}), which means that Fujikawa's method can be applied to derive the $\theta$ term in the FKMH model.
It follows that \cite{Sekine2014}
\begin{align}
\theta=\frac{\pi}{2}[1+\mathrm{sgn}(\delta t_1)]-\sum_{f=1,2,3}\tan^{-1}\left(\frac{Un_f}{\delta t_1}\right).
\label{Expression-theta-FKMH}
\end{align}
Here, note that this expression for $\theta$ is valid only when the symmetry-breaking mass $Un_f$ ($f=1,2,3$) is small so that the condition $Un_f\ll 2\lambda$ is satisfied.
In other words, the Dirac Hamiltonian of the form~(\ref{Effective-Hamiltonian-FKMH}) must be derived as the effective Hamiltonian of the system.

A comparison of the analytical result [Eq.~(\ref{Expression-theta-FKMH})] with a numerical result obtained from Eq.~(\ref{exact-theta1}) in Ref.~\onlinecite{Essin2009} has been made \cite{Sekine2014}.
In the numerical result (Fig.~\ref{Fig-Essin2009}), in which the N\'{e}el vector is set to be in the [111] direction as $n_x=n_y=n_z\equiv h/U$, the value of $\theta$ has a linear dependence on $\beta\propto h/\delta t_1$ when $Un_f/\delta t_1\ll 1$ (i.e., around $\beta=0$ or $\beta=\pi$).
Thus, the analytical result [Eq.~(\ref{Expression-theta-FKMH})] is in agreement with the numerical result when the deviation from the quantized value ($0$ or $\pi$) is small, since in Eq.~(\ref{Expression-theta-FKMH}) $\tan^{-1}(Un_f/\delta t_1)\approx Un_f/\delta t_1$ when $Un_f/\delta t_1\ll 1$.

\subsection{Values of $\theta$ in real materials from first principles}
\begin{figure}[!t]
\centering
\includegraphics[width=0.9\columnwidth]{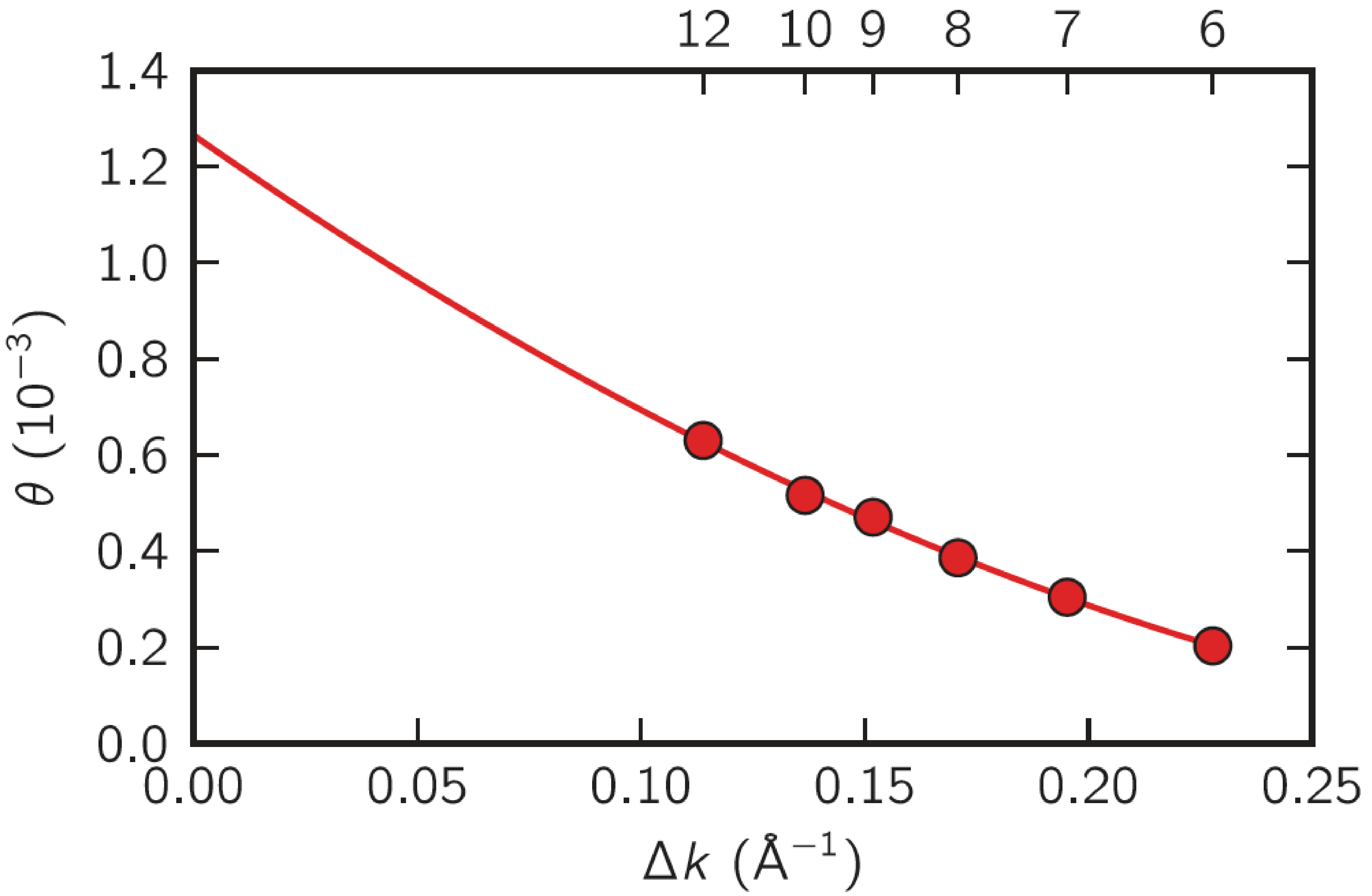}
\caption{Value of $\theta$ in Cr$_2$O$_3$ obtained from a first-principles calculation as a function of the nearest-neighbor distance on the momentum-space mesh.
The line indicates the second-order polynomial extrapolation to an infinitely dense mesh ($\Delta k\to 0$).
Adapted from Ref.~\onlinecite{Coh2011}.
}
\label{Fig-Coh2011}
\end{figure}
In real materials, there are two contributions to the linear magnetoelectric coupling: electronic and ionic (i.e., lattice) contributions.
These contributions can be further decomposed in to spin and orbital parts.
Among the electronic contribution, Eq.~(\ref{exact-theta1}) represents on an electronic orbital contribution to the {\it isotropic} linear magnetoelectric coupling.
Here, note that there exist two additional electronic orbital (but non-topological) contributions to the isotropic linear magnetoelectric coupling \cite{Essin2010,Malashevich2010}.
Cr$_2$O$_3$ is an antiferromagnetic insulator with broken time-reversal and inversion symmetries, and is well known as a material that exhibits a linear magnetoelectric effect with $\alpha_{xx}=\alpha_{yy}$ and $\alpha_{zz}$.
Figure~\ref{Fig-Coh2011} shows the value of $\theta$ in Cr$_2$O$_3$ obtained from a first-principles calculation as a function of the nearest-neighbor distance on the momentum-space mesh $\Delta k$ \cite{Coh2011}.
The value of $\theta$ extrapolated in the $\Delta k=0$ limit is $\theta=1.3\times 10^{-3}$ which corresponds to $\alpha_{ii}=0.01\ \mathrm{ps/m}$ ($i=x,y,z$).
This value is about two orders of magnitude smaller than the experimentally observed value (i.e., full response) of the linear magnetoelectric tensor in Cr$_2$O$_3$.
The values of $\theta$ in other conventional magnetoelectrics have also been evaluated in Ref.~\onlinecite{Coh2011} as $\theta=0.9\times 10^{-4}$ in BiFeO$_3$ and $\theta=1.1\times 10^{-4}$ in GdAlO$_3$, which are both very small compared to the quantized value $\pi$.
As a different approach, it has been proposed that the value of $\theta$ may be extracted from experimental observed parameters \cite{Hehl2008,Hehl2008a}.

What are the conditions for larger values of $\theta$ in real materials?
It was also shown in Ref.~\onlinecite{Coh2011} the value of $\theta$ in Cr$_2$O$_3$ is approximately proportional to the spin-orbit coupling strength, which implies that materials with strong spin-orbit coupling can have large values of $\theta$.
In addition, as we have seen in previous sections, the breaking of both time-reversal and inversion symmetries are necessary to induce the deviation of $\theta$ from the quantized values $\pi$ or $0$.
The value of $\theta$ changes continuously from $\pi$ [see Fig.~\ref{Fig-Essin2009} and Eq.~(\ref{Expression-theta-m_5-mass})].
Therefore, a system that lies near a topological insulator phase such as magnetically doped topological insulators can be one of good candidate systems.
It is notable that if a material has a large value of $\theta(\sim \pi)$, then it will exhibit a significantly large magnetoelectric effect of $\alpha_{ii}=e^2\theta/[(4\pi^2\hbar c)(c\mu_0^2)]\sim 24\ \mathrm{ps/m}$.

\section{Dynamical axion field in topological magnetic insulators \label{Sec-Dynamical-Axion-Field}}
So far, we have seen the ``static'' expressions for $\theta$ in insulators.
In other words, we have not considered what happens in a system with a $\theta$ term when the system is excited by external forces.
In general, the total value of $\theta$ can be decomposed into the sum of the static part (the ground-state value) $\theta_0$ and the dynamical part $\delta\theta(\bm{r},t)$ as
\begin{align}
\theta(\bm{r},t)=\theta_0+\delta\theta(\bm{r},t).
\end{align}
The dynamical part $\delta\theta(\bm{r},t)$ is often referred to as the {\it dynamical axion field} \cite{Li2010}, since the $\theta$ term has exactly the same form as the action describing the coupling between a hypothetical elementary particle, axion, and a photon.
Namely, $\theta(\bm{r},t)$ in condensed matter can be regarded as a (pseudoscalar) field for axion quasiparticles.
In this section, first we derive the action of axion quasiparticles in topological antiferromagnetic insulators.
Then, we consider the consequences of the realization of the dynamical axion field in condensed matter.

\subsection{Derivation of the action of axion quasiparticles}
\begin{figure}[!t]
\centering
\includegraphics[width=0.85\columnwidth]{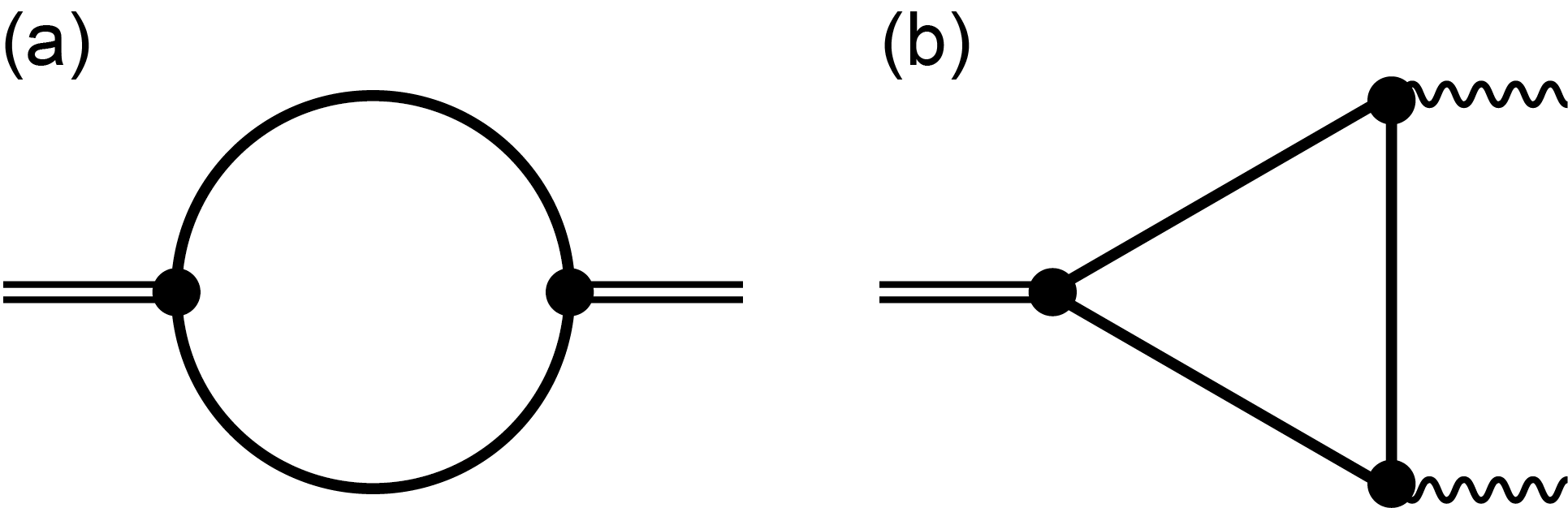}
\caption{Schematic of (a) a bubble-type Feynman diagram and (b) a triangle-type Feynman diagram.
The solid lines, wavy lines, and double lines indicate the Green's function $G_0$, the electromagnetic field $A$, and the N\'{e}el field $\delta m_5$, respectively.
}\label{Fig-diagrams}
\end{figure}

Here, following Refs.~\onlinecite{Li2010,Sekine2016}, we derive the action of axion quasiparticles in topological antiferromagnetic insulators whose effective Hamiltonian is given by a massive Dirac Hamiltonian~(\ref{Effective-Hamiltonian-alpha_5}), which is applicable to magnetically doped Bi$_2$Se$_3$ and the Fu-Kane-Mele-Hubbard model as we have seen.
In this case, the presence of the mass term $m_5\alpha_5$ that breaks time-reversal and inversion symmetries results in nonquantized values of $\theta$.
Here, let us consider the fluctuation of $m_5$ (which corresponds to the fluctuation of the N\'{e}el field) denoted by $m_5+\delta m_5$, and derive the action for $\delta m_5$.
For this purpose, it is convenient to adopt a perturbative method.
The action of the antiferromagnetic insulator phase in the presence of an external electromagnetic potential $A_\mu$ is written as [see Eq.~(\ref{Effective-action-alpha_5})]
\begin{align}
S=\int dtd^3 r\, \bar{\psi}(\bm{r},t)\left[i\gamma^\mu D_\mu-m_0+i\gamma^5(m_5+\delta m_5)\right]\psi(\bm{r},t),\label{Action-gamma5}
\end{align}
where $D_\mu=\partial_\mu-ieA_\mu$ with $e>0$ being the magnitude of the electron charge.
By integrating out the fermionic field $\psi$, we obtain the effective action $W_{\rm eff}$ for $\delta m_5$ and $A_\mu$ as
\begin{align}
Z&=\int\mathcal{D}[\psi,\bar{\psi}]\, e^{iS}\equiv e^{iW_{\rm eff}[\delta m_5, A_\mu]}\nonumber\\
&=\exp\left\{\mathrm{Tr}\ln\left[G_0^{-1}(1+G_0V)\right]\right\}\nonumber\\
&=\exp\left[\mathrm{Tr}\left(\ln G_0^{-1}\right)-\sum_{n=1}^\infty\frac{1}{n}\mathrm{Tr}\left(-G_0V\right)^n\right].
\end{align}
In order to obtain the action of the low-energy spin-wave excitation, i.e., the antiferromagnetic magnon, we set the Green's function of the unperturbed part as $G_0=(i\gamma^\mu\partial_\mu-m_0+i\gamma^5m_5)^{-1}$, and the perturbation term as $V=e\gamma^\mu A_\mu+i\gamma^5\delta m_5$.
Note that we have used that $i\gamma^\mu D_\mu-m_0+i\gamma^5(m_5+\delta m_5)=G_0^{-1}+V$.
In the random phase approximation, the leading-order terms read
\begin{align}
iW_{\rm eff}[\delta m_5, A_\mu]=& -\frac{1}{2}\mathrm{Tr}\left(G_0i\gamma^5\delta m_5\right)^2\nonumber\\
&+\mathrm{Tr}\left[\left(G_0e\gamma^\mu A_\mu\right)^2\left(G_0i\gamma^5\delta m_5\right)\right],\label{Eq-S4}
\end{align}
where the first and second terms on the right-hand side correspond to a bubble-type diagram and a triangle-type digram, respectively (see Fig.~\ref{Fig-diagrams}).

To compute the traces of the gamma matrices we use the following identities:
$\mathrm{tr}(\gamma^\mu)=\mathrm{tr}(\gamma^5)=0$, $\mathrm{tr}(\gamma^\mu\gamma^\nu)=4g^{\mu\nu}$, $\mathrm{tr}(\gamma^\mu\gamma^\nu\gamma^5)=0$, and $\mathrm{tr}(\gamma^\mu\gamma^\nu\gamma^\rho\gamma^\sigma\gamma^5)=-4i\epsilon^{\mu\nu\rho\sigma}$.
The first term in Eq. (\ref{Eq-S4}) is given explicitly by
\begin{align}
W_1&=\int\frac{d^4q}{(2\pi)^4}\, \Pi(q)\delta m_5(q)\delta m_5(-q)\nonumber\\
&\approx i J\int dtd^3r\, \left[(\partial_t \delta m_5)^2-(v_i\partial_i \delta m_5)^2-m^2(\delta m_5)^2\right].
\label{action-Neel-field}
\end{align}
Here, $J$, $v_i$, and $m$ are the stiffness, velocity, and mass of the spin-wave excitation mode, which are given respectively by \cite{Li2010}
\begin{align}
J=\left.\frac{\partial^2\Pi(q)}{\partial q_0^2}\right|_{q\to 0}=\int_{\mathrm{BZ}}\frac{d^3k}{(2\pi)^3}\, \frac{\sum_{i=1}^4 R_i^2}{16|R|^5},\\
Jm^2=\left.\Pi(q)\right|_{q\to 0}=m_5^2\int_{\mathrm{BZ}}\frac{d^3k}{(2\pi)^3}\, \frac{1}{4|R|^3},
\end{align}
where $|R|=\sqrt{\sum_{a=1}^5 R_a^2}$ and $q\to 0$ indicates the limit of both $q_0\to 0$ and $\bm{q}\to\bm{0}$.
The second term in Eq. (\ref{Eq-S4}) is the so-called triangle anomaly, which gives the $\theta$ term.
The final result is \cite{Nagaosa1996,Hosur2010}
\begin{align}
W_2=i\int dtd^3 r\, \frac{e^2}{4\pi^2 \hbar}\left[-\frac{\delta m_5(\bm{r},t)}{m_0}\right] \bm{E}\cdot\bm{B},
\label{action-theta-term}
\end{align}
from which we find that the fluctuation of the $m_5\alpha_5$ mass term behaves just as a dynamical axion field.

For concreteness, let us consider the antiferromagnetic insulator phase of Bi$_2$Se$_3$ family doped with magnetic impurities such as Fe \cite{Li2010}.
In this case, the direction of the N\'{e}el field $\bm{n}$ in the ground state is along the $z$ axis: $m_5=-(2/3)Un_z$ and $n_x=n_y=0$, where $U$ is the on-site electron-electron interaction strength.
Defining $\delta\theta(\bm{r},t)=-\delta m_5(\bm{r},t)/m_0=(2/3)U\delta n_z/m_0$ and substituting this into Eqs.~(\ref{action-Neel-field}) and (\ref{action-theta-term}), we finally arrive at the action of the axion quasiparticle:
\begin{align}
S_{\rm axion}=&\ g^2J\int dtd^3r\left[(\partial_t \delta \theta)^2-(v_i\partial_i \delta \theta)^2-m^2\delta \theta^2\right]\nonumber\\
&+\int dtd^3 r\, \frac{e^2}{4\pi^2 \hbar}\delta\theta(\bm{r},t) \bm{E}\cdot\bm{B},
\label{Action-of-axion}
\end{align}
where $g^2=m_0^2$.
Finally, we mention briefly the case of the FKMH model.
We find from Eq.~(\ref{Effective-Hamiltonian-FKMH}) that there exist three $m_{5,f}\alpha_5$ mass terms with $m_{5,f}=Un_f$ ($f=1,2,3$).
Namely, all the three spatial components of the N\'{e}el field $\bm{n}$ is contained in the kinetic part of the action of the axion field, which means that the kinetic part is described by the nonlinear sigma model for antiferromagnets \cite{Haldane1983}.
This is interesting because an effective action of an antiferromagnet is naturally derived although our original action~(\ref{Action-gamma5}) does not explicitly indicate that the mass $m_5$ corresponds to a component of the N\'{e}el field.

\subsection{Emergent phenomena from axion electrodynamics \label{Sec-Emergent-phenomena-axion}}
In the following, we consider the consequences of the realization of a dynamical axion field in condensed matter.
Among several theoretical studies on the emergent phenomena from a dynamical axion field \cite{Li2010,Ooguri2012,Sekine2016,Sekine2016a,Taguchi2018,Imaeda2019}, we particularly focus on three studies on the responses of topological antiferromagnetic insulators with a dynamical axion field $\delta\theta(\bm{r},t)$ to external electric and magnetic fields.

\subsubsection{Axionic polariton \label{Sec-Axionic-polariton}}
\begin{figure}[!t]
\centering
\includegraphics[width=\columnwidth]{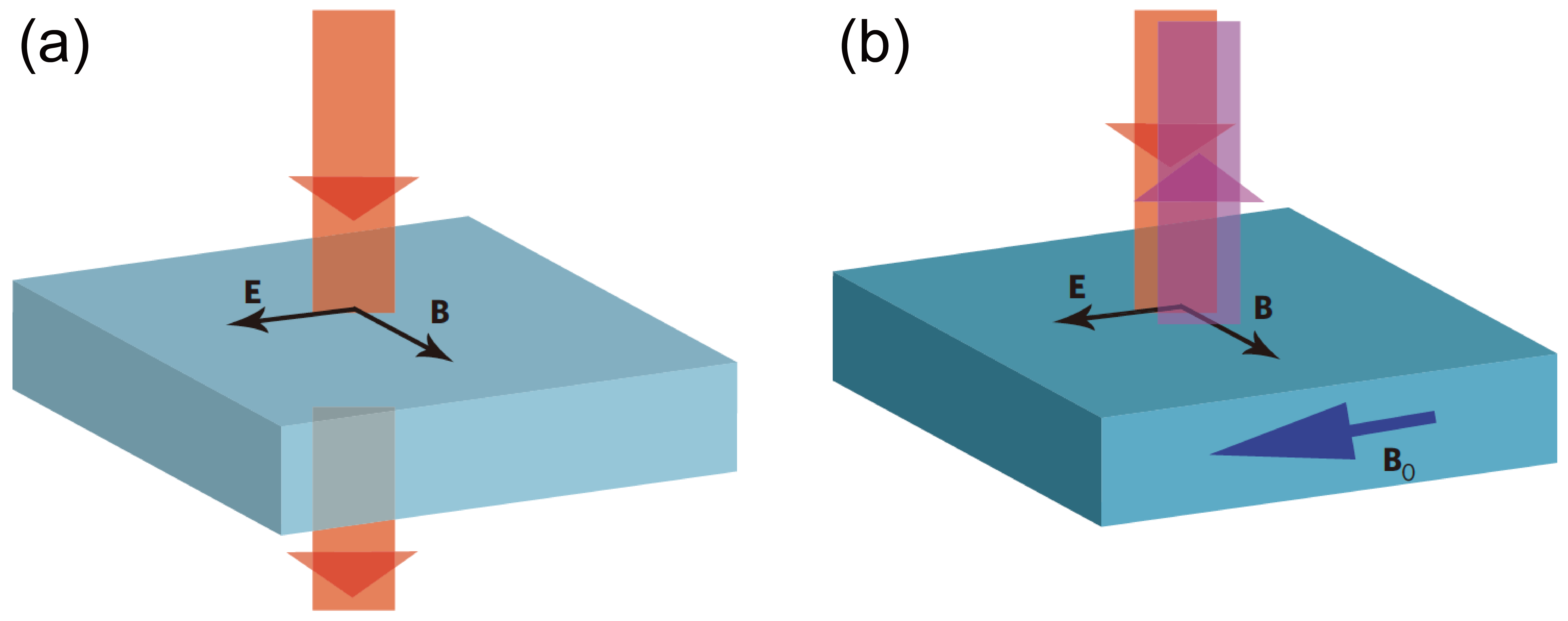}
\caption{Axionic polariton phenomenon. (a) In the absence of a static magnetic field, the incident light can transmit through the media.
(b) In the presence of a static magnetic field parallel to the electric field of light, a total reflection of incident light occurs when the incident light frequency is in the gap.
Adapted from Ref.~\onlinecite{Li2010}.
}\label{Fig-Li2010}
\end{figure}
It has been proposed that the presence of a dynamical axion field can lead to a new type of polariton, the axionic polariton \cite{Li2010}. 
To see this, we start with the total action involving an axion field $\delta\theta$ [Eq.~(\ref{Action-of-axion})] and an electromagnetic field $A_\mu=(A_0,-\bm{A})$, which is given by
\begin{align}
S=&\ g^2J\int dtd^3r\left[(\partial_\mu \delta \theta)(\partial^\mu \delta \theta)-m^2\delta \theta^2\right]\nonumber\\
&+\int dtd^3 r\, \frac{\alpha}{4\pi^2}\delta\theta \bm{E}\cdot\bm{B}-\frac{1}{16\pi}\int dtd^3 r\, F_{\mu\nu}F^{\mu\nu},
\end{align}
where $\alpha=e^2/\hbar c\simeq 1/137$ is the fine-structure constant and $F_{\mu\nu}=\partial_\mu A_\nu-\partial_\nu A_\mu$ is the electromagnetic field tensor.
Note that $\bm{E}\cdot\bm{B}=(1/8)\epsilon^{\mu\nu\rho\lambda}F_{\mu\nu}F_{\rho\lambda}$ and $F_{\mu\nu}F^{\mu\nu}=2(\bm{B}^2/\mu_0-\epsilon_0\bm{E}^2)$.
Here, recall that the classical equation of motion for a field $\phi$ is generically obtained from the Euler-Lagrange equation:
\begin{align}
\frac{\delta S}{\delta\phi}=\frac{\partial\mathcal{L}}{\partial \phi}-\partial_\mu\left(\frac{\partial\mathcal{L}}{\partial(\partial_\mu\phi)}\right)=0,
\label{Euler-Lagrange-Eq}
\end{align}
where $\mathcal{L}$ is the Lagrangian density of the system.
We consider the case of a constant magnetic field $\bm{B}=\bm{B}_0$.
Then, the equations of motion for the axion and electromagnetic fields are obtained from Eq.~(\ref{Euler-Lagrange-Eq}) as
\begin{align}
&\frac{\partial^2\bm{E}}{\partial t^2}-c'^2\nabla^2\bm{E}-\frac{\alpha}{\pi\epsilon}\bm{B}_0\frac{\partial^2\delta\theta}{\partial t^2}=0, \nonumber\\
&\frac{\partial^2\delta\theta}{\partial t^2}-v^2\nabla^2\delta\theta+m^2\delta\theta-\frac{\alpha}{8\pi^2 g^2J}\bm{B}_0\cdot\bm{E}=0,
\end{align}
where $c'$ is the speed of light in the media and $\epsilon$ is the dielectric constant.
Neglecting the dispersion of the axion field compared to the electric field $\bm{E}$, the dispersion of the electric field, i.e., the axionic polariton, $\omega_\pm(k)$, is given by \cite{Li2010}
\begin{align}
2\omega_\pm(k)=\ &c'^2k^2+m^2+b^2\nonumber\\
&\pm\sqrt{(c'^2k^2+m^2+b^2)^2-4c'^2k^2m^2}
\end{align}
with $b^2=\alpha^2\bm{B}_0^2/8\pi^3\epsilon g^2J$.
The photon dispersion in the absence of the axion field is just $\omega(k)=c'k$.
In the presence of the axion field, the photon dispersion $\omega_\pm(k)$ has two branches separated by a gap between $m$ and $\sqrt{m^2+b^2}$.
As shown in Fig.~\ref{Fig-Li2010}, this gap gives rise to a total reflection of incident light in the case when the incident light frequency is in the gap.
The point is the tunability of the axionic polariton gap by the external magnetic field $\bm{B}_0$.

\subsubsection{Dynamical chiral magnetic effect and anomalous Hall effect \label{DynamicalCME-and-AHE}}
Next, we consider an electric current response in insulators with a dynamical axion field.
To this end, we rewrite the $\theta$ term in the Chern-Simons form, which procedure becomes possible when a dynamical axion field is realized:
\begin{align}
S_\theta=-\int dt d^3r\, \frac{e^2}{8\pi^2 \hbar}\epsilon^{\mu\nu\rho\lambda}[\partial_\mu \theta(\bm{r},t)] A_\nu\partial_\rho A_\lambda.
\end{align}
Then, the induced four-current density $j^\nu$ can be obtained from the variation of the above action with respect to the four-potential $A_\nu$ as
$
j^\nu=\delta S_\theta/\delta A_\nu=-(e^2/4\pi^2 \hbar)\epsilon^{\mu \nu\rho\lambda}[\partial_\mu \theta(\bm{r},t)]\partial_\rho A_\lambda.
$
The induced electric current density and charge density are given by \cite{Wilczek1987}
\begin{align}
\bm{j}(\bm{r},t)&=\frac{\delta S_\theta}{\delta \bm{A}}=\frac{e^2}{4\pi^2 \hbar}\left[\dot{\theta}(\bm{r},t)\bm{B}+\nabla\theta(\bm{r},t)\times\bm{E}\right],\nonumber\\
\rho(\bm{r},t)&=\frac{\delta S_\theta}{\delta A_0}=-\frac{e^2}{4\pi^2 \hbar}\nabla\theta(\bm{r},t)\cdot\bm{B},
\label{Induced-current-Eq}
\end{align}
where $\dot{\theta}=\partial \theta(\bm{r},t)/\partial t$.
The magnetic-field induced current is the so-called chiral magnetic effect, which was first studied in nuclear physics \cite{Fukushima2008}.
The electric-field induced current is the anomalous Hall effect, since it is perpendicular to the electric field.
Note that the electric current [Eq.~(\ref{Induced-current-Eq})] is a bulk current that can flow in insulators \cite{Sekine2016}: the magnetic-field induced and electric-field induced currents are respectively understood as a polarization current $\partial\bm{P}/\partial t=e^2/(4\pi^2 \hbar)\dot{\theta} \bm{B}$ and a magnetization current $\nabla\times\bm{M}=e^2/(4\pi^2 \hbar)\nabla\theta\times \bm{E}$, where $\bm{P}$ and $\bm{M}$ are directly obtained from the $\theta$ term [see Eq.~(\ref{topological-ME-effect})].

\begin{figure}[!t]
\centering
\includegraphics[width=\columnwidth]{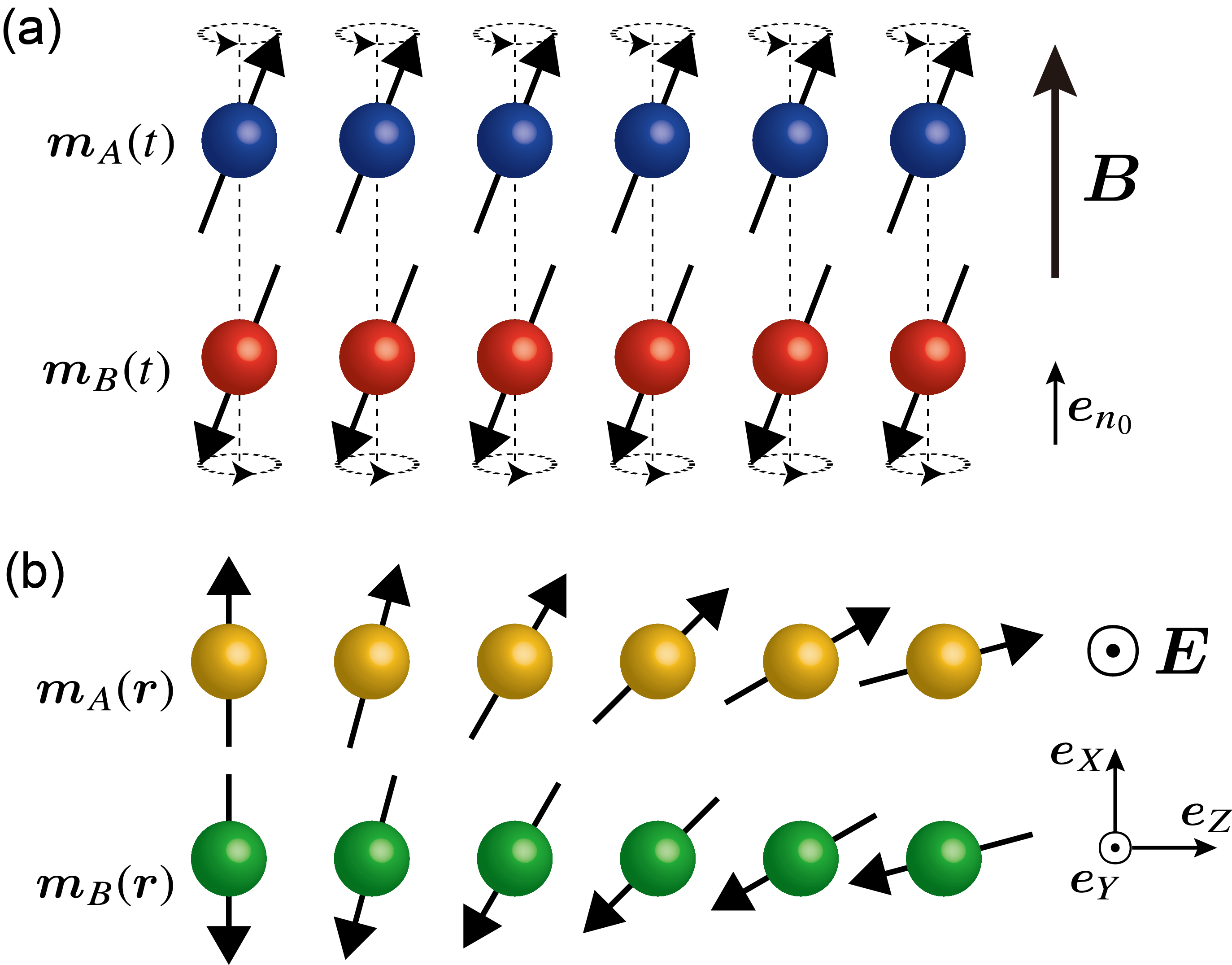}
\caption{Schematic figures of (a) an antiferromagnetic resonance state and (b) a 1D antiferromagnetic domain wall.
}\label{Fig-AF-spintexture}
\end{figure}
The electric current given by Eq.~(\ref{Induced-current-Eq}) has been studied in the antiferromagnetic insulator phase of the FKMH model \cite{Sekine2016}.
As we have seen in Eq.~(\ref{Expression-theta-FKMH}), the dynamical axion field can be realized in the FKMH model by the fluctuation of the antiferromagnetic order parameter, i.e., by the antiferromagnetic spin excitation.
The magnetic-field induced current in Eq.~(\ref{Induced-current-Eq}), i.e., the dynamical chiral magnetic effect, emerges due to the time dependence of the antiferromagnetic order parameter.
The simplest situation is the antiferromagnetic resonance.
The dynamics of the sublattice magnetizations $\langle \bm{S}_{i'A}\rangle=\bm{m}_A$ and $\langle \bm{S}_{i'B}\rangle=\bm{m}_B$ can be phenomenologically described by \cite{Keffer1952}
\begin{align}
\dot{\bm{m}}_A&=\bm{m}_A\times\left\{-\omega_J\bm{m}_B+\left[g\mu_B B+\omega_A(\bm{m}_A\cdot\bm{e}_{n_0})\right]\bm{e}_{n_0}\right\},\nonumber\\
\dot{\bm{m}}_B&=\bm{m}_B\times\left\{-\omega_J\bm{m}_A+\left[g\mu_B B+\omega_A(\bm{m}_B\cdot\bm{e}_{n_0})\right]\bm{e}_{n_0}\right\},
\label{Antiferromagnetic-dynamics}
\end{align}
where $\omega_J$ and $\omega_A$ are the exchange field and anisotropy field, respectively.
Here, we have considered the case where a microwave (i.e., ac magnetic field) of frequency $\omega_{\mathrm{rf}}$ is irradiated and a static magnetic field $\bm{B}=B\bm{e}_{n_0}$ is applied along the easy axis of the antiferromagnetic order.
In the antiferromagnetic resonance state which is realized when $\omega_{\mathrm{rf}}=\omega_\pm$, the antiferromagnetic order parameter is described as the precession around the easy axis \cite{Keffer1952}:
\begin{align}
\bm{n}_\pm(t)\equiv[\bm{m}_A(t)-\bm{m}_B(t)]/2\approx n_0\bm{e}_{n_0}+\delta\bm{n}_\pm e^{i\omega_\pm t},\label{n(t)}
\end{align}
where $\omega_\pm=g\mu_B B\pm\sqrt{(2\omega_J+\omega_A)\omega_A}$ are the resonance frequencies.
Schematic illustration of the dynamics of $\bm{m}_A$ and $\bm{m}_B$ in the antiferromagnetic resonance state is shown in Fig. \ref{Fig-AF-spintexture}(a).
Substituting the solution~(\ref{n(t)}) into the first term in Eq.~(\ref{Induced-current-Eq}), a simplified expression for the dynamical chiral magnetic effect is obtained around the phase boundary where $Un_f/M_0\ll 1$ as \cite{Sekine2016}
\begin{align}
\bm{j}_{\rm CME}(t)
=\frac{e^2}{4\pi^2 \hbar}\frac{UD_1}{M_0}\bm{B}\sum_{a=\pm}\omega_a\delta n_a\sin\left(\omega_a t+\alpha\right),\label{expression-j-CME}
\end{align}
where $D_1$ is a constant and $\delta n_\pm$ is a Lorentzian function of $\omega_{\mathrm{rf}}$.
Equation (\ref{expression-j-CME}) means that an alternating current is induced by the antiferromagnetic resonance.
The maximum value of the dynamical chiral magnetic effect (\ref{expression-j-CME}) $|j_{\rm CME}|_{\rm max}=\frac{e^2}{4\pi^2 \hbar}\frac{U|D_1|}{|M_0|}B\omega_\pm\delta n_\pm$ is estimated as $|j_{\rm CME}|_{\rm max}\sim 1\times 10^4\ \mathrm{A/m^2}$, which is experimentally observable.
It should be noted that there is no energy dissipation due to Joule heat in the dynamical chiral magnetic effect, unlike the conventional transport regime under electric fields.

The electric-field induced current in Eq.~(\ref{Induced-current-Eq}), i.e., the anomalous Hall effect, emerges due to the spatial dependence of the antiferromagnetic order parameter.
As an example, we consider a 1D antiferromagnetic spin texture of length $L$ along the $Z$ direction, an orientational domain wall \cite{Bode2006,Tveten2013}.
As shown in Fig. \ref{Fig-AF-spintexture}(b), the antiferromagnetic order parameter $\bm{n}(\bm{r})=[\bm{m}_A(\bm{r})-\bm{m}_B(\bm{r})]/2$ at the two edges has a relative angle $\delta$, resulting in $\theta(Z=0)=\theta_0$ and $\theta(Z=L)=\theta_0+\delta$ in the original spherical coordinate.
A simplified expression for the anomalous Hall effect is obtained around the phase boundary where $Un_f/M_0\ll 1$ as \cite{Sekine2016}
\begin{align}
J^X_{\rm AHE}&=\int_0^L dZ\, j^X_{\rm AHE}(Z)=\frac{e^2}{4\pi^2 \hbar}\frac{UD_2}{M_0}E_Y,
\end{align}
where $D_2(\delta)=\sum_f[n_f(\theta_0+\delta)-n_f(\theta_0)]$ is a constant and a static electric filed $\bm{E}$ is applied perpendicular to the antiferromagnetic order as $\bm{E}=E_Y\bm{e}_{Y}$.
The Hall conductivity is estimated as $\sigma_{XY}=\frac{e^2}{4\pi^2 \hbar}\frac{UD_2}{M_0}\sim 1\times 10^{-2}\  e^2/h$, which is experimentally observable.
Note that $D_2=0$ when $\delta=0$, which means that this anomalous Hall effect does not arise in uniform ground states.

\subsubsection{Inverse process of the dynamical chiral magnetic effect}
In Eq.~(\ref{expression-j-CME}) we have seen that ac current is generated by the antiferromagnetic resonance.
It is natural to consider the inverse process of the dynamical chiral magnetic effect, i.e., a realization of the antiferromagnetic resonance induced by ac electric field \cite{Sekine2016a}.
To this end, we study a continuum model of an antiferromagnet whose free energy is given by \cite{LL-book,Hals2011}
\begin{align}
F_0=\int d^3r \left[\frac{a}{2}\bm{m}^2+\frac{A}{2}\sum_{i=x,y,z}(\partial_i\bm{n})^2-\frac{K}{2}n_z^2-\bm{H}\cdot\bm{m}\right],
\end{align}
where $a$ and $A$ are the homogeneous and inhomogeneous exchange constants, respectively, and $K$ is the easy-axis anisotropy along the $z$ direction.
$\bm{n}$ and $\bm{m}$ are the N\'{e}el vector and small net magnetization satisfying the constraint $\bm{n}\cdot\bm{m}=0$ with $|\bm{n}|=1$ and $|\bm{m}|\ll 1$.
The fourth term is the Zeeman coupling with $\bm{H}=g\mu_B\bm{B}$ being an external magnetic field.
For concreteness, we consider the antiferromagnetic insulator phase of the FKMH model (see Sec.~\ref{Sec-FKMH-model}).
The $\theta$ term can be written in the free energy form [see also Eq.~(\ref{Free-energy-TME})]
\begin{align}
F_\theta=-\frac{e^2}{4\pi^2 \hbar}\frac{\sqrt{3}Un_0}{M_0}\int d^3r\, (\bm{n}\cdot\bm{e}_{[111]})\bm{E}\cdot\bm{B},
\label{F-theta}
\end{align}
where we have used the fact that $\sum_{f=1,2,3}n_f=\sqrt{3}\bm{n}\cdot\bm{e}_{[111]}$ with $\bm{e}_{[111]}$ being the unit vector along the [111] direction of the original diamond lattice in the FKMH model.

Phenomenologically, the antiferromagnetic spin dynamics can be described by the Landau-Lifshitz-Gilbert equation.
From the total free energy of the system $F_{\rm AF}=F_0+F_\theta$, the effective fields for $\bm{n}$ and $\bm{m}$ are given by $\bm{f}_n=-\delta F_{\rm AF}/\delta\bm{n}$ and $\bm{f}_m=-\delta F_{\rm AF}/\delta\bm{m}$.
The Landau-Lifshitz-Gilbert equation is given by \cite{Hals2011,Sekine2016a}
\begin{align}
\dot{\bm{n}}&=(\gamma\bm{f}_m-G_1\dot{\bm{m}})\times\bm{n},\nonumber\\
\dot{\bm{m}}&=(\gamma\bm{f}_n-G_2\dot{\bm{n}})\times\bm{n}+(\gamma\bm{f}_m-G_1\dot{\bm{m}})\times\bm{m}+\tau_{\mathrm{SP}},
\label{LLG-Eq}
\end{align}
where $\gamma=1/\hbar$, $G_1$ and $G_2$ are dimensionless Gilbert-damping parameters, and $\tau_{\mathrm{SP}}=-G_{\mathrm{SP}}(\dot{\bm{n}}\times\bm{n}+\dot{\bm{m}}\times\bm{m})$ is the additional damping torque with a spin pumping parameter $G_{\mathrm{SP}}$ \cite{Cheng2014,Takei2014}.

\begin{figure}[!t]
\centering
\includegraphics[width=\columnwidth]{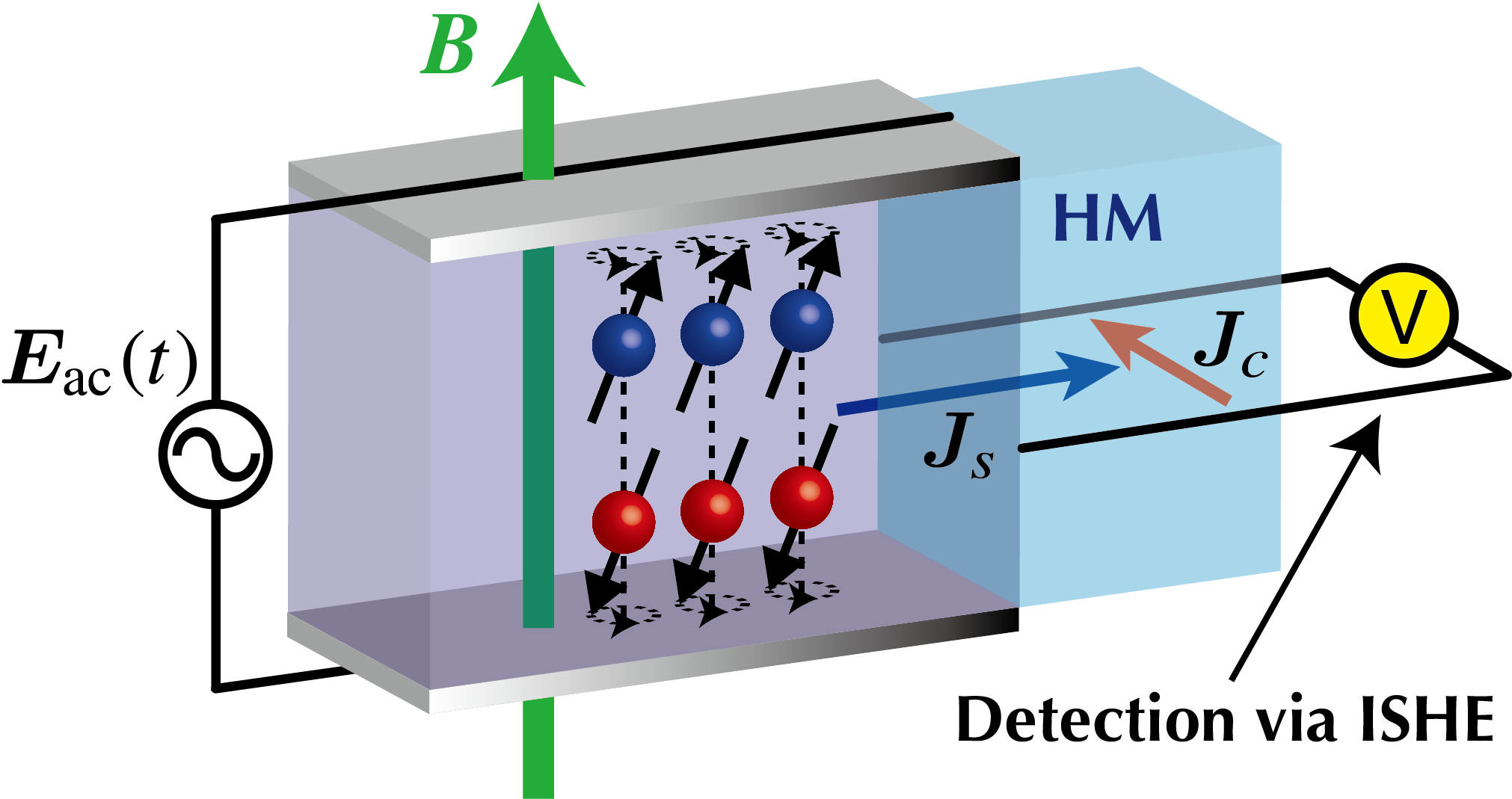}
\caption{Schematic figure of the electric-field induced antiferromagnetic resonance and its detection.
An ac electric field $\bm{E}_{\mathrm{ac}}(t)$ induces the antiferromagnetic resonance.
A dc pure spin current $\bm{J}_{s}$ generated by the spin pumping into the attached heavy metal (HM) such as Pt can be detected through the inverse spin Hall effect (ISHE) as a direct current $\bm{J}_{c}$ (i.e., the voltage $V_{\rm SP}$).
}\label{Fig-ElectricAFMR}
\end{figure}
Let us consider a case where an ac electric field $\bm{E}_{\mathrm{ac}}(t)=E_{\mathrm{ac}}e^{i\omega_{0}t}\bm{e}_z$ and a static magnetic field $\bm{B}=B\bm{e}_z$ are both applied along the easy axis.
Assuming the dynamics of the N\'{e}el field $\bm{n}(t)=\bm{e}_z+\delta\bm{n}(t)$ and the net magnetization $\bm{m}(t)=\delta \bm{m}(t)$ and solving the above Landau-Lifshitz-Gilbert equation, it is shown that the antiferromagnetic resonance can be realized by the ac electric field $\bm{E}_{\mathrm{ac}}(t)$.
The resonance frequencies are \cite{Sekine2016a}
\begin{align}
\omega_\pm=\omega_{H} \pm\sqrt{\omega_{a}\omega_{K}},
\end{align}
where $\omega_{H}=\gamma g\mu_B B$, $\omega_{a}=\gamma a$, and $\omega_{K}=\gamma K$.
The essential point is the coupling of the N\'{e}el field and the electric field through the $\theta$ term, as is readily seen in Eq.~(\ref{F-theta}).
Note that these resonance frequencies are not dependent on the parameters of the $\theta$ term.
This is because the $\theta$ term acts only as the driving force to cause the resonance.

As shown in Fig. \ref{Fig-ElectricAFMR}, in the resonance state a pure dc spin current $\bm{J}_{s}$ generated by the spin pumping is injected into the attached heavy-metal layer through the interface \cite{Cheng2014}.
The spin current is converted into an electric voltage across the transverse direction via the inverse spin Hall effect \cite{Saitoh2006}: $V_{\mathrm{SP}}(\omega_{0})\propto \alpha_{\mathrm{SH}}J_{s}(\omega_{0})$, where $\alpha_{\mathrm{SH}}$ is the spin Hall angle.
For example, in the case of $B=0.1\ \mathrm{T}$ and $E_{\mathrm{ac}}=1\ \mathrm{V/m}$ with possible (typical) values of the parameters, the magnitude of $V_{\mathrm{SP}}$ in the resonance state is found to be $V_{\mathrm{SP}}(\omega_\pm)\sim 10\ \mu\mathrm{V}$ \cite{Sekine2016a}, which is experimentally observable.
Furthermore, it should be noted that the above value of the ac electric field, $E_{\mathrm{ac}}=1\ \mathrm{V/m}$, is small.
Namely, from the viewpoint of lower energy consumption, the spin current generation using topological antiferromagnets with the $\theta$ term has an advantage compared to conventional ``current-induced'' methods that require such high-density currents as $\sim 10^{10}\ \mathrm{A/m^2}$ \cite{Brataas2012}.

\section{Topological Response of Weyl Semimetals \label{Sec-Weyl-Semimetal}}
So far we have focused on the axion electrodynamics in 3D insulators.
In this section, we overview topological responses of Weyl semimetals to external electric and magnetic fields which are described by the $\theta$ term.
Although a number of novel phenomena have been proposed theoretically and observed experimentally in Weyl semimetals \cite{Hosur2013,Burkov2015,Armitage2018}, we here focus on the very fundamental two effects, the anomalous Hall effect and chiral magnetic effect, starting from the derivation of the $\theta$ term.
We also discuss the negative magnetoresistance effect that arises as a consequence of the condensed-matter realization of the chiral anomaly.

\subsection{Derivation of the $\theta$ term in Weyl semimetals \label{Theta-term-WSM}}
The Weyl semimetals have nondegenerate gapless linear dispersions around band-touching points (Weyl nodes).
The low-energy effective Hamiltonian around a Weyl node is written as
\begin{align}
\mathcal{H}_{\mathrm{Weyl}}(\bm{k})=Q\hbar v_{\mathrm{F}}\bm{k}\cdot\bm{\sigma},
\label{3D-Weyl-Hamiltonian}
\end{align}
where $Q=\pm1$ indicates the chirality, $v_{\mathrm{F}}$ is the Fermi velocity, and $\sigma_i$ are Pauli matrices.
The two energy eigenvalues are $\pm \hbar v_{\mathrm{F}}\sqrt{k_x^2+k_y^2+k_z^2}$.
In contrast to 2D Weyl fermions such as those on the topological insulator surfaces, the 3D Weyl fermions described by Eq.~(\ref{3D-Weyl-Hamiltonian}) cannot acquire the mass, i.e., cannot be gapped, since all the three Pauli matrices are already used.
This indicates the stableness of a single Weyl node.
Because the sum of the chiralities of the Weyl nodes (or equivalently the monopoles in momentum space) in a system must be zero, the simplest realization of a Weyl semimetal is one with two Weyl nodes of opposite chiralities.
Note that the minimal number of Weyl nodes in Weyl semimetals with broken inversion symmetry is four \cite{Murakami2007}, while it is two in Weyl semimetals with broken time-reversal symmetry.

For concreteness, we consider a $4\times4$ continuum model Hamiltonian for two-node Weyl semimetals with broken time-reversal symmetry \cite{Burkov2011,Vazifeh2013,Sekine2014a,Burkov2015}
\begin{align}
\mathcal{H}_0(\bm{k})=\hbar v_{\mathrm{F}}(\tau_z\bm{k}\cdot\bm{\sigma}+\Delta\tau_x+\bm{b}\cdot\bm{\sigma}),
\label{WSM-H_eff}
\end{align}
where $\tau_i$ and $\sigma_i$ are the Pauli matrices for Weyl-node and spin degrees of freedom, respectively, and $\Delta$ is the mass of 3D Dirac fermions.
The term $\bm{b}\cdot\bm{\sigma}$ represents a magnetic interaction such as the exchange interaction between conduction electrons and magnetic impurities or the Zeeman coupling with an external magnetic field.
Note that the Hamiltonian with $\bm{b}=0$ describes a topological or normal insulator depending on the sign of $\Delta$ [see Eq.~(\ref{Hamiltonian-3DTI})].
Therefore, the above Hamiltonian~(\ref{WSM-H_eff}) can be regarded as a model Hamiltonian describing a magnetically doped (topological or normal) insulator.
Without loss of generality, we may set $\bm{b}=(0,0,b)$.
In this case the Weyl semimetal phase is realized when $|b/\Delta|>1$, and the Weyl nodes are located at $(0,0,\pm\sqrt{b^2-\Delta^2})$ \cite{Burkov2015}.

\begin{figure}[!t]
\centering
\includegraphics[width=\columnwidth]{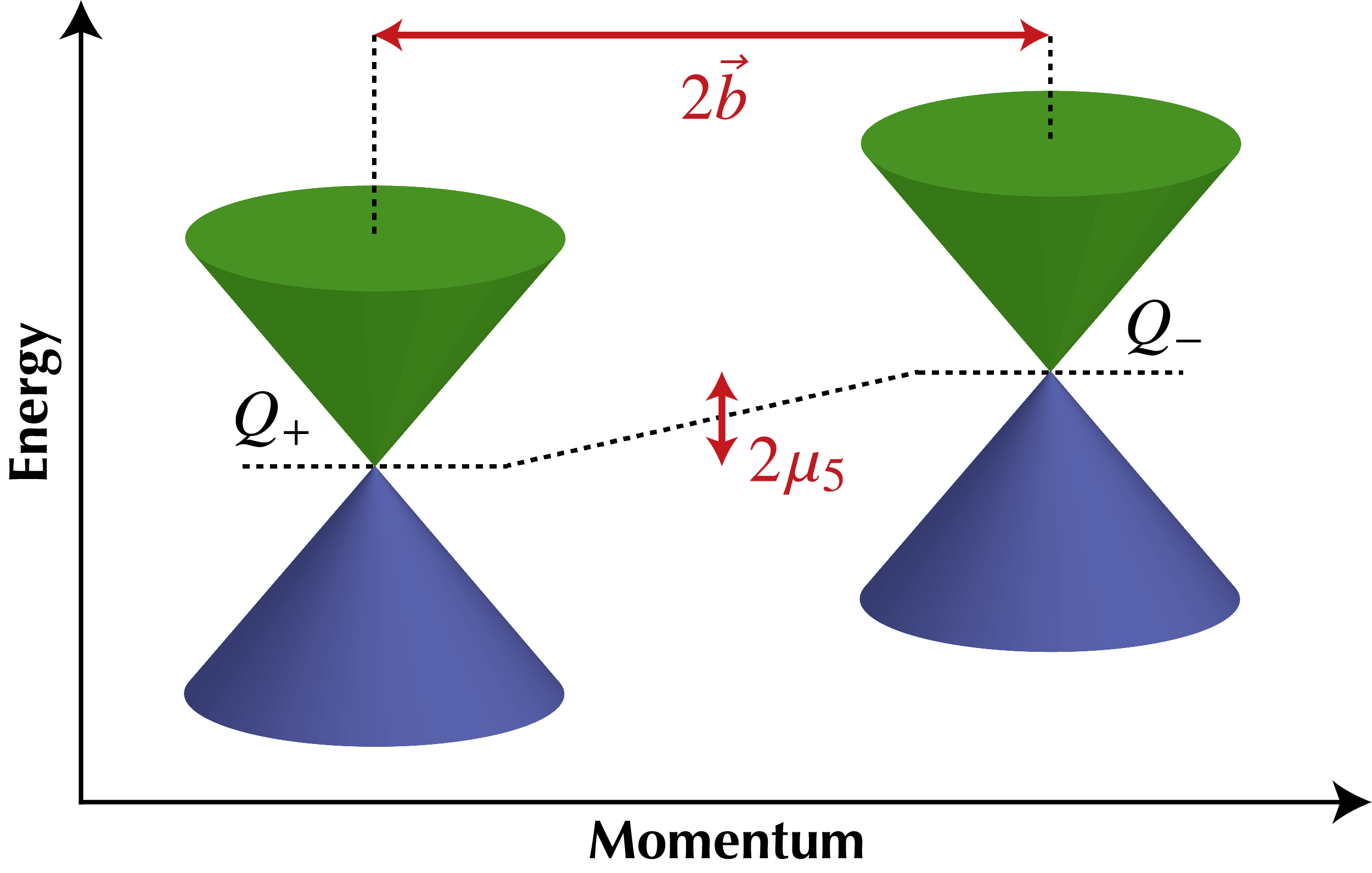}
\caption{Schematic illustration of a Weyl semimetal with two Weyl nodes.
$2\bm{b}$ and $2\mu_5$ are the momentum-space distance and the chemical potential difference between the Weyl nodes, respectively.
$Q_\pm=\pm1$ are the chiralities of the Weyl nodes.
}\label{Fig-WSM}
\end{figure}
Here, we outline the derivation of the $\theta$ term from the microscopic four-band model (\ref{WSM-H_eff}).
In order to describe a more generic Weyl semimetal, we add the term $\mu_5\tau_z$ to the Hamiltonian, which generates a chemical potential difference $2\mu_5$ between the two Weyl nodes.
Note that this term breaks inversion symmetry.
We also set $\Delta=0$ for simplicity, so that the momentum-space distance between the Weyl nodes are $2\bm{b}$.
Figure~\ref{Fig-WSM} shows a schematic illustration of the Weyl semimetal we consider.
The action of the system in the presence of external electric and magnetic fields with the four potential $A_\mu=(A_0,-\bm{A})$ is given by [see also Eq.~(\ref{action-Dirac-fermion})]
\begin{align}
S&=\int dtd^3r\, \psi^\dag\left\{i(\partial_t-ieA_0)-[\mathcal{H}_0(\bm{k}+e\bm{A})-\mu_5\tau_z]\right\}\psi \nonumber\\
&=\int dtd^3r\, \bar{\psi}i\gamma^\mu(\partial_\mu-ieA_\mu-ib_\mu\gamma^5)\psi,
\end{align}
where $e>0$, $\psi$ is a four-component spinor, $\bar{\gamma}=\psi^\dag\gamma^0$, $\gamma^0=\tau_x$, $\gamma^j=\tau_x\tau_z\sigma_j=-i\tau_y\sigma_j$, $\gamma^5=i\gamma^0\gamma^1\gamma^2\gamma^3=\tau_z$, and $b_\mu=(\mu_5, -\bm{b})$.
Now, we apply Fujikawa's method \cite{Fujikawa1979,Fujikawa1980} to the action.
The procedure is the same as that in the case of topological insulators presented in Sec.~\ref{Sec-3DTI-FujikawaMethod}.
Performing an infinitesimal gauge transformation for infinite times such that
\begin{align}
\psi\rightarrow \psi'= e^{-id\phi\theta(\bm{r},t)\gamma^5/2}\psi,\ \ \ \ \ \ \bar{\psi}\rightarrow \bar{\psi}'=\bar{\psi}e^{-id\phi\theta(\bm{r},t)\gamma^5/2},
\end{align}
with $\theta(\bm{r},t)=-2x^\mu b_\mu=2(\bm{b}\cdot\bm{r}-\mu_5 t)$ and $\phi\in [0,1]$, the action of the system becomes \cite{Zyuzin2012}
\begin{align}
S=\, &\int dtd^3r\, \bar{\psi}[i\gamma^\mu(\partial_\mu-ieA_\mu)]\psi\nonumber\\
&+\frac{e^2}{2\pi^2 \hbar}\int dtd^3 r\, (\bm{b}\cdot\bm{r}-\mu_5 t)\bm{E}\cdot\bm{B},
\label{S_effective-WSM}
\end{align}
where the first term represents the (trivial) action of massless Dirac fermions and the second term is nothing else but a $\theta$ term [Eq.~(\ref{S_theta_realtime})] with $\theta(\bm{r},t)=2(\bm{b}\cdot\bm{r}-\mu_5 t)$.
It should be noted that nonzero, nonquantized expression for $\theta$ is due to the time-reversal symmetry breaking by $\bm{b}$ and the inversion symmetry breaking by $\mu_5$.

\subsection{Anomalous Hall effect and chiral magnetic effect}
Next, let us consider the consequences of the presence of a $\theta$ term in Weyl semimetals.
As we have also seen in the case of insulators with a dynamical axion field, an electric current is induced in the presence of a $\theta$ term.
The induced electric current density and charge density are given by \cite{Wilczek1987}
\begin{align}
\bm{j}(\bm{r},t)&=\frac{\delta S_\theta}{\delta \bm{A}}=\frac{e^2}{4\pi^2 \hbar}\left[\dot{\theta}(\bm{r},t)\bm{B}+\nabla\theta(\bm{r},t)\times\bm{E}\right],\nonumber\\
\rho(\bm{r},t)&=\frac{\delta S_\theta}{\delta A_0}=-\frac{e^2}{4\pi^2 \hbar}\nabla\theta(\bm{r},t)\cdot\bm{B}.
\end{align}
In the present case of $\theta(\bm{r},t)=2(\bm{b}\cdot\bm{r}-\mu_5 t)$, we readily obtain a static current of the form
\begin{align}
\bm{j}=\frac{e^2}{2\pi^2 \hbar}\left(\bm{b}\times\bm{E}-\mu_5\bm{B}\right),
\label{Current-Eq-WSM}
\end{align}
in the ground state.
The electric-field induced and magnetic-field induced terms are the anomalous Hall effect and chiral magnetic effect, respectively \cite{Zyuzin2012,Son2012,Grushin2012,Wang2013,Goswami2013,Zhou2013,Burkov2015,Fukushima2008,Vazifeh2013}.

To understand the occurrence of the anomalous Hall effect in Weyl semimetals [the first term in Eq.~(\ref{Current-Eq-WSM})], let us consider a 2D plane in momentum space which is perpendicular to the vector $\bm{b}$.
For clarity, we set $\bm{b}=(0,0,b)$ and $\Delta=0$.
In this case, performing a canonical transformation, Eq.~(\ref{WSM-H_eff}) can be rewritten in a block-diagonal form with two $2\times2$ Hamiltonians given by \cite{Burkov2015}
\begin{align}
\mathcal{H}_{\pm}(\bm{k})=\hbar v_{\mathrm{F}}(k_x\sigma_x+k_y\sigma_y)+m_\pm(k_z)\sigma_z
\label{H_pm}
\end{align}
with $m_\pm(k_z)=\hbar v_{\mathrm{F}}(b\pm |k_z|)$.
The two Weyl nodes are located at $(0,0,\pm b)$.
It can be seen readily that $m_+(k_z)$ is always positive, and that $m_-(k_z)$ is positive when $-b \le k_z\le b$ and otherwise negative.
As we have seen in Eq.~(\ref{Surface-QHE}), the Hall conductivity of 2D massive Dirac fermions of the form~(\ref{H_pm}) is given by $\sigma_{xy}^\pm(k_z)=-\mathrm{sgn}[m_\pm(k_z)]e^2/2h$.
Therefore, we find that the total 2D Hall conductivity is nonzero in the region $-b\le k_z\le b$ and otherwise zero, which gives the 3D Hall conductivity as
\begin{align}
\sigma^{\mathrm{3D}}_{xy}=\int_{-b}^{b}\, \frac{dk_z}{2\pi} \left[\sigma_{xy}^+(k_z)+\sigma_{xy}^-(k_z)\right]=\frac{b e^2}{\pi h}.
\label{AHE-two-node-Weyl}
\end{align}
This value is exactly the same as that of the first term in Eq.~(\ref{Current-Eq-WSM}).
The expression for the anomalous Hall conductivity can be generalized straightforwardly to the case of multi-node Weyl semimetals  \cite{Yang2011}.
The anomalous Hall conductivity in two-node Weyl semimetals [Eq.~(\ref{AHE-two-node-Weyl})] is robust against disorder in the sense that the vertex correction in the ladder-diagram approximation is absent as long as the chemical potential lies sufficiently close to the Weyl nodes \cite{Burkov2014,Sekine2017}.

The chiral magnetic effect in Weyl semimetals [the second term in Eq.~(\ref{Current-Eq-WSM})] looks like a peculiar phenomenon.
The chiral magnetic effect indicates that a direct current is generated along a static magnetic field even in the absence of electric fields, when there exists a chemical potential difference $\delta\mu=2\mu_5$ between the two Weyl nodes.
If the static chiral magnetic effect exists in real materials, there will be substantial possible applications.
The existence of the static chiral magnetic effect is, however, ruled out in crystalline solids as discussed in Ref.~\onlinecite{Vazifeh2013}, which is also consistent with our understanding that static magnetic fields do not generate equilibrium currents.
As shall be discussed in detail below, the chiral magnetic effect can be realized under nonequilibrium circumstances, i.e., when the system is driven from equilibrium, for example, by the combined effect of electric and magnetic fields, which has been experimentally observed as the negative magnetoresistance in Weyl semimetals.
Another possible situation for realizing the chiral magnetic effect is applying oscillating (low-frequency) magnetic field \cite{Chang2015,Goswami2015,Ma2015,Zhong2016}.
A related current generation by oscillating magnetic field is the gyrotropic magnetic effect (natural optical activity) \cite{Ma2015,Zhong2016}, which is governed by the orbital magnetic moment of the Bloch electrons on the Fermi surface.
This is in contrast to the chiral magnetic effect which is driven by the chiral anomaly and governed by the Berry curvature \cite{Son2012}.
Finally, we note that the dynamical chiral magnetic effect in topological antiferromagnetic insulators shown in Sec.~\ref{DynamicalCME-and-AHE} is also one of the dynamical realizations of the chiral magnetic effect.

\subsection{Chiral anomaly and the negative magnetoresistance}
\begin{figure}[!t]
\centering
\includegraphics[width=\columnwidth]{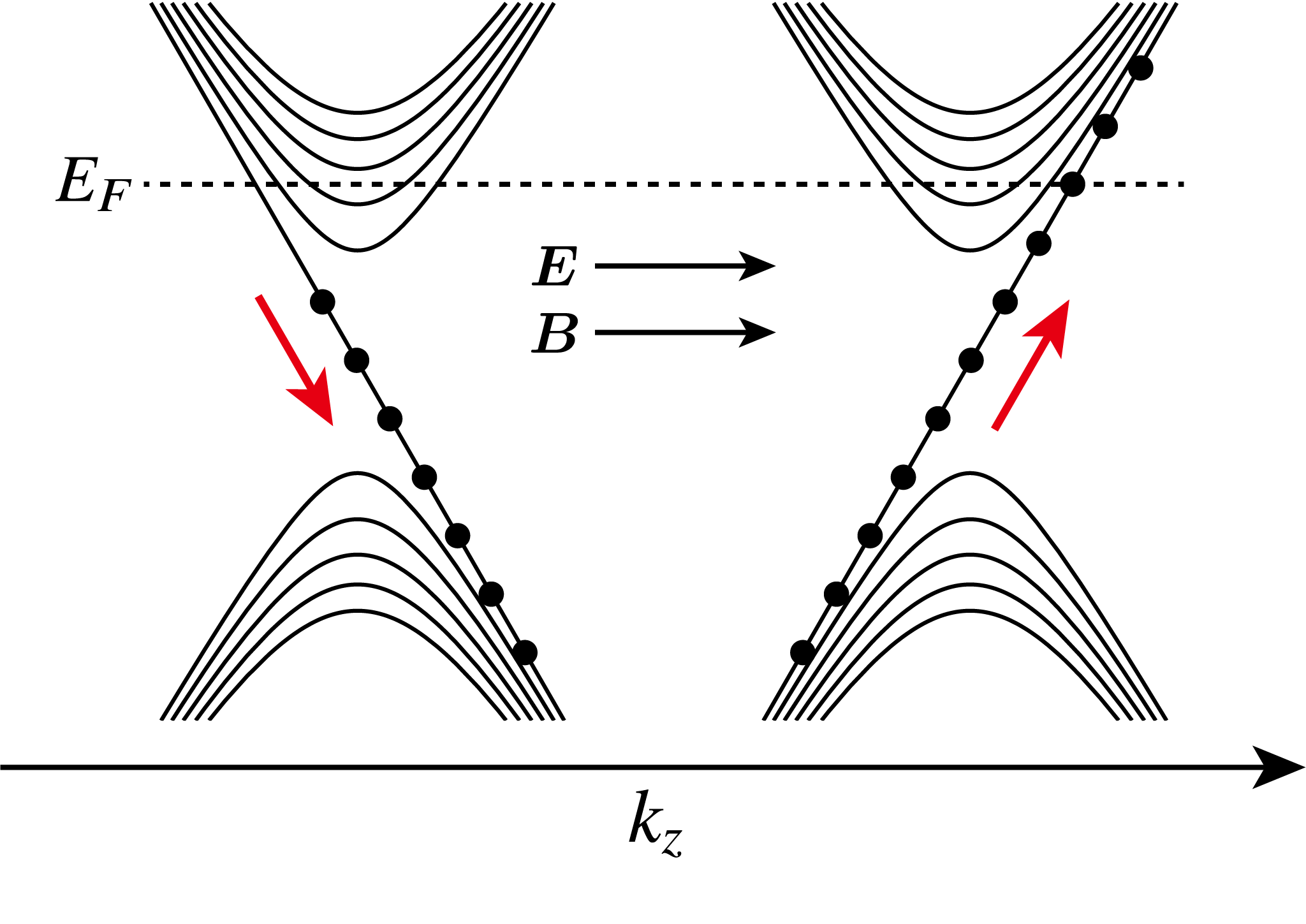}
\caption{Electron pumping due to the chiral anomaly in a Weyl semimetal under parallel electric and magnetic fields along the $z$ direction.}
\label{Fig-WSM-ChiralAnomaly}
\end{figure}
As we have seen above, the chiral magnetic effect does not occur in equilibrium.
This means that a chemical potential difference between Weyl nodes $\delta\mu=2\mu_5$ needs to be generated dynamically in order for the chiral magnetic effect to be realized in Weyl semimetals.
In the case of Weyl semimetals, such a chemical potential difference can be generated by the so-called {\it chiral anomaly}.
The chiral anomaly in Weyl semimetals is referred to as the electron number nonconservation in a given Weyl cone under parallel electric and magnetic fields, in which the rate of pumping of electrons is given by \cite{Son2012,Son2013,Sekine2017}
\begin{align}
\frac{\partial N_i}{\partial t} =  Q_i \, \frac{e^2}{4\pi^2\hbar^2 c} \bm{E}\cdot\bm{B},
\label{CA-conventional}
\end{align}
where $i$ is a valley (Weyl node) index and
\begin{align}
Q_i=\int\frac{d^3 k}{2 \pi\hbar}\frac{\partial f_{0}(\varepsilon_{\bm{k}}^m)}{\partial \varepsilon^m_{\bm{k}}}\bm{v}^m_{\bm{k}} \cdot \bm{\Omega}^m_{\bm{k}}
\label{CA-conventional}
\end{align}
is the chirality of the valley.
Here, $\varepsilon^m_{\bm{k}}$ is the energy of Bloch electrons with momentum $\bm{k}$ in band $m$ in a given valley $i$, $f_{0}(\varepsilon_{\bm{k}}^m)$ is the Fermi distribution function, $\bm{v}^m_{\bm{k}}$ is the group velocity, and $\bm{\Omega}^m_{\bm{k}}$ is the Berry curvature.
The difference of the total electron number between the Weyl nodes leads to the difference of the chemical potential between the Weyl nodes $\delta\mu$.
As shown in Fig.~\ref{Fig-WSM-ChiralAnomaly}, this electron pumping can also be understood by the electron flow through the zeroth Landau level connecting Weyl nodes of opposite chiralities induced by a magnetic field.
It should be noted here that electron pumping also occurs in parallel temperature gradient and magnetic field \cite{Spivak2016,Sekine2020}:
\begin{align}
\frac{\partial N_i}{\partial t}
=\frac{e\bm{B}\cdot\nabla T}{4\pi^2\hbar^2 c}\int \frac{d^3 k}{2 \pi\hbar}\frac{\varepsilon^m_{\bm{k}}-\mu}{T}\frac{\partial f_{0}(\varepsilon_{\bm{k}}^m)}{\partial \varepsilon^m_{\bm{k}}}\bm{v}^m_{\bm{k}} \cdot \bm{\Omega}^m_{\bm{k}},
\label{dN/dt-thermal}
\end{align}
which can be termed the thermal chiral anomaly.
Here, $T$ is the (unperturbed) temperature and $\mu$ is the chemical potential.

\begin{figure}[!t]
\centering
\includegraphics[width=0.85\columnwidth]{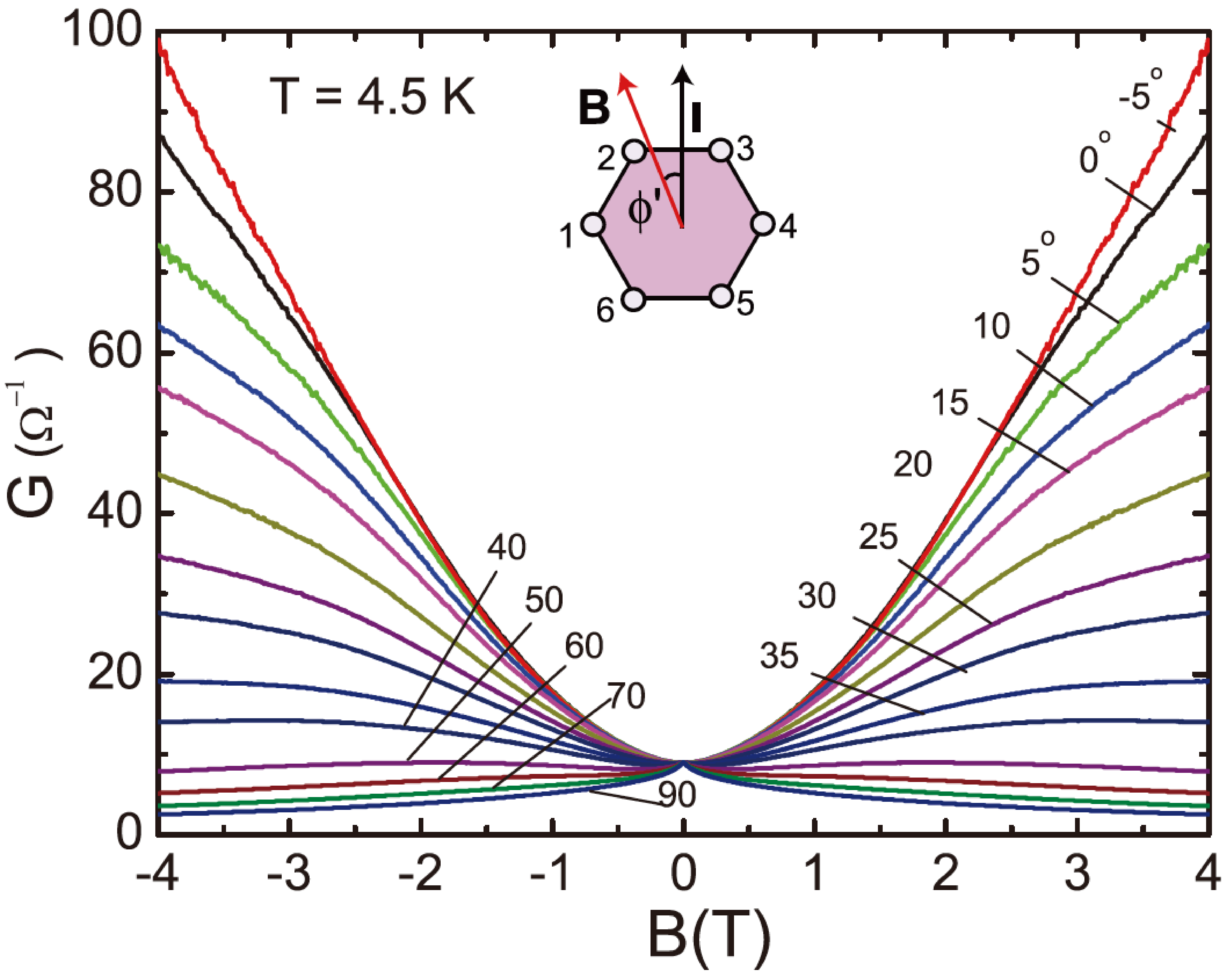}
\caption{Magnetic field dependence of the longitudinal conductance in the Dirac semimetal Na$_3$Bi.
The conductance shows a quadratic dependence on the magnetic field strength when the angle $\phi'$ between the applied current and magnetic field is small, as expected from Eq.~(\ref{Sigma_zz-ChiralAnomaly}).
Adapted from Ref.~\onlinecite{Xiong2015}.}
\label{Fig-Xiong2015}
\end{figure}
A phenomenon manifested by the chiral anomaly is a negative magnetoresistance (or equivalently positive magnetoconductance) quadratic in magnetic field for parallel electric and magnetic fields in Weyl and Dirac semimetals \cite{Son2013,Burkov2014,Spivak2016,Sekine2017}.
Here, note that the usual magnetoresistance due to Lorentz force is positive.
For concreteness, we consider the case of electric and magnetic fields along the $z$ direction.
The positive quadratic magnetoconductivity arising from the chiral anomaly reads \cite{Son2013,Burkov2014,Spivak2016,Sekine2017}
\begin{align}
\sigma_{zz}(B_z^2)=\frac{e^2}{4\pi^2\hbar c^2}\frac{(eB_z)^2v_{\mathrm{F}}^3}{\mu^2}\tau_{\mathrm{inter}},
\label{Sigma_zz-ChiralAnomaly}
\end{align}
where $\mu$ is the equilibrium chemical potential and $\tau_{\mathrm{inter}}$ is the intervalley scattering time.
This unusual magnetoconductivity holds in the low-field limit $B_z\to 0$, since it is derived from a semiclassical approach where the Landau quantization can be neglected.
Expression~(\ref{Sigma_zz-ChiralAnomaly}) is understood as coming from $\bm{j}\propto (\bm{E}\cdot\bm{B}\tau_{\mathrm{inter}})\bm{B}$, which indicates that it is a consequence of the chiral magnetic effect [the second term in Eq.~(\ref{Current-Eq-WSM})].
It has been shown that the vertex correction in the ladder-diagram approximation is absent in the positive quadratic magnetoconductivity [Eq.~(\ref{Sigma_zz-ChiralAnomaly})] \cite{Sekine2017}.
Such an unusual negative magnetoresistance has recently been experimentally observed in the Dirac semimetals Na$_3$Bi \cite{Xiong2015}, Cd$_3$As$_2$ \cite{Li2015,Li2016}, and ZrTe$_5$ \cite{Li2016a}, and in the Weyl semimetals TaAs \cite{Huang2015} and TaP \cite{Arnold2016}.
As shown in Fig~\ref{Fig-Xiong2015}, the observed conductance is positive and proportional to $B^2$ in the low-field limit as expected from Eq.~(\ref{Sigma_zz-ChiralAnomaly}).
Also, we can see that the enhancement of the conductance is largest when the angle between the applied current and magnetic field is zero (i.e., when they are parallel), which is in agreement with the theoretical prediction.
However, it must be noted here that those experimental observations of the negative magnetoresistance is now generally understood to be an artifact of ``current jetting'' \cite{Armitage2018}, which can be large in high-mobility semimetals.
The point is that disentangling precisely the intrinsic quantum effect of the chiral anomaly from the extrinsic classical effect of current jetting is not easy in experiments \cite{Liang2018}, although its presence is manifested theoretically.

\section{Gravitational Response of Topological Superconductors \label{Sec-Topological-Superconductor}}
In this section, we discuss topological responses of 3D topological superconductors and superfluids that can regarded as the thermodynamic analogue of the axion electromagnetic responses of topological insulators and Weyl semimetals.
A well-known example of 3D topological superfluids is the superfluid $^3$He $B$ phase \cite{Schnyder2008}.
The topological nature of such topological superconductors and superfluids will manifest itself in thermal transport properties, such as the quantization of the thermal Hall conductivity \cite{Read2000}, since charge and spin are not conserved while energy is still conserved.

\subsection{Derivation of a gravitational $\theta$ term \label{Derivation-Gravitational-Theta}}
The systematic classification of topologically nontrivial insulators and superconductors has been established in terms of symmetries and dimensionality, and has clarified that topologically nontrivial superconductors and superfluids with time-reversal symmetry are also realized in three dimensions \cite{Schnyder2008,Kitaev2009,Ryu2010}.
From the bulk-boundary correspondence, there exist topologically protected gapless surface states in topological superconductors.
In particular, the superconductivity infers that the gapless surface states are their own antiparticles, and thus Majorana fermions \cite{Schnyder2008}.
Because of the fact that Majorana fermions are charge neutral objects, an electric-transport study such as quantum Hall measurement cannot characterize their topological nature of topological superconductors.
Instead, since the energy is still conserved, thermal transport, especially the thermal Hall conductivity, reflects the topological character of topological superconductors as 
\begin{align}
\kappa_{\mathrm{H}}= \mathrm{sgn}(m) \frac{\pi^2}{6} \frac{k_{\mathrm{B}}^2}{2h}T
\label{Majorana-THE}
\end{align}
for the massive Majorana fermion with mass $m$ \cite{Read2000,Nomura2012}.

A spatial gradient in energy is related to a temperature gradient, as one can infer from the thermodynamic equality $dU = T dS$ as follows.
Here, $U$ is the internal energy, $S$ is the entropy, and $T$ is the temperature.
For simplicity, let us first divide the total system into two subsystems (subsystem 1 and 2). 
The equilibrium of the total system is achieved when the total entropy is maximized: $dS=dS_1+dS_2=0$.
Since the energy is conserved, $dE_2=-dE_1$, and hence $dS_1/dE_1-dS_2/dE_2=0$, i.e., $T_1=T_2$. 
Let us now turn on a gradient in the ``gravitational potential'', so that the gravitational potential felt by subsystems 1 and 2 differs by $\delta\phi_g$.
In this case, we have $dE_2=-dE_1(1+\delta\phi_g)$.
This suggests the generation of a temperature difference $T_2 = T_1(1 + \delta\phi_g)$.
In other words, we can view the “electric” field $\bm{E}_g$ associated with the gradient of $\phi_g$, which we call a “gravitoelectric field”, as a temperature gradient \cite{Luttinger1964}:
\begin{align}
\bm{E}_g = -\nabla\phi_g = -T^{-1}\nabla T.
\end{align}

In analogy with electromagnetism, let us next consider the following quantity described in terms of a vector potential $\bm{A}_g$, which we call a ``gravitomagnetic field'':
\begin{align}
\bm{B}_g=\nabla\times\bm{A}_g.
\end{align}
For example, in a system rotating with the angular velocity $\Omega^z$ around the $z$ axis, $\bm{A}_g$ can be expressed as $\bm{A}_g=(1/v)\Omega^z\bm{e}_z\times\bm{r}$ \cite{Lynden-Bell1998,Volovik-book}, which gives $\bm{B}_g=(2/v)\Omega^z\bm{e}_z$.
Here, $v$ is the Fermi velocity of the system.
Therefore, the gravitomagnetic field $\bm{B}_g$ can be understood as an angular velocity vector.
A gravitomagnetic field $\bm{B}_g$ can also be introduced as a quantity which is conjugate to the energy magnetization (momentum of energy current) $\bm{M}_E$ in the free energy of a Lorentz-invariant system \cite{Nomura2012}.
It follows that $\bm{M}_E=(v/2)\bm{L}$ with $\bm{L}$ the angular momentum in Lorentz-invariant systems, which also leads to $\bm{B}_g=(2/v)\bm{\Omega}$ \cite{Nomura2012,Volovik-book}.

\begin{figure}[t]
\centering
\includegraphics[width=\columnwidth]{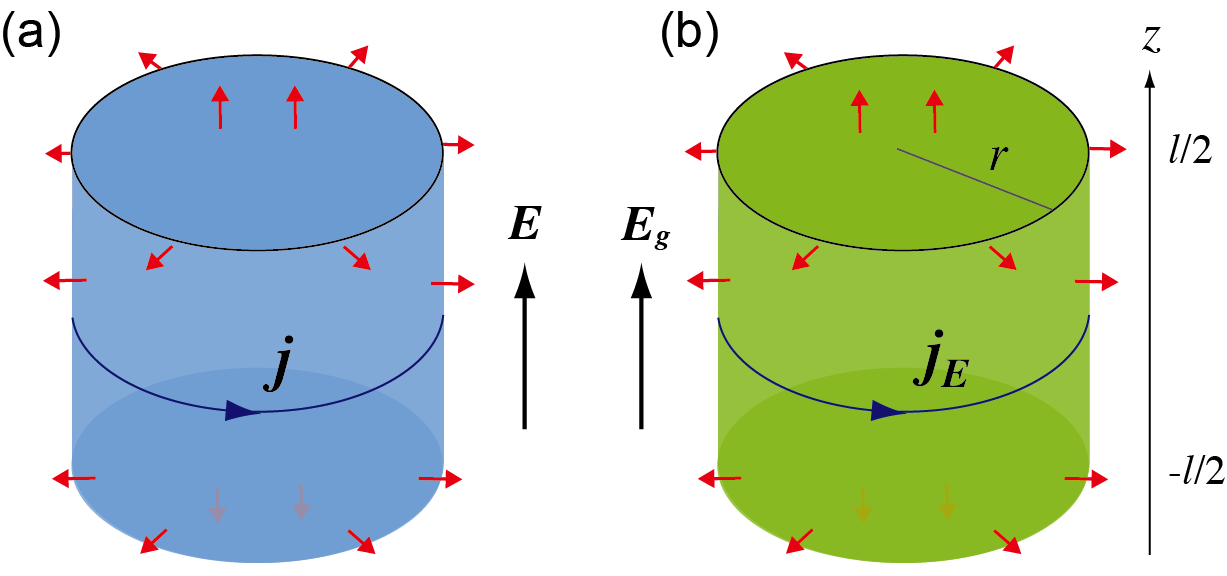}
\caption{
Electromagnetic responses in (a) 3D topological insulators and thermal and mechanical (rotation) responses in (b) 3D topological superconductors.
In (a), an electric field $\bm{E}$ induces the surface Hall current $\bm{j}$.
In (b), a temperature gradient $\bm{E}_g$ induces the surface thermal Hall current $\bm{j}_E$.
A uniform mass gap is induced in the surface fermion spectra by doping magnetic impurities near the surface of the 3D topological insulator (a) and topological superconductor (b) such that the magnetization directions are all perpendicular to the surfaces (as indicated by red arrows).
}
\label{Fig-TSC}
\end{figure}
Now, we study the responses of 3D topological superconductors to a temperature gradient $\bm{E}_g$ and a mechanical rotation $\bm{B}_g$.
For simplicity, we consider a sample in a cylindrical geometry with height $\ell$ and radius $r$ as illustrated in Fig.~\ref{Fig-TSC}(b).
We assume that magnetic impurities are doped near the surface and the magnetization directions are all perpendicular to the surface so that a uniform mass gap is formed in the surface Majorana state.
Let us first introduce a temperature gradient in the $z$ direction, which generates the energy current $j_E = \kappa_{\mathrm{H}} \partial_z T$ on the surface.
Since $j_E/v^2$ corresponds to the momentum per unit area, total momentum due to the surface energy current is 
$P_{\varphi}= (2\pi r\ell) j_E/v^2$ and thus the induced orbital angular momentum per volume is given by
\begin{align}
\left. L^z \right|^{\ }_{\Omega^z}=\frac{r P_{\varphi}}{(\pi r^2\ell)}=\frac{2}{v^2}\kappa_{\mathrm{H}}\partial_zT.
\label{14}
\end{align}
Similarly, upon rotating the cylinder with $\bm{\Omega}=\Omega^z\bm{e}_z$ (without a temperature gradient), we obtain the induced thermal energy density (the induced entropy change) localized on the top and bottom surfaces \cite{Nomura2012}
\begin{align}
\left.\Delta Q(z)\right|^{\ }_T=
\frac{2T\Omega^z }{v^2}
\left[
\kappa_{\mathrm{H}}^{\mathrm{t}}\delta(z-\ell/2)
+\kappa_{\mathrm{H}}^{\mathrm{b}}\delta(z+ \ell/2)
\right],
\label{surface Q}
\end{align}
where $\kappa_{\mathrm{H}}^\mathrm{t}$ ($\kappa_{\mathrm{H}}^\mathrm{b}$) is the thermal Hall conductivity on the top (bottom) surface given by Eq.~(\ref{Majorana-THE}).
Here, $\kappa_{\mathrm{H}}^\mathrm{t}=-\kappa_{\mathrm{H}}^\mathrm{b}$ because the magnetization directions on the top and bottom surfaces are opposite to each other, resulting in different signs of $m$ [see Fig.~\ref{Fig-TSC}(b)].

\begin{table}[t]
\caption{
Comparison between cross correlations in topological insulators (TIs) and topological superconductors (TSCs) in two and three spatial dimensions.
In topological superconductors, the orbital angular momentum $\bm{L}$ (momentum of energy current $\bm{M}_E$) and the entropy $S$ (thermal polarization $\bm{P}_E$ in three dimensions) are generated by a temperature gradient $\bm{E}_g=-T^{-1}\nabla T$ and by a mechanical rotation with angular velocity vector $\bm{\Omega}=(v/2)\bm{B}_g$.
In analogy with the orbital magnetoelectric polarizability $\chi_{\theta}^{ab}=\delta_{ab}e^2/(4\pi \hbar c)$ in 3D topological insulators, the gravitomagnetoelectric polarizability $\chi_{\theta,g}^{ab}=\delta_{ab}\pi k_{\mathrm{B}}^2T^2/(24\hbar v)$ can be introduced in 3D topological superconductors.
Note that the relations for topological superconductors applies also to the thermal response of topological insulators.}
\begin{ruledtabular}
\begin{tabular}{ccc}
& TI   &   TSC  \\
\hline
2D & $\displaystyle \sigma_{\mathrm{H}}=ec\frac{\partial M^z}{\partial \mu}=ec\frac{\partial N}{\partial B^z}$
& 
$\displaystyle \kappa_{\mathrm{H}}=\frac{v^2}{2}\frac{\partial L^z}{\partial T}=\frac{v^2}{2}\frac{\partial S}{\partial \Omega^z}$ 
\\
\\
3D & $\displaystyle \chi_{\theta}^{ab}=\frac{\partial M^a}{\partial E^b} =\frac{\partial P^a}{\partial B^b}$
&
$\displaystyle \chi^{ab}_{\theta,g}=\frac{\partial M_E^a}{\partial E^b_g}=\frac{\partial P^a_E}{\partial B_g^b}$
\end{tabular}
\end{ruledtabular}
\label{Table-TSC}
\end{table}
In terms of the gravitoelectric field $\bm{E}_g=-T^{-1}\nabla T$ and the momentum of the energy current (i.e., energy magnetization) $\bm{M}_E$, Eq.~(\ref{14}) can be written as $\bm{M}_E= (T\kappa_{\mathrm{H}}/v) \bm{E}_g$ from the relation $\bm{M}_E=(v/2)\bm{L}$. 
Furthermore, introducing the thermal polarization $\bm{P}_E$ by $\Delta Q=-\nabla\cdot \bm{P}_E$, Eq.~(\ref{surface Q}) can be written similarly as $\bm{P}_E= (T\kappa_{\mathrm{H}}/v) \bm{B}_g$.
Combining these, we find the correspondence between topological insulators and topological superconductors,
\begin{align}
\mathrm{TI:}\  \ 
\frac{\partial M^a}{\partial E^b}
=
\frac{\partial P^a}{\partial B^b}
\ \ \ 
 \Leftrightarrow
\ \ \ 
\mathrm{TSC:}\ \ 
\frac{\partial M_{E}^a}{\partial E_{g}^b}
=
\frac{\partial P^a_E}{\partial B_{g}^b}.
\end{align}
Since the orbital angular momentum is obtained from the internal energy functional as $L^a=-\delta U_{\theta}/\delta \Omega^a$, the coupling energy of the temperature gradient and angular velocity is written as \cite{Nomura2012}
\begin{align}
 U_{\theta}^g= \int d^3x\,  \frac{2}{v^2}\kappa_{\mathrm{H}}\nabla T\cdot\bm{\Omega}
=\int d^3x\,  \frac{ k_{\mathrm{B}}^2T^2}{24\hbar v} \theta_g \bm{E}_g\cdot\bm{B}_g.
\label{Internal-Energy}
\end{align}
This is analogous to the axion electromagnetic response with $e^2/\hbar c \leftrightarrow (\pi k_{\mathrm{B}} T)^2/6\hbar v$ and $\theta_g=\pi$ playing the same role as $\theta$ in the $\theta$ term.
Here, note that we have considered the contribution from one Majorana fermion to the internal energy~(\ref{Internal-Energy}).
In general, 3D time-reversal invariant (class DIII) topological superconductors with topological invariant $N$ possesses $N$ gapless Majorana fermions localized at the surface \cite{Schnyder2008}.
When uniform mass gaps (of the same sign) are induced in these Majorana fermions, each Majorana fermion gives rise to the half-integer thermal Hall effect [Eq.~(\ref{Majorana-THE})] \cite{Wang2011PRB,Shiozaki2013}.
Therefore, it follows that $\theta_g=N\pi$ in Eq.~(\ref{Internal-Energy}) for this generalized case.

In the case of 2D topological superconductors, the corresponding term is written as
\begin{align}
U=\int d^2x\,  (2/v^2)T\kappa_{\mathrm{H}}\, \phi\, \Omega^z.
\end{align}
This is the thermodynamical analogue of the Chern-Simons term.
A similar term has been derived in the context of 3D $^3$He $A$ phase with point nodes \cite{Volovik2000}, where the current flows parallel to the $\bm{\Omega}$ vector.
A Comparison between cross correlations in topological insulators and topological superconductors in two and three spatial dimensions is summarized in Table~\ref{Table-TSC}.

\subsection{Gravitational instanton term}
Here, we overview a topological field theory approach to the gravitational (thermal) response of 3D topological superconductors and superfluids \cite{Wang2011PRB,Ryu2012}.
In the previous section, we have introduced gravitoelectric and gravitomagnetic fields that are written in terms of (fictitious) scalar and vector potentials.
Strictly speaking, the presence of a gravitational background should be described as a curved spacetime.
Let us consider the Bogoliubov–de Gennes Hamiltonian of the $^3$He $B$ phase:
\begin{align}
\mathcal{H}_{\rm BdG}(\bm{k})=(\Delta_p/k_F)\bm{k}\cdot\bm{\alpha}+\xi_{\bm{k}}\alpha_4,
\label{BdG-3He-B}
\end{align}
where $k_F$ is the Fermi wave number, $\Delta_p$ is the $p$-wave pairing amplitude, $\xi_{\bm{k}}=\hbar^2 k^2/2m-\mu$ with $\mu$ the chemical potential is the kinetic energy, and $4\times4$ matrices $\alpha_\mu$ satisfy the Clifford algebra $\{\alpha_\mu,\alpha_\nu\}=2\delta_{\mu\nu}$.
Clearly Eq.~(\ref{BdG-3He-B}) is a massive Dirac Hamiltonian.
When $\mu>0$ ($\mu<0$), the system is topologically nontrivial (trivial) \cite{Schnyder2008,Qi2009}.
In the presence of such a gravitational background, the action of a 3D topological superconductor such as the $^3$He $B$ phase is written as \cite{Nakahara-book}
\begin{align}
S&=\int d^4 x\, \sqrt{-g}\mathcal{L},\nonumber\\
\mathcal{L}&=\bar{\psi}e_a^\mu i\gamma^a\left(\partial_\mu-\frac{i}{2}\omega_\mu^{ab}\Sigma_{ab}\right)\psi-m\bar{\psi}\psi,
\label{Action-Gravitational-field}
\end{align}
where $\mu=0,1,2,3$ is a spacetime index, $a,b=0,1,2,3$ is a flat index, $\sqrt{-g}=\sqrt{-\mathrm{det}(g)}$ with $g_{\mu\nu}$ the metric tensor, $e_a^\mu$ is the vielbein, $\omega_\mu^{ab}$ is the spin connection, and $\Sigma_{ab}=[\gamma_a,\gamma_b]/(4i)$ is the generator of Lorentz transformation.
As in the case of topological insulators (Sec.~\ref{Sec-3DTI-FujikawaMethod}) and Weyl semimetals (Sec.~\ref{Theta-term-WSM}), we can apply Fujikawa's method to the action~(\ref{Action-Gravitational-field}), in which the topological term of a system comes from the Jacobian.
After a calculation, we arrive at a gravitational effective action \cite{Wang2011PRB,Ryu2012}
\begin{align}
S_g=\frac{1}{1536\pi^2}\int d^4x\, \theta \epsilon^{\mu\nu\rho\sigma}\mathcal{R}^\alpha_{\beta\mu\nu}\mathcal{R}^\beta_{\alpha\rho\sigma},
\label{Gravitational-Instanton}
\end{align}
where $\theta=\pi$ and $\mathcal{R}^\alpha_{\beta\mu\nu}$ is the Riemannian curvature tensor.

It follows that the coefficient $\theta$ in Eq.~(\ref{Gravitational-Instanton}) is $\theta=0$ or $\pi$ (mod $2\pi$) due to time-reversal symmetry.
However, in 3D time-reversal invariant topological insulators with topological number $N$, topological actions should have $\theta=N\pi$.
This is because the Hamiltonian of a noninteracting 3D time-reversal invariant topological insulator with topological number $N$ can be decomposed into $N$ copies of the Hamiltonian of the form~(\ref{BdG-3He-B}).
The gravitational effective action~(\ref{Gravitational-Instanton}) provides only a $\mathbb{Z}_2$ classification of 3D time-reversal invariant topological insulators, which is weaker than the $\mathbb{Z}$ classification that they have.

\subsection{Emergent phenomena from a dynamical gravitational axion field}
As we have seen in Sec.~\ref{Derivation-Gravitational-Theta}, the derivation of the internal energy term [Eq.~(\ref{Internal-Energy})] for 3D topological superconductors and superfluids is not a microscopic derivation but a heuristic one based on the surface thermal Hall effect.
It has been suggested that the fluctuation of $\theta_g$ in a $p+is$-wave superconductor can be written as a function of the relative phase between the two superconducting gaps \cite{Goswami2014,Shiozaki2014}.
Such a fluctuation of a relative phase is known as the Leggett mode and can depend on time.
Also, in analogy with 3D topological insulators, it is expected that the internal energy term can be extended to the form of an action (see Sec.~\ref{Phenomenological-derivation} for the derivation of the $\theta$ term in 3D topological insulators).
Therefore, it would be appropriate to consider the action of the form \cite{Shiozaki2014,Sekine2016TSC}
\begin{align}
S^g_\theta=\frac{k_B^2 T_0^2}{24\hbar v}\int dt d^3 r\, \theta_g(\bm{r},t)\bm{E}_g\cdot\bm{B}_g
\label{S_theta_g}
\end{align}
for non-quantized and dynamical values of $\theta_g$, instead of the internal energy $U_g^\theta$ [Eq.~(\ref{Internal-Energy})].

In order to induce the deviation of $\theta_g$ from the quantized value $N\pi$ (with $N$ the topological number of the system), time-reversal symmetry of the bulk needs to be broken, as in the case of insulators.
It has been shown theoretically that the imaginary $s$-wave pairing in class DIII topological superconductors such as the $^3$He $B$ phase leads to the deviation of the value of $\theta_g$ from $\pi$ such that $\theta_g=\pi+\mathrm{tan}^{-1}(\Delta^{\rm Im}_s/\mu)$ with $\Delta^{\rm Im}_s$ the imaginary $s$-wave pairing amplitude \cite{Goswami2014}.
Such an imaginary $s$-wave pairing term in a Bogoliubov–de Gennes Hamiltonian corresponds to the chiral symmetry breaking term (which also breaks time-reversal symmetry) $\Gamma=\Theta\Xi$, where $\Theta$ and $\Xi$ are the time-reversal and particle-hole operators, respectively \cite{Shiozaki2014,Wang2011,Shiozaki2013}.
Therefore, the resulting superconducting state belongs to the class D \cite{Schnyder2008,Ryu2010,Altland1997}.
When we take into account the superconducting fluctuations $\Delta^{\rm Im}_s=|\Delta^{\rm Im}_s|e^{i\theta_s(\bm{r},t)}$ and $\Delta_p=|\Delta_p|e^{i\theta_p(\bm{r},t)}$, the relative phase fluctuation $\theta_r(\bm{r},t)\equiv \theta_s(\bm{r},t)-\theta_t(\bm{r},t)$, i.e., the Leggett mode, gives rise to a dynamical gravitational axion field, as $\delta\theta_g(\bm{r},t)\propto\delta\theta_r(\bm{r},t)$ \cite{Goswami2014,Shiozaki2014}.

Let us briefly consider the consequences of the realization of a dynamical gravitational axion field in 3D topological superconductors.
In the presence of a dynamical gravitational axion field $\delta\theta_g(\bm{r},t)$, a bulk heat current is obtained from the action~(\ref{S_theta_g}) as \cite{Sekine2016TSC}
\begin{align}
\bm{j}_T(\bm{r},t)=\frac{k_B^2 T_0^2}{12\hbar v}\left[\dot{\theta}_g(\bm{r},t) \bm{B}_g+v\nabla\theta_g(\bm{r},t).\times\bm{E}_g\right].
\label{thermalcurrent}
\end{align}
This expression should be compared with an electric current~(\ref{Induced-current-Eq}) obtained from the $\theta$ term in insulators.
The first term in Eq.~(\ref{thermalcurrent}) indicates that a heat current is induced in the bulk of a 3D superconductor by a gravitomagnetic field, i.e., by a mechanical rotation.
This phenomenon is called the {\it chiral gravitomagnetic effect} \cite{Sekine2016TSC}, and can be understood as the thermal analogue of the chiral magnetic effect.
The second term in Eq.~(\ref{thermalcurrent}) indicates that a heat current is induced in the bulk by a gravitoelectric field, i.e., by a temperature gradient, which is the anomalous thermal Hall effect since this current is perpendicular to the temperature gradient.

\section{Summary and Outlook \label{Sec-Summary}}
In this tutorial, we have overviewed the responses of 3D condensed-matter systems to external fields, which are described by the topological terms in their low-energy effective actions.
We have seen microscopically that the so-called $\theta$ term, which originally appeared in particle theory, is derived in topological insulators and Weyl semimetals.
In the case of insulators, the coefficient $\theta$ in the $\theta$ term takes the quantized value $\pi$ or $0$ in the presence of either time-reversal or inversion symmetry, and it can be arbitrary in the absence of both symmetries.
The $\theta$ term with $\theta=\pi$ leads to a hallmark response of topological insulators, the topological magnetoelectric effect.
We note that, in spite of intensive experimental efforts, the direct observation of the topological magnetoelectric effect, i.e., observing the electric polarization induced by a magnetic field or the magnetization induced by an electric field, in topological insulators is yet to be realized.
We have also seen that a dynamical axion field $\delta\theta(\bm{r},t)$, the deviation of $\theta$ from the ground-state value $\theta_0$, can be realized by the antiferromagnetic spin fluctuation in a class of antiferromagnetic insulators with a $\theta$ term.
In general, it is possible that the fluctuation of order parameters other than the antiferromagnetic order parameter also realizes a dynamical axion field.
In the case of Weyl semimetals, the expression for $\theta$ has a simpler form given in terms of the distance in momentum space and the energy difference between Weyl nodes.
The $\theta$ term leads to a realization of the chiral anomaly in condensed-matter systems, which has been experimentally observed in Weyl and Dirac semimetals through the negative magnetoresistance effect due to the chiral magnetic effect.

In Sec.~\ref{Sec-Axion-Insulators} we have focused on recent experimental realizations of the axion insulator state where $\theta=\pi$ due to an ``effective'' time-reversal symmetry in MnBi$_2$Te$_4$ family of materials.
The MnBi$_2$Te$_4$ family of materials are layered van der Waals compounds and thus the synthesis of few-layer thin films that can realize exotic phases and phenomena is possible.
Especially, because of the intrinsic ferromagnetism of the MnBi$_2$Te$_4$ septuple layer, the anomalous Hall conductivity of even-layer (odd-layer) thin films is zero (quantized).
Such a magnetization configuration with zero anomalous Hall conductivity is indeed the situation that has been pursued for the observation of the topological magnetoelectric effect.
Therefore, an experimental observation of the topological magnetoelectric effect might be achieved in the near future.

As we have seen in Sec.~\ref{Sec-Emergent-phenomena-axion}, the dynamical chiral magnetic effect and its inverse effect in insulators have an important feature that they are energy-saving.
The dynamical chiral magnetic effect in insulators is an ac electric current generation by a magnetic field and therefore does not cause energy dissipation due to Joule heat, although the dynamical axion field needs to be excited by external forces (which may cause energy loss).
Its inverse effect is an electrical excitation of a dynamical axion field and the applied ac electric field does not cause energy dissipation due to Joule heat because the system is insulating.
These effects might be utilized for low-energy consumption devices.

Recently, it was proposed that topological antiferromagnetic insulators with a dynamical axion field can be utilized to detect (true) axion as dark matter \cite{Marsh2019}, which will be certainly an interesting possible application of such topological antiferromagnetic insulators.
The outline of the proposal is as follows.
Inside the topological antiferromagnetic insulator, the (true) axion couples to the axionic polaritons (i.e., electric field) which are generated in the presence of axion quasiparticles (see also Sec.~\ref{Sec-Axionic-polariton}).
At the topological antiferromagnetic insulator dielectric boundary, the axionic polaritons convert to propagating photons which are finally detected in the THz regime.
Such conversion process is resonantly enhanced when the the axion frequency is equal to the axionic polariton frequency.

A microscopic derivation of the gravitational $\theta$ term [Eqs.~(\ref{Internal-Energy}) and (\ref{S_theta_g})] in the bulk of 3D topological superconductors remains an important open issue, since the derivation outlined in Sec.~\ref{Derivation-Gravitational-Theta} is based on the surface thermal Hall effect of Majorana fermions.
Similarly, it has been suggested that Weyl superconductors can exhibit the anomalous thermal Hall effect \cite{Goswami2015a} and that a gravitational $\theta$ term should also be derived in Weyl superconductors \cite{Sekine2016TSC}, considering the fact that topological insulators and Weyl semimetals are both described by the $\theta$ term.
When treating gravitoelectric and gravitomagnetic fields microscopically, we might need to introduce a torsion field \cite{Shitade2014,Gromov2015,Bradlyn2015}.

\acknowledgements
A.S. acknowledges valuable discussions with Koji Ishiwata and Makoto Naka.
A.S. is supported by the Special Postdoctoral Researcher Program of RIKEN.
K.N. is supported by JST CREST Grant No. JP-MJCR18T2 and JSPS KAKENHI Grant No. JP20H01830.



\end{document}